\documentclass[manuscript,screen]{acmart}
\usepackage{tikz}
\usetikzlibrary{shapes, arrows.meta, positioning}
\usepackage{svg}
\usepackage{adjustbox}
\usepackage[utf8]{inputenc}
\usepackage{comment}
\usepackage{array, multirow,makecell}
\usepackage{titlesec}
\graphicspath{{img/}} 
\usepackage{placeins}
\usepackage{makecell}
\usepackage{array}
\usepackage[most]{tcolorbox}
\usepackage{geometry}
\usepackage{import}
\usepackage{pifont}
\usepackage{tabularx}
\usepackage{tikz}
\usepackage{textcomp}
\usepackage{makecell}
\usepackage{xspace}
\usepackage{amsmath}
\usepackage{enumitem}
\usepackage{booktabs}
\usepackage{ragged2e}
\usepackage{rotating}
\usepackage{fontawesome5}
\usepackage{caption}
\usepackage{colortbl}
\usepackage{multirow}
\usepackage{graphicx}
\usepackage{mdframed}
\usepackage{needspace}
\usepackage{hyphenat}
\usepackage[table,xcdraw]{xcolor}
\usepackage{multirow}
\usepackage{booktabs}
\usepackage{longtable}
\usepackage{titlesec}
\usepackage{makecell,xcolor,colortbl,longtable}

\newcommand{\paintFL}{%
  \multicolumn{1}{|>{\raggedright\arraybackslash\cellcolor{FunctionLevel!14}}p{2.5cm}|}{}%
}

\newcommand{\paintRL}{%
  \multicolumn{1}{|>{\raggedright\arraybackslash\cellcolor{RepoLevel!28}}p{2.5cm}|}{}%
}

\newcommand{\paintCL}{%
  \multicolumn{1}{|>{\raggedright\arraybackslash\cellcolor{CommitLevel!18}}p{2.5cm}|}{}%
}

\titlespacing*{\subsubsection}{0pt}{1em}{1em}
\acmJournal{TOSEM}
\AtBeginDocument{%
  }

\newcolumntype{L}{>{\raggedright\arraybackslash}X}

\newcommand{\cmark}{\textcolor{green!70!black}{\ding{51}}}
\newcommand{\xmark}{\textcolor{red!80!black}{\ding{55}}}

\newcolumntype{M}[1]{>{\centering\arraybackslash}m{#1}} 
\newcolumntype{C}[1]{>{\centering\arraybackslash}m{#1}} % horizontal+vertical center
\newcolumntype{M}[1]{>{\raggedright\arraybackslash}m{#1}} % left-align+vertical center

\newcommand{\ie}{\emph{i.e.,}\xspace}
\newcommand{\eg}{\emph{e.g.,}\xspace}

\newcommand{\etal}{\emph{et~al.}\xspace}
\newcommand{\secref}[1]{Section~\ref{#1}\xspace}

\newcommand{\figref}[1]{Fig.~\ref{#1}\xspace}

\newcommand{\tabref}[1]{Table~\ref{#1}\xspace}

\newboolean{showcomments}
\setboolean{showcomments}{true}

\ifthenelse{\boolean{showcomments}}
{\newcommand{\nb}[2]{
		\fbox{\bfseries\sffamily\scriptsize#1}
		{\sf\small$\blacktriangleright$\textit{#2}$\blacktriangleleft$}
	}
	
}
{\newcommand{\nb}[2]{}
	
}

\definecolor{FunctionLevel}{RGB}{13,17,100}
\definecolor{CommitLevel}{RGB}{100,13,95}
\definecolor{ModuleLevel}{RGB}{234,34,100}
\definecolor{RepoLevel}{RGB}{247,141,96}

\newcommand{\RQ}[1]{\textit{RQ$_{#1}$}}

\begin{document}

\title{Prompt-Driven Code Summarization: A Systematic Literature Review}

\author{Afia Farjana}
\email{afarjana@wm.edu}
\affiliation{%
  \institution{William \& Mary}
  \city{Williamsburg}
  \state{Virginia}
  \country{USA}
}

\author{Zaiyu Cheng}
\email{zcheng06@wm.edu}
\affiliation{%
  \institution{William \& Mary}
  \city{Williamsburg}
  \state{Virginia}
  \country{USA}
}

\author{Antonio Mastropaolo}
\affiliation{%
  \institution{William \& Mary}
  \city{Williamsburg}
  \state{Virginia}
  \country{USA}
}
\email{amastropaolo@wm.edu}

\renewcommand{\shortauthors}{Farjana \etal}

\begin{abstract}
Software documentation is essential for program comprehension, developer onboarding, code review, and long-term maintenance. Yet producing quality documentation manually is time-consuming and frequently yields incomplete or inconsistent results. Large language models (LLMs) offer a promising solution by automatically generating natural language descriptions from source code, helping developers understand code more efficiently, facilitating maintenance, and supporting downstream activities such as defect localization and commit message generation. However, the effectiveness of LLMs in documentation tasks critically depends on how they are prompted. Properly structured instructions can substantially improve model performance, making prompt engineering--the design of input prompts to guide model behavior--a foundational technique in LLM-based software engineering. Approaches such as few-shot prompting, chain-of-thought reasoning, retrieval-augmented generation, and zero-shot learning show promise for code summarization, yet current research remains fragmented. There is limited understanding of which prompting strategies work best, for which models, and under what conditions. Moreover, evaluation practices vary widely, with most studies relying on overlap-based metrics that may not capture semantic quality. This systematic literature review consolidates existing evidence, categorizes prompting paradigms, examines their effectiveness, and identifies gaps to guide future research and practical adoption.
\end{abstract}





\maketitle

\section{Introduction}

Software Engineering (SE) is experiencing a significant shift driven by the increasing adoption of Artificial Intelligence (AI) and, in particular, Large Language Models (LLMs) in everyday development workflows. These technologies have made it possible to automate tasks that were once thought to require exclusively human expertise and ingenuity, reshaping how software is designed, documented, and maintained \cite{hou2024large}.  

At the forefront of this shift are LLMs--large neural models trained on vast corpora of source code and natural language--that exhibit remarkable ability to capture code semantics and produce context-aware outputs across a wide spectrum of SE activities. Among the various \emph{generative tasks} enabled by LLMs are code generation \cite{li2022competition}, bug fixing \cite{sobania2023analysis}, and program translation \cite{huang2023program}, as well as many others.
For a comprehensive overview of LLM adoption in software engineering, we refer readers to the systematic literature review by Watson \etal \cite{watson2022systematic}and the more recent SLR by Hou \etal.\cite{hou2024large}. These tasks are commonly organized into three broad categories: \emph{code2code}, which involves transforming one code artifact into another (\eg program repair, refactoring, or translation across programming languages); \emph{NL2Code}, where natural language instructions are mapped to executable code (\eg generating implementations from problem descriptions or API specifications); and \emph{Code2NL}, which translates source code into natural language representations (\eg code summarization, documentation generation, or explanation). This taxonomy captures the main modalities through which LLMs mediate between code and natural language and highlights their versatility in supporting different aspects of the software development lifecycle.  

Within the Code2NL category, \emph{code summarization}--the automatic generation of succinct, natural language descriptions of code snippets, functions, or entire classes--stands out for its remarkable effectiveness and practical utility. It not only demonstrates the models' ability to comprehend complex code structures but also directly supports critical development activities such as documentation, on-boarding, and maintenance \cite{moreno2013automatic,zhu2019automatic}.  

Beyond traditional documentation, advances in LLM-driven summarization are increasingly applied to related activities such as documentation generation \cite{allamanis2016convolutional}, code review \cite{tufano2024deep}, and defect analysis \cite{white2015toward}. Empirical evidence highlights that effective summaries can streamline processes that were once highly manual and error-prone \cite{buse2010automatically}, improve developer productivity \cite{zitouni2013simulated}, and ultimately enhance software quality across diverse projects and programming languages \cite{li2023commit}.  

Despite these advances, harnessing the full potential of LLMs for code summarization remains a non-trivial challenge \cite{sun2024source}. While general-purpose LLMs--such as GPT-4 \cite{achiam2023gpt}, Gemini \cite{team2024gemma}, and Claude \cite{enis2024llm}--are capable of producing fluent and context-aware explanations, the relevance, accuracy, and clarity of their outputs depend heavily on the prompts used to guide them. Unlike conventional models designed for a single fixed task, LLMs are fundamentally prompt driven: their behavior is highly sensitive to the wording, context, and structure of the input prompt \cite{webson2022prompt,lu2024prompts,zhou2022large}. Crucially, no parameters are updated during the inference phase (\ie no backpropagation procedure takes place); instead, the model (re)-calibrates its behavior ``on the fly'' according to the specifications described in the prompt, which effectively serves as a natural language description of the task at hand.

Unlike conventional models designed for a single fixed task, LLMs are fundamentally multi-task learners \cite{radford2019language}; as a result, the prompt effectively determines what task the model believes it should perform and how it should respond. Consequently, even subtle variations in wording, context, or structure can shift the model's focus and assumptions, substantially altering its behavior \cite{webson2022prompt,lu2024prompts,zhou2022large}.

This sensitivity has given rise to \emph{prompt engineering}, a rapidly evolving discipline in AI and SE research that focuses on designing, optimizing, and evaluating prompts to unlock more reliable and effective model performance. As the field moves toward increasingly capable models--with GPT-5 \cite{roziere2023code} and its contemporaries promising even greater fluency, reasoning, and code understanding--the role of prompt engineering becomes even more critical. It serves as the key interface between human intent and model output, determining whether these advanced systems can be harnessed effectively for code summarization and related documentation tasks.  

Prompt engineering encompasses a wide range of strategies for steering LLMs, including zero-shot prompting, few-shot prompting, chain-of-thought reasoning, retrieval-augmented prompting, and template-based designs \cite{brown2020gpt3,parvez2021retrieval,sun2024source}. Each presents distinct opportunities and trade-offs, influencing both the quality of generated summaries and their alignment with human intent and domain-specific requirements \cite{geroimenko2025key,chen2025unleashing}.  

Yet, the research landscape remains fragmented. Prompt design is often treated as a secondary concern, with little systematic synthesis of what works, for which models, or under which conditions. The absence of standardized benchmarks, evaluation metrics, and reporting practices further complicates comparisons across studies~\cite{roy2021reassessing,haque2022semantic,mastropaolo2024evaluating} leaving both researchers and practitioners uncertain about how to adapt and assess prompt strategies in practice.

These challenges are amplified by the heterogeneity of LLM architectures--encoder-only (\eg CodeBERT \cite{feng2020codebert}, GraphCodeBERT \cite{guo2020graphcodebert}), encoder-decoder (\eg CodeT5 \cite{wang2021codet5}, PLBART \cite{,ahmad2021unified}), and decoder-only (\eg GPT-4 \cite{achiam2023gpt}, CodeLlama \cite{roziere2023code}, DeepSeek-Coder \cite{guo2024deepseek})--each with distinct capabilities and constraints. Further complexity arises from variations in code granularity, domain specificity, and the demand for reproducibility in settings where non-determinism is the norm rather than the exception. Together, these factors constitute a clear \textit{call to action} for a systematic synthesis of prompt engineering strategies for LLM-based code summarization.  

The research presented in this manuscript provides the first systematic literature review dedicated to prompt engineering for code summarization with LLMs. We consolidate fragmented evidence across the field, categorize the paradigms proposed thus far, and examine their effectiveness and evaluation practices. By distilling best practices and identifying open challenges, our study highlights not only what is currently known but also where significant gaps remain, thereby providing a structured foundation to guide future research efforts and inform practical adoption in software engineering contexts \cite{sun2024source,xu2022systematic}.

\begin{itemize}
    \item \RQ{1}: How do code granularity and prompting paradigm choices influence prompt-based code summarization performance?
% Which granularity levels have been explored in prompt-based code summarization research?
    
     \item \RQ{2}:  How are prompt engineering paradigms instantiated in LLM-driven code summarization, and which design dimensions characterize their implementations?

    \item \RQ{3}: Are certain large language models consistently adopted across different code-summarization tasks and prompting paradigms?
    
    \item \RQ{4}: Which benchmarks and metrics have been used to assess the quality of prompt-based code summarizers, and how do these measures align with human and LLM evaluations? 
        
    \item \RQ{5}: What is the state of reproducibility and resource sharing in prompt-based code summarization research?  
    % \item \RQ{6}: How do prompt-based summaries align with human judgments of quality, usefulness, or readability in real development contexts?  
\end{itemize}

The remainder of this paper is organized as follows. \secref{sec:background} provides background on LLM-based code summarization and prompt engineering. \secref{sec:methodology} describes our SLR methodology, including search strategy, study selection, and data extraction. \secref{sec:results} presents our findings organized by research question, each concluding with key takeaways and directions for future research. 

All metadata, extraction forms, Python scripts, and visualization code supporting this SLR are publicly available at: \url{https://github.com/afia2023/prompt-engineering}
\section{Background}
\label{sec:background}
In this section, we first review recent advancements in LLMs and their application to code summarization, highlighting how these models have transformed automated documentation and software comprehension. We then present an in-depth discussion of prompt engineering techniques, detailing their development, methodological diversity, and significance in optimizing LLM performance for code summarization tasks.

\subsection{Evolution of Code Summarization}

In today's software development landscape, rising source code complexity increases the demand for tools that support documentation and comprehension. Code summarization--the automatic generation of concise natural language descriptions for code--has become essential for improving maintainability, streamlining onboarding, and aiding code review \cite{zhang2024review}.

Over the past two decades, research in code summarization has progressed through three major phases. First, classical approaches relied on rule-based and information-retrieval techniques--using heuristics, templates, TF-IDF, and BM25 models--to surface relevant tokens or phrases from code \cite{zhang2022survey}. These methods, while interpretable, were largely extractive and limited in generality.

The neural era followed, with sequence-to-sequence models leveraging RNNs (such as LSTMs and GRUs) reframing summarization as a translation task, and generating more fluent, semantically relevant summaries \cite{iyer2016summarizing}. Graph-based neural models subsequently advanced this field by incorporating code structure via ASTs or program graphs using GNNs, capturing dependencies such as control flow and data flow more effectively \cite{leclair2020improved,ahmad2020transformer}.

The advent of Transformer-based architectures further revolutionized code summarization. Transformers excel at modeling long-range dependencies and allow parallel computation. They have been adapted for code by incorporating subtoken embeddings, structure-aware attention, and specialized handling of code modalities, leading to state-of-the-art performance \cite{ahmad2020transformer,liu2020retrieval}.

For this review, our scope begins with the neural phase onward--covering seq2seq, graph-based, Transformer, and LLM-based methods--focusing on the automatic generation of natural language summaries for code snippets and commit changes. Readers interested in the classical pre-neural methods may refer to the systematic literature review by Zhu and Pan \etal which surveyed automatic code summarization up to January 2019 \cite{zhu2019automatic}.

\subsubsection{Large Language Models and Code-Specific Architectures}
The most recent and profound advances in code summarization are driven by LLMs pre‑trained on massive, mixed corpora of source code and natural language. Through this pre‑training, LLMs acquire structural, semantic, and contextual representations that are crucial for producing high‑quality summaries \cite{brown2020gpt3,fang2025enhanced}. Unlike earlier neural approaches that required gradient‑based fine‑tuning on task‑specific $\langle\text{code},\text{comment}\rangle$ pairs--often on relatively small datasets--LLMs such as GPT‑3, Codex \cite{safavi2020codex}, and StarCoder \cite{li2023starcoder} can be adapted to new programming tasks, including summarization, by \emph{prompting} alone \cite{sun2024source}--allowing an ``on-the-fly'' calibration of the model's behaviour.

In this study, we focus on prompt-based LLMs summarization, which can be viewed as a form of ``training by prompting'' performed entirely at inference time. In this setting, the model's parameters remain fixed--no backpropagation or gradient updates are involved--and adaptation arises solely from the way information is framed in the prompt (\eg through instructions, few-shot demonstrations, retrieved examples, or step-by-step reasoning). Accordingly, we exclude approaches that involve updating model parameters for summarization (such as task-specific fine-tuning, PEFT/LoRA \cite{shi2025efficient}, or RLHF \cite{christiano2017deep}) and instead focus on how prompt design choices and retrieval strategies shape the performance and behavior of LLM-based code summarizers.

These models can be broadly categorized into three architectural families:
\begin{itemize}
  \item \textbf{Encoder‑only models} (\eg CodeBERT, GraphCodeBERT) learn rich, bidirectional representations of code tokens and structure. They excel at extraction and matching tasks such as code search, clone detection, and retrieval‑based summarization pipelines \cite{feng2020codebert,guo2020graphcodebert}.
  \item \textbf{Encoder–decoder models} (\eg CodeT5, PLBART, T5) combine bidirectional encoding with autoregressive decoding, making them well‑suited to sequence‑to‑sequence summarization and translation. They have achieved strong performance on generating concise, human‑like summaries that aid comprehension and onboarding \cite{wang2021codet5,ahmad2021plbart,raffel2020exploring,hu2018deepcom,leclair2020code,lu2021codexglue}.
  \item \textbf{Decoder‑only models} (\eg GPT‑4, StarCoder) specialize in generative tasks and can produce summaries, completions, and contextual explanations directly from prompts, enabling zero‑ and few‑shot summarization without any parameter updates \cite{achiam2023gpt,li2023starcoder}.
\end{itemize}

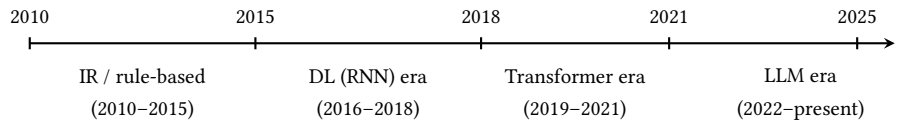
\begin{figure}[h]
  \centering
  \begin{tikzpicture}[>=stealth, line width=0.8pt, font=\small]

    % main arrow: extend slightly past last tick
    \draw[->] (0,0) -- (11.5,0);

    % year ticks and labels
    \foreach \x/\year in {0/2010,3/2015,6/2018,8.5/2021,11/2025}{
      \draw (\x,0.08) -- (\x,-0.08);
      \node[above=2pt] at (\x,0.08) {\year};
    }

    % era labels under the line
    \node[below=6pt]  at (1.5,0) {IR / rule-based};
    \node[below=18pt] at (1.5,0) {(2010--2015)};

    \node[below=6pt]  at (4.5,0) {DL (RNN) era};
    \node[below=18pt] at (4.5,0) {(2016--2018)};

    \node[below=6pt]  at (7.25,0) {Transformer era};
    \node[below=18pt] at (7.25,0) {(2019--2021)};

    \node[below=6pt]  at (10.25,0) {LLM era};
    \node[below=18pt] at (10.25,0) {(2022--present)};

  \end{tikzpicture}
  \caption{Evolution timeline of code summarization paradigms (2010--2025), based on Zhu \& Pan~\cite{zhu2019automatic} and later model milestones (CodeBERT, CodeT5, GPT, StarCoder, CodeLlama).}
  \label{fig:timeline}
\end{figure}

\figref{fig:timeline} illustrates the evolution of code summarization research from 2010 to 2025. Building on Zhu and Pan’s \cite{zhu2019automatic} review of 41 studies covering 2010 to 2019, which traced a shift from rule-based and information retrieval (IR) methods to neural RNN models. After 2016, we extend the timeline to capture later paradigm transitions. Early studies between 2010 and 2018 were dominated by rule-based and RNN methods using encoder-only architectures that processed code sequentially without generative decoding. Transformer-based code summarization emerged in 2019 with the work of Ahmad \etal \cite{ahmad2020transformer}, which introduced self-attention architectures for capturing long-range dependencies in code. This shift was further advanced by pretrained transformer models such as CodeBERT \cite{feng2020codebert} and CodeT5 \cite{wang2021codet5}, which produced more context-aware and fluent summaries. From 2022 onward, decoder-only LLMs such as GPT \cite{achiam2023gpt}, StarCoder \cite{li2023starcoder}, and CodeLlama \cite{roziere2023code} drove a rapid expansion, surpassing earlier paradigms. Overall, the timeline highlights a steady progression from encoder-only and RNN methods to encoder–decoder transformers and finally to LLM-based architectures with greater generative capability in code summarization.

% \begin{figure}[h]
% \centering
% \resizebox{0.7\textwidth}{!}{%
% \ttfamily
% \begin{tabular}{l}
% 2010 ─────────┬────────────┬──────────────┬──────────────▶ \\
% \quad IR-driven \hspace{1em} DL-Based dominance begins \hspace{1em} Transformer Era \hspace{1em} LLM Era \\
% \quad (2010–2015) \hspace{2em} (2016–2018) \hspace{2em} (2019–2021) \hspace{2em} (2022–present)
% \end{tabular}
% }
% \caption{Evolution timeline of code summarization paradigms (2010–2025) based on Zhu \& Pan (2019) and later model milestones (CodeBERT, Codex, CodeLlama).}
% \label{fig:timeline}
% \end{figure}

% in preamble (if not already there)

While all three architectural types demonstrate remarkable code comprehension capabilities, decoder-only models are particularly well-suited for prompt-based code summarization. Their autoregressive structure inherently matches the linear nature of both programming code and human language. These models have become predominant in the LLM field, largely due to their ability to effectively model long-distance relationships between tokens across varying contexts, including single functions to complete codebases. These characteristics have established decoder-only architectures as the ``de facto'' standard foundation for supporting prompt-driven source code documentation.

Prominent examples of applications that have been built on the backbone of large decoder models are: Codex \cite{chen2021evaluating}, which powers GitHub Copilot\footnote{GitHub Copilot is a widely used AI pair programmer that suggests code and documentation in real time within IDEs such as Visual Studio Code and JetBrains.}, as well as newer models like Gemini \cite{team2024gemma} and StarCoder \cite{li2023starcoder}, which leverage massive corpora to achieve competitive performance in summarization and related tasks \cite{brown2020gpt3,team2024gemma}. Recent evidence indicates that decoder-only models increasingly compete with, and in some cases surpass, encoder-decoder architectures for code summarization \cite{touvron2023llama}, further strengthened by advances in adaptation techniques such as instruction tuning, domain specialization, and prompt engineering \cite{wei2021finetuned,ouyang2022training,wang2022selfinstruct}.  

% \begin{figure}[ht]
%     \centering
%     \includesvg[width=0.92\textwidth]{img/publication_trends.svg}
%     \caption{Estimated publication trends in code summarization research by model type from 2015 to 2025.\textsuperscript{*}}
%     \label{fig:Code-Summarization-Evaluation}
% \end{figure}

% \footnotetext{* Publication counts are estimated by synthesizing results from key systematic literature reviews, manual searches of digital libraries (DBLP, arXiv, IEEE Xplore), and analysis of milestone papers in code summarization. The data illustrate overall research trends rather than exact totals.}

Despite these advances, to push the boundaries in code summarization activities, prompts, and thus natural language instruction informing the model on the task at hand, have largely influenced the performances. In other words, the design of prompts has been shown to play a pivotal role in the context of source code documentation, particularly code summarization.
In this context, prompt design has emerged as a critical factor in enabling LLMs to generate high-quality code summaries. As a result, prompt engineering has become a key research direction for improving robustness, reproducibility, and human alignment in automated code summarization \cite{marvin2023prompt,kilic2024source}.

% In summary, the evolution of code summarization reveals a clear trajectory: from rule-based and statistical approaches, through neural architectures, to today's large pre-trained models, where performance is tightly coupled with prompt engineering. Fig.~\ref{fig:Code-Summarization-Evaluation} reflects this trajectory, showing the RAGid surge of LLM-based approaches in the last five years, underscoring both the progress made and the pressing need for a systematic, evidence-based synthesis of prompt engineering techniques--a gap this study seeks to fill.

\subsection{The Role of Prompt Engineering in LLM-based Code Summarization}
While earlier deep learning approaches were typically constrained to a single predefined task, large language models exhibit flexibility and responsiveness to input phrasing, contextual cues, and the inclusion of examples or instructions. This adaptability makes the design of prompts a decisive factor in shaping both the fluency and correctness of generated summaries, positioning prompt engineering as an increasingly impactful area of research \cite{wang2024advanced,wang2023codet5}. At its core, prompt engineering comprises a suite of strategies aimed at optimizing the interaction between LLMs and the task at hand--code summarization in this case--by carefully crafting the model's input.

This process can be formalized by describing the prompt as a function $P(C)$, where $C$ denotes the source code snippet and $P$ encodes the prompt construction strategy. The LLM, represented as $f_\theta$, generates a summary $S$ such that $S = f_\theta(P(C))$.  In practice, it is convenient to expose the auxiliary context explicitly and write $P(C;A)$ where $A$ bundles instructions, templates, exemplars, and retrieved snippets. The quality and style of S depend on the particular choice of $A$ and the template used by $P$.

Building on these insights, researchers have explored a broad spectrum of prompt engineering strategies for code summarization, ranging from very simple approaches such as zero-shot setups to more structured, retrieval-augmented, and template-driven methods.
Before detailing the forefront of prompt-driven code summarization, let us clarify the terminology used in the paper.

\noindent \textbf{Prompting Technique.} A concrete way of choosing $A$ and the template in $P(C;A)$---\eg zero-shot, few-shot examples, or retrieval-augmented context. Two prompting techniques are considered different if they produce different input token sequences for $f_\theta$.

\noindent \textbf{In-context learning (ICL).} The model's ability to adapt its predictions at inference time based solely on the tokens in $P(C;A)$, without updating $\theta$. ICL is the underlying mechanism by which $f_\theta$ exploits contextual cues or examples; prompting techniques are the surface-level designs that supply those cues.

\noindent \textbf{Core Notation}.
\begin{description}
\item[$S$]: The generated summary.
\item[$f_\theta$]: The large language model (LLM).
\item[$C$]: The code to summarize.
\item[$P_{XY}(C)$]: The prompt formed by concatenating the instruction and code snippet. With $XY$, we identify a generic prompting strategy as defined below.
\end{description}

\subsubsection{\textbf{Zero-Shot prompting}}
\label{zero-shot}
One of the easiest strategies to distill knowledge from LLMs in code summarization task requires providing the model with ``only'' a generic task instruction, such as ``Summarize the following code \$CODE\_PLACEHOLDER''. Formally, we can define zero-shot prompting as follows:
% requires \usepackage{amsmath}
\begin{equation}\tag{3}\label{eq:zero-shot}
S = f_\theta\big(P_{ZS}(C)\big) = f_\theta\big([\text{Instruction};\,C]\big)
\end{equation}

This approach evaluates the LLM's capacity to generalize based on its comprehension of the task and its pre-training knowledge. Because no labeled examples or domain-specific cues are provided, the model must rely entirely on its interpretation of the natural-language instruction. Unlike in a few-shot setting (\secref{sec:few-shot})--where exemplars can guide the response--the absence of such hints leaves the model without external anchors. In this configuration, the model faces its greatest challenge.

Although zero-shot prompting can--to a good extent--generalize to unseen codebases without task-specific data, several studies have highlighted its shortcomings. Sun \etal \cite{sun2023prompt} and Hou \etal \cite{hou2024large} report that zero-shot summaries often lack detail and fail to capture project-specific terminology, particularly in complex or domain-focused settings. Similarly, Brown \etal \cite{brown2020language} observed that zero-shot methods frequently yield overly generic outputs that do not align with developer expectations. Thus, while zero-shot approaches promise scalability and minimal setup effort, these advantages are often outweighed by their limitations.

\subsubsection{\textbf{Few-Shot prompting}}
\label{sec:few-shot}
Access to a small set of $\langle\text{code},\text{summary}\rangle$ demonstrations can substantially improve an LLM's code summaries by letting the model infer the mapping from in-prompt context, without parameter updates.  If the context consists of $n$ example pairs $\{(C_1,S_1),\ldots,(C_n,S_n)\}$ for target code $C$, we write
\[
\mathrm{Ex}(n)=\{(C_1,S_1),\ldots,(C_n,S_n)\},\qquad
S=f_\theta\!\big(\text{``Summarize this code:''}\,\Vert\,\mathrm{Ex}(n)\,\Vert\,C\big),\ \ n\ge 1.\tag{4}
\]
Brown \etal\ \cite{brown2020language} showed that in-prompt demonstrations can improve performance without training; for code summarization, Sun \etal\ \cite{sun2023prompt} and Hou \etal\ \cite{hou2024large} report more accurate, project-style-aligned summaries when examples match the domain. However, the approach has notable challenges. For example non-standardized exemplar selection also challenges reproducibility and fair comparison \cite{vatsal2024survey}. Despite these issues, few-shot prompting offers a practical trade-off between adaptation capability and annotation cost when $n$ is small and examples are well chosen.

\subsubsection{\textbf{Retrieval-Augmented prompting:}}
\label{sec:Retrieval-augmented}

This strategy provides the model with information from a trusted knowledge base, helping it rely on external data instead of only its pre-training.
When high-quality resources are provided, the LLM benefits from targeted, domain-specific guidance that supports a more accurate and aligned calibration towards context-aware summaries \cite{yang2025raxcs,zhang2020retrieval}. Formally, a retrieval function $\mathcal{R}(C)$ identifies relevant supporting data for a given context $C$ and appends them to the input prompt, yielding:
\begin{equation}
P_{\text{RAG}}(C) = \text{``Summarize this code:''} \;\Vert\; C \;\Vert\; \mathcal{R}(C)
\end{equation}

\begin{equation}
S = f_\theta(P_{\text{RAG}}(C))
\end{equation}
where $S$ is the generated summary, $f_\theta$ is the large language model, and $P_{\text{RAG}}(C)$ is the retrieval-augmented prompt formed by concatenating a summarization instruction, the code $C$, and the retrieved auxiliary information $\mathcal{R}(C)$. Here, $\mathcal{R}(C)$ denotes the retrieved context relevant to $C$ (such as documentation or similar code). This approach leverages the retrieved context to enrich the model's background knowledge, improving both the factuality and technical accuracy of summaries, especially for complex or unfamiliar code \cite{parvez2021retrieval, shi2022evaluation}. Parvez \etal \cite{parvez2021retrieval} demonstrated that retrieval-augmented prompting substantially reduces hallucinations and increases specificity by grounding the summary in relevant, real-world examples. Similarly, Liu \etal\cite{liu2023summary} and Vatsal \etal \cite{vatsal2024survey} found that providing contextualized evidence enhances model performance in large, heterogeneous codebases. The primary strength of this method is its ability to integrate up-to-date, domain-specific, or project-specific knowledge that may not be captured during model pre-training. However, retrieval-augmented prompting also presents practical challenges: the overall summary quality is highly contingent on the precision and relevance of the retrieved examples. Noisy, redundant, or irrelevant context can mislead the model and degrade performance \cite{liu2023summary}. Additionally, implementing efficient and accurate retrieval mechanisms at scale increases system complexity and computational overhead, requiring sophisticated retrieval algorithms and infrastructure. Despite these limitations, retrieval-augmented prompting is a promising direction for advancing LLM-based code summarization in real-world development environments.

\subsubsection{\textbf{Chain-of-Thought prompting:}}
\label{sec:Chain-of-thought}

Chain-of-Thought prompting is an advanced prompt engineering technique that encourages LLMs to articulate intermediate reasoning steps before producing a final code summary, mimicking human-like explanation. The CoT prompt is typically structured as:
\begin{equation}
P_{\text{CoT}}(C) = \text{``Reason step-by-step about what the code does:''} \;\Vert\; C
\end{equation}
with the output summary given by:
\begin{equation}
(R, S) = f_\theta(P_{\text{CoT}}(C))
\end{equation}

where $R$ is the intermediate reasoning and $S$ is the generated summary.  Wei \etal \cite{wei2021finetuned} established the value of CoT prompting for complex reasoning tasks in LLMs, and its adaptation to code summarization has yielded notable improvements in accuracy, logical coherence, and interpretability for multi-step functions or code with intricate logic \cite{zhang2022automatic,ahmad2021unified}. For example, Zhang \etal \cite{zhang2022automatic} report that CoT prompting enhances summary quality for functions involving nested loops, conditionals, or data structure manipulations by ensuring the model first explains intent and sequence before abstracting the summary. A key advantage of CoT prompting is increased transparency, allowing both researchers and practitioners to inspect the model's reasoning path, which is valuable for error analysis and model debugging. However, CoT also brings trade-offs: it can generate overly verbose or redundant outputs when applied to simple functions, and the additional reasoning tokens may increase computational costs and context window usage. Furthermore, crafting effective CoT prompts for code remains less explored than for natural language reasoning, representing an ongoing area for empirical innovation and standardization, Vatsal \etal \cite{vatsal2024survey}.

\subsection{Influence of Code Granularity on Prompt Engineering}

The granularity of code artifacts, such as methods, classes, or files, influences both the complexity of the code summarization task and the selection of effective prompt engineering strategies \cite{ahmad2021unified,wang2023codet5,leclair2019neural}. 
Function/Method-level summarization is the most commonly studied in the literature, due to its relatively contained context and the availability of benchmark datasets \cite{leclair2019neural, allamanis2016convolutional}. 
In contrast, class-level and file-level summarization require the model and the prompt to manage substantially more context--and tokens--technically speaking, often necessitating the use of a combination of prompting approaches to maintain relevance and coherence in the generated summaries \cite{sun2025commenting,rukmono2023achieving}  
Recent studies have indicated that prompt engineering techniques--including chain-of-thought and retrieval-augmented prompting, may be particularly beneficial for larger code units, where the reasoning process becomes more complex and context management is essential. Therefore, the selection and design of prompts for LLM-based code summarization should carefully consider the code granularity of the code component being summarized to maximize summary quality and practical utility \cite{liu2020retrieval,dong2024promptexp,yang2025docagent}.

The rapid expansion of large language models has led to huge advances in automated code summarization. Prior research has explored individual prompting techniques~\cite{wei2020retrieve, wang2023codet5}, but often in isolation, without a comprehensive synthesis of how prompt design, model architecture, and code granularity interact to affect summarization quality. \tabref{tab:slr-prompt-code-summ} lists prior secondary studies in chronological order. Secondary studies are overviews of prior research (surveys, mappings, or SLRs) that synthesize results from many primary papers instead of proposing a new technique. In our table, we report each study's main goal, total corpus they used, and the year of publication. Early surveys such as Song \etal \cite{song2019survey}, classify algorithms and workflows for automatic code comments, but the study was conducted before the rise of LLM-based prompting and does not report a fixed number of included papers. Zhu \etal \cite{zhu2019automatic} conducts an SLR of 41 studies on automatic code summarization, detailing extraction, generation, and evaluation, but without a prompt-based lens. Zhang \etal \cite{zhang2024review} presents a broad, up-to-date review of automatic code summarization, but it mainly focuses on traditional IR-based retrieval and Neural Network-based generation approaches, rather than prompting with LLMs. Hou \etal \cite{hou2024large} presents a broad SLR on LLMs for Software Engineering, analyzing 395 papers across many SE tasks (summarization included) but treat prompting as one optimization among others rather than a primary focus. Vatsal \etal \cite{vatsal2024survey} synthesize 44 papers on prompt engineering for NLP tasks; their taxonomy is transferable but not SE-specific. Budiardjo \etal \cite{syahputri2025unlocking} presents an SLR of prompt engineering within SE with 42 papers, identifying clusters (\eg RAG, CoT, automated prompt tuning), but they do not analyze prompt-based code summarization in depth. Our work targets this gap; we synthesize prompt-based code summarization specifically, organizing results by task granularity, prompting paradigms, evaluation, and reproducibility.

\begin{table}[htbp]
\centering
\caption{Previous studies on prompt engineering and code summarization, indicating secondary (SLR/Survey) and primary studies.}
\label{tab:slr-prompt-code-summ}
\resizebox{0.7\textwidth}{!}{%
\begin{tabular}{@{}l p{7cm} c l@{}}
\toprule
\textbf{Reference} & \textbf{Main Goal} & \textbf{Corpus Number} & \textbf{Year} \\
\midrule
Song \etal \cite{song2019survey} & Classifying algorithms and techniques for automatic code comment generation and summarizing evaluation practices & - & 2019 \\
Zhu and Pan \cite{zhu2019automatic} & Systematically reviewing approaches to automatic code summarization & 41 & 2019 \\
Hou \etal \cite{hou2024large} & Mapping the use of LLMs across software engineering—models, data handling, optimization/evaluation, and task applications & 395 & 2024 \\
Vatsal \etal \cite{vatsal2024survey} & Summarizing prompt-engineering methods across NLP tasks and organizing them with a task-oriented taxonomy & 44 & 2024 \\
Zhang \etal \cite{zhang2024review} & Documenting advances in automatic source code summarization—datasets, code modeling, generation methods, and evaluation—and outlining future directions & - & 2024 \\
Budiardjo \etal \cite{syahputri2025unlocking} & Synthesizing prompt engineering in software engineering, identifying research clusters and gaps, and proposing a framework/roadmap & 42 & 2025 \\
\midrule
\textbf{Our work} & \textbf{Comprehensive SLR of prompt engineering techniques, benchmarks, and reproducibility in LLM-based code summarization} & 29 & \textbf{2025} \\
\bottomrule
\end{tabular}
}
\end{table}

In summary, the evolution of code summarization has reached a stage where LLMs, guided through prompt engineering, play a decisive role in shaping output quality and applicability. While empirical studies highlight the potential of carefully crafted prompts, existing research remains scattered across tasks, models, granularity, and evaluation setups, making it difficult to identify consistent best practices. The heterogeneity of architectures and experimental designs further exacerbates this fragmentation. These challenges underscore the need for a systematic literature review that categorizes prompt engineering strategies, and clarifies their impact on code summarization. Our study responds to this call by providing the first comprehensive synthesis of prompt engineering methods in this domain.

\section{Methodology}
\label{sec:methodology}

This systematic literature review is conducted in accordance with established guidelines for evidence-based software engineering research, as articulated by \cite{kitchenham2007guidelines}. The review process encompasses rigorous study identification, selection, extraction, and synthesis procedures to ensure transparency, reproducibility, and comprehensive coverage of the literature relevant to prompt engineering techniques in code summarization with LLMs.

\subsection{Research  Questions}
The principal objective of this SLR is to explore, categorize, and critically evaluate prompt engineering strategies employed for code summarization in the context of LLMs. Specifically, the review is guided by the following research questions:

\begin{itemize}[leftmargin=*] % removes extra indent

\item\textbf{RQ$_{1}$: How do code granularity and prompting paradigm choices influence prompt-based code summarization performance?}.\
This research question investigates the adoption of prompt-based code summarization techniques across varying granularity levels--function/method-level, file/class/module-level, project/repository-level, and code changes-level--while exploring how prompt-based methods are adapted to each of these. By analyzing primary studies, the intention is to construct a granularity-level taxonomy that systematically organizes these contexts, enabling clearer comparisons, highlighting methodological patterns, and identifying underexplored areas for future work.

\vspace{0.5em}

\item \textbf{RQ$_{2}$: How are prompt engineering paradigms instantiated in LLM-driven code summarization, and which design dimensions characterize their implementations?}\
This research question aims to identify, describe, and systematically classify the concrete instantiations of each prompt engineering paradigm introduced in RQ$_1$. Specifically, we examine how these paradigms are realized in practice by analyzing: (i) the prompt components and interaction patterns that operationalize each paradigm, and (ii) the design dimensions along which implementations vary. Key dimensions include the number and provenance of examples; example selection strategies (\eg fixed vs.\ adaptive); the use of structured templates and role-based instructions; reasoning or planning guidance; incorporation of retrieved or contextual information; and robustness techniques such as constraint enforcement, self-checking, and prompt perturbation.

\vspace{0.5em}

\item \textbf{RQ$_{3}$: Are certain large language models consistently adopted across different code-summarization tasks and prompting paradigms?}\
This research question examines the distribution and prominence of LLMs used across existing studies on prompt-based code summarization. The goal is to identify whether specific models--such as GPT variants, Codex, CodeT5, or CodeLlama--are recurrently adopted across multiple granularity levels and prompting paradigms (\eg Zero-shot, Few-shot, Retrieval-Augmented, Chain-of-Thought), suggesting community convergence or de facto standards. This analysis considers both the frequency of model usage and the diversity of tasks to which each model has been applied, thereby uncovering potential trends in model preference, capability, and generalization. 

\vspace{0.5em}

\item\textbf{RQ$_{4}$: Which benchmarks and metrics have been used to assess the quality of prompt-based code summarizers, and how do these measures align with human and LLM evaluations?}
%of quality, usefulness, and readability in real development contexts?}
This research question investigates the evaluation landscape of prompt-based code summarization by examining both the resources and metrics employed and their correspondence with human and LLM-based assessments. On one hand, the analysis maps benchmark datasets (\eg CodeXGLUE \cite{lu2021codexglue}) to the automatic evaluation metrics applied, ranging from lexical overlap measures such as BLEU \cite{papineni2002bleu}, ROUGE \cite{lin2004rouge}, and METEOR \cite{banerjee2005meteor}, to semantic or embedding-based measures (\eg BERTScore, SentenceBERT). On the other hand, it considers how these automated scores compare with human-centered evaluations (\eg readability, informativeness, usefulness) and LLM-as-judge protocols (\eg GPT-4–based scoring), which have emerged to capture qualitative aspects not reflected by traditional metrics. The goal is to provide a comprehensive benchmark–metric–judgment mapping that reveals patterns in dataset and metric choices, highlights strengths and shortcomings of automatic measures against human/LLM evaluations, and informs the design of more reliable future evaluation methodologies.

\vspace{0.5em}

\item \textbf{RQ$_{5}$: What is the state of reproducibility and resource sharing in prompt-based code summarization research?}  
This research question investigates reproducibility practices and the extent of resource sharing in existing studies. The focus is on evaluating the accessibility and completeness of key artifacts--such as datasets, prompt templates, trained or fine-tuned model checkpoints, and source code--while also accounting for the inherent non-determinism of LLM outputs. The goal is to provide a comprehensive reproducibility table that examines both artifact availability and reporting quality, explicitly noting whether each study’s repository is accessible, provided but non-accessible, or not provided at all. This framework enables systematic comparison across the literature and highlights areas where improved transparency and open access could strengthen replicability.
\end{itemize}

\noindent We present the structure of our systematic literature review, following the guidelines of Kitchenham \etal \cite{kitchenham2007guidelines}. Our review focuses on studies that apply prompt engineering techniques to code summarization using large language models (LLMs).
It is guided by a set of key Research Questions (RQs) that direct our analysis and help organize the reviewed studies.

\begin{enumerate}
    \item \textbf{Identifying Primary Studies}
    \begin{enumerate}[label=(\alph*)] % nested enumerate for a, b, c
        \item Executed broad search queries across six major digital libraries (IEEE Xplore, ACM Digital Library, SpringerLink, ScienceDirect, Wiley Online Library, and Scopus) using finalized search strings.
        \item Applied database-specific adjustments to handle platform constraints (\eg Boolean limits, category exclusions).
        \item Adopted a two-phase retrieval process: Phase 1: Broad code summarization keyword search
Phase 2: Automated prompt-engineering content screening using a Python script.
    \end{enumerate}

    \item \textbf{Study Selection and Eligibility Filtering}
    \begin{enumerate}[label=(\alph*)]
        \item \textbf{Publication Type Filtering} – Retained only peer-reviewed journal and full conference papers; excluded non-research formats.
        \item \textbf{Venue Quality Assessment} – Prioritized inclusion from reputable conferences and journals based on CSRankings and field consensus.
        \item \textbf{Snowballing and Manual Addition} – Performed backward snowballing by reference analysis and manually added seminal works meeting the inclusion criteria.
    \end{enumerate}

    \item \textbf{Data Extraction, Quality Assessment, and Synthesis}
    \begin{enumerate}[label=(\alph*)]
        % \item \textbf{Data Extraction} – Used a structured extraction form capturing bibliographic data, prompt paradigm, model/dataset use, metrics, artifact availability, and human evaluation details.
        % \item \textbf{Quality Assessment} – Applied a quality checklist covering methodological clarity, reproducibility, evaluation robustness, and human validation.
        % \item \textbf{Data Synthesis} – Organized findings according to the five RQs; generated summary tables, visualizations, and a PRISMA flow diagram.
         \item \textbf{Quality Assessment} – Developed a structured questionnaire to evaluate methodological clarity and reporting transparency across studies.
        \item \textbf{Data Extraction} – Used the questionnaire as a guide to extract bibliographic information, granularity level, prompt paradigms, model/dataset usage, evaluation metrics, artifact availability, and human evaluation details.
        \item \textbf{Data Synthesis} – Summarized and compared findings according to the five RQs; generated taxonomy table, diagrams for visualizations, to highlight overall research trends and reproducibility patterns.
    \end{enumerate}
\end{enumerate}

\begin{figure}[ht]
  \centering
  \includegraphics[width=0.95\textwidth]{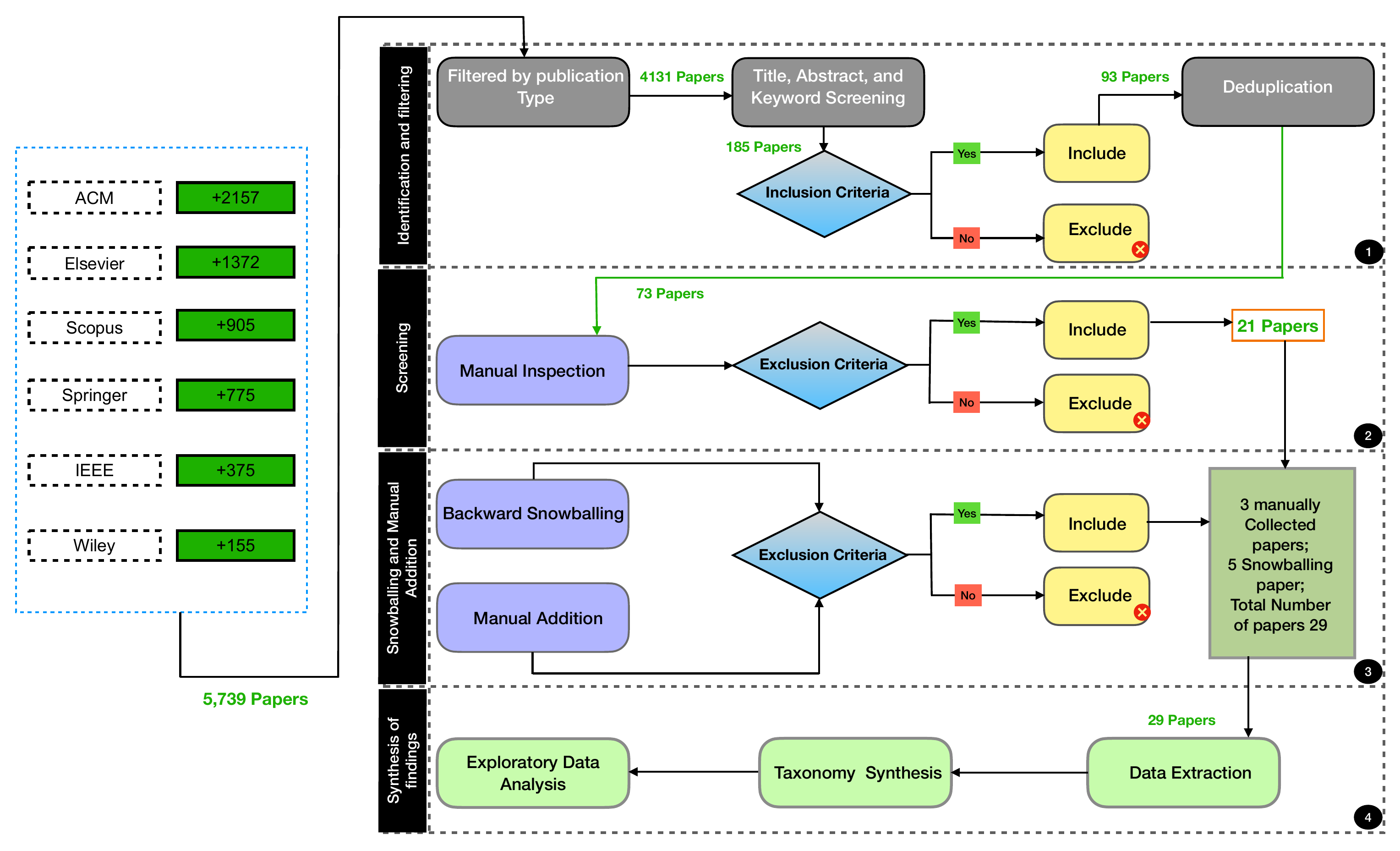}
  \caption{Study selection process for the systematic review on prompt-based code summarization. 
  The pipeline includes four stages: (i) identification and filtering,
  (ii) screening, (iii) snowballing/manual addition, and (iv) synthesis of findings}
  \label{fig:Study-Selection-Process}
\end{figure}

\figref{fig:Study-Selection-Process} illustrates our study selection workflow, outlining the progression from the initial retrieval of records to the final set of papers included in our SLR. The detailed methodology for each step in this process is described in \secref{sec:identifying-studies}. (Study Selection and Filtering Process), including database-specific search adjustments, screening procedures, and the application of inclusion and exclusion criteria, as well as snowballing and manual addition. Data Extraction and Analysis part has been explained in \secref{sec:Data Extraction and Analysis}

\subsection{Identifying Primary Studies}\label{sec:identifying-studies}

\subsubsection{Search Strategy.}

\begin{figure}[ht]
    \centering
    \includegraphics[width=0.65\textwidth]{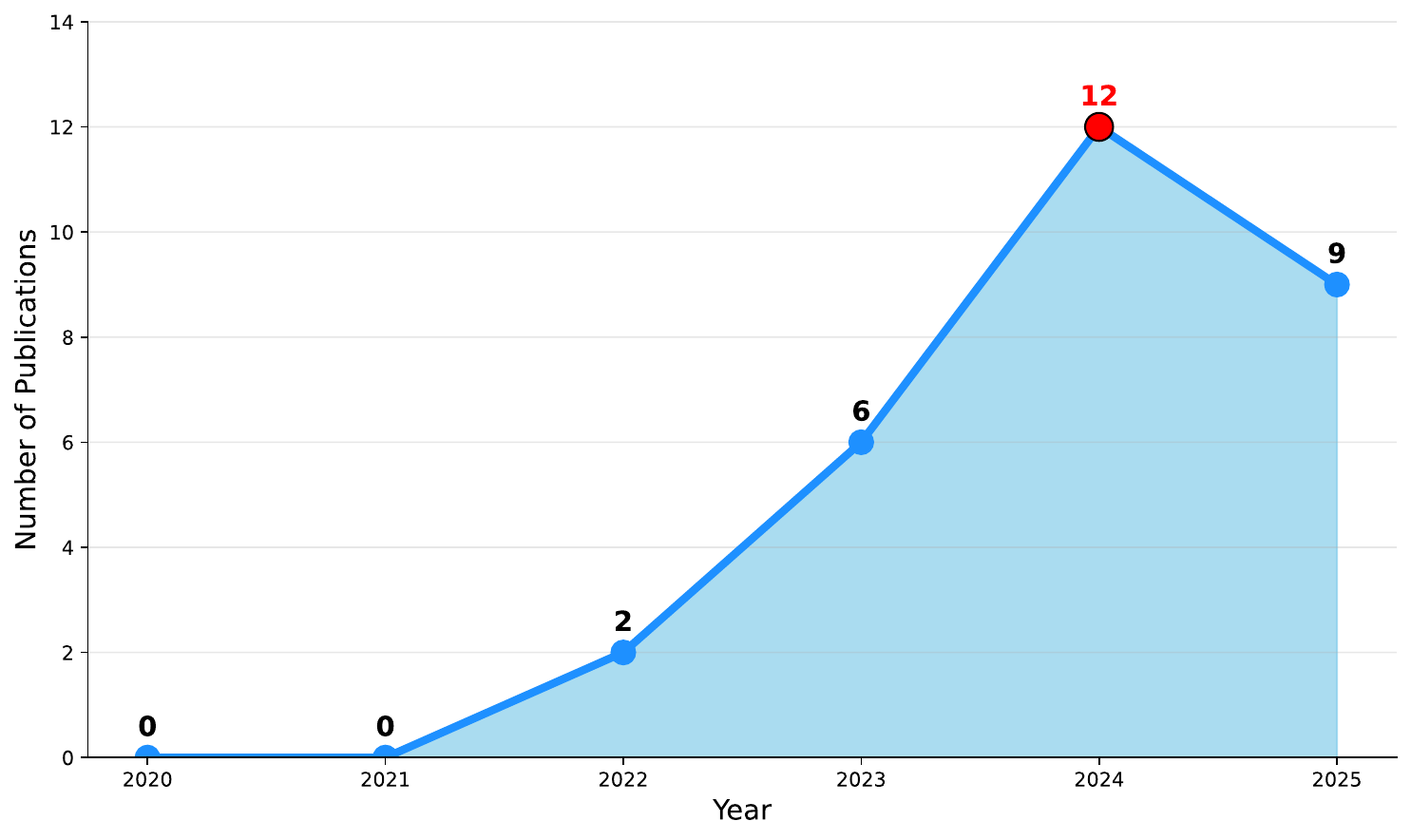}
    \caption{Publication Year}
    \label{fig: Publication-Year-Trend}
\end{figure}

To answer our RQs, we focused on primary studies published between 2020 and 2025, a period that coincides with the emergence and rapid evolution of prompt-based code summarization techniques in conjunction with large language models. \figref{fig: Publication-Year-Trend} illustrates the yearly distribution of the selected studies, showing a sharp increase from 2022 onward, with peaks in 2024. This upward trend highlights both the recency and growing research interest in this area, justifying our choice of the timeframe and emphasizing the timely nature of our review.

Next, we began our search by querying six major digital libraries: IEEE Xplore\cite{ieee}, ACM Digital Library\cite{acm}, SpringerLink\cite{springer}, ScienceDirect\cite{elsevier}, Wiley Online Library\cite{wiley}, and Scopus\cite{scopus}. We did not query Google Scholar due to the limitations documented by Halevi \etal \cite{halevi2017suitability} (\eg lack of quality control, missing support for data download). 

\begin{figure}[ht]
    \centering
    \includegraphics[width=0.65\textwidth]{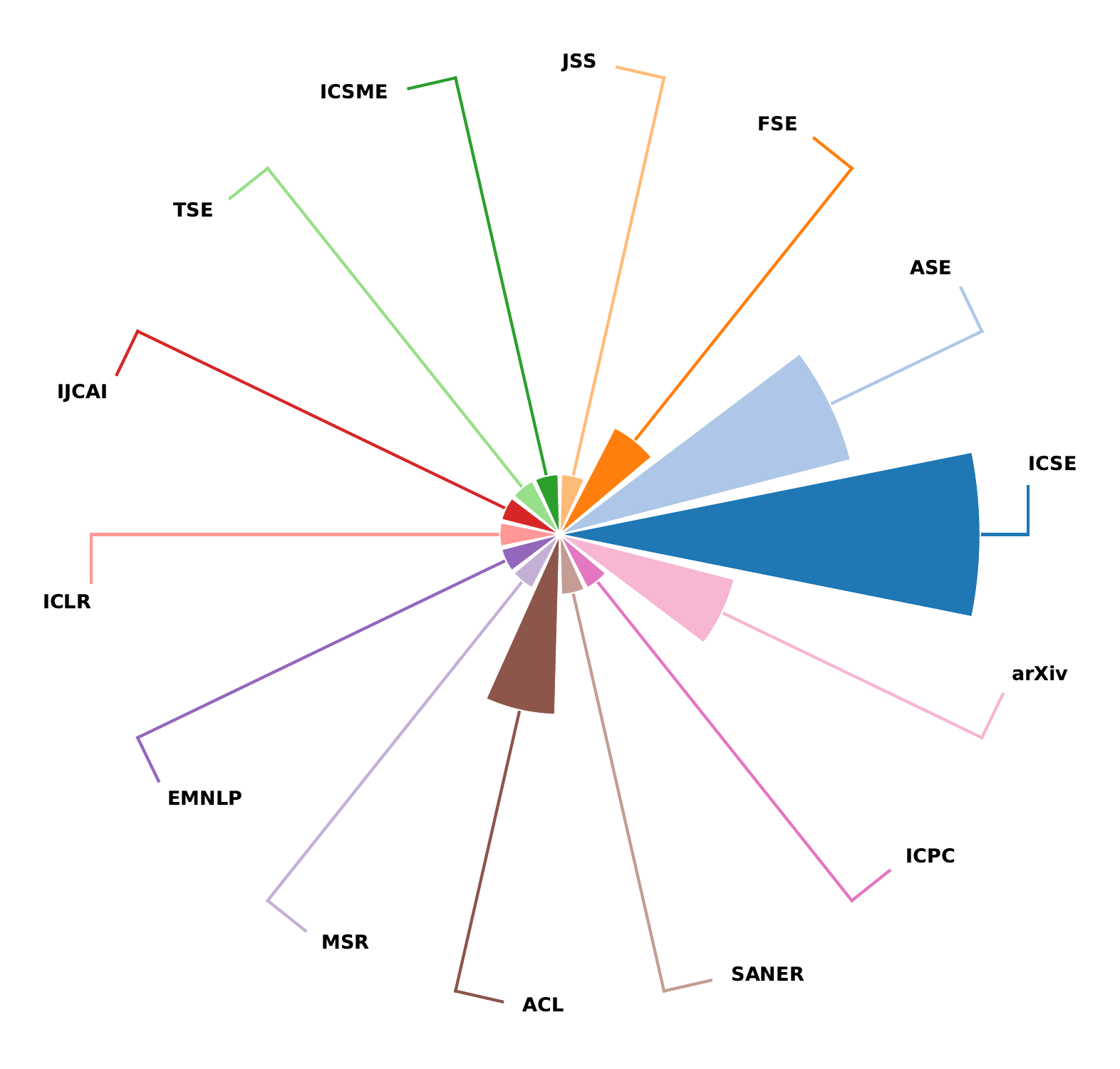}
    \caption{Venue Distribution of Prompt Engineering Techniques in Code Summarization}
    \label{fig:venue-distribution}
\end{figure}

In terms of publication venues, \figref{fig:venue-distribution} illustrates the distribution of selected studies across conferences and journals. As shown in \figref{fig:venue-distribution} , our final collection of papers spans 14 venues, with the \textbf{ASE, ICSE,ACL} contributing the largest number of relevant studies. While this approach may exclude some potentially relevant papers from less prominent venues, it ensures our review emphasizes high-quality, peer-reviewed research from both top-ranked and widely recognized publication forums. We also observed contributions in established journals such as TSE and TOSEM, alongside early-stage works disseminated via arXiv. This spread of venues highlights the interdisciplinary nature of prompt engineering for code summarization, spanning both core SE venues and broader AI/NLP outlets. 

To define the query for identifying works related to code summarization, two authors followed a trial-and-error procedure. It quickly became evident that searching paper titles for keywords such as ``prompt engineering,'' ``large language model,'' ``context learning,'' and their variations was not a viable strategy, as it would exclude several relevant studies. Therefore, we adopted a more conservative and inclusive approach, aiming to capture all studies on code summarization that employ prompt-based methods, even if prompt-related terminology does not explicitly appear in the title (\eg ``Retrieve and Refine: Exemplar-based Neural Comment Generation'' \cite{wei2020retrieve}). To this end, we focused exclusively on the task of code summarization, designing a query that leveraged a comprehensive set of synonyms and related expressions. We adopted the keyword taxonomy in Zhu \etal~\cite{zhu2019automatic} as a starting point and tailored it to our scope through iterative synonym expansion. \tabref{tab:dl-counts} summarizes the number of articles retrieved from each digital library before screening and duplicate removal.

\

We adopted a two-phase search strategy.  
\faArrowCircleRight~ \textbf{Phase 1:} A broad query was designed to identify all code summarization studies, regardless of whether prompting terminology was explicitly mentioned.  
\faArrowCircleRight~ \textbf{Phase 2:} An automated Python-based script scores each study based on the presence of relevant terms within the title, abstract, and keywords metadata, enabling the detection of both explicit cases (\eg ``Automatic Semantic Augmentation of Language Model Prompts (for Code Summarization)''~\cite{ahmed2024automatic}). This approach enabled us to collect all prompt-engineering-related code summarization studies, including both explicit and implicit ones.
  
As part of this process, two authors reached consensus on the set of prompt-engineering–related terms to be used as inclusion keywords, which formed the basis for developing the Python script. The script facilitated the screening and prioritization of the initial study corpus. For studies without author-supplied keywords, manual annotation was performed using title, abstract, and methodological content, following the guidelines of \cite{kitchenham2007guidelines}. This process yielded 4 candidate papers for manual analysis. Both the Python script and the complete prompt-engineering terms used in our search are provided in our replication package.
\textit{Thresholding rule:} we retained all records with \texttt{total\_score} $\geq$1 (\ie at least one keyword match in any field) and excluded records with \texttt{total\_score} $= 0$.

Among the queried search engines, Elsevier, Scopus, Springer, and Wiley allowed specification of a discipline of interest, which we set to ``Computer Science,'' with language restricted to English, to minimize false positives. To further ensure study quality, only peer-reviewed publications were retained. The search strategy aimed to maximize coverage of relevant primary studies, with inclusion and exclusion decisions made during the subsequent review.   

To refine relevance and manage technical limitations, we excluded certain categories and adapted queries per platform. Specifically, papers categorized under ``Lisp'' and ``Sequence annotation'' were removed from SpringerLink, as they primarily concerned language-specific or bioinformatics contexts unrelated to prompt engineering for code summarization. This exclusion also ensured compliance with SpringerLink's bulk download limit of approximately 1,000 items. Furthermore, Elsevier's ScienceDirect database restricted the use of Boolean operators to a maximum of eight, requiring us to split our comprehensive query into two parts. This approach guaranteed complete coverage without exceeding platform constraints. The exact queries used are documented in the replication package accompanying this study.

\begin{itemize}
  \item (``Code Summarization'' OR ``Source Code Summarization'' OR ``Program Summarization'' OR ``Code Documentation'' OR ``Automatic Comment Generation'' OR ``Code To Text'' OR ``Code Documentation Generation'' OR ``Code Summary'')
  \item (``Source Code Summary'' OR ``Comment Generation'' OR ``Code Explanation'' OR ``Code Description'' OR ``Documentation Generation'' OR ``Automated Source Code Documentation'' OR ``Code Comment Generation'')
\end{itemize}

We acknowledge that some relevant articles published in related fields (\eg artificial intelligence) may not have been captured by our query. To partially mitigate this limitation, we incorporated a snowballing process, as described later in \secref{sec:Study-Selection}. Nevertheless, we would like to reiterate that all included studies were required to meet our predefined inclusion criteria; papers failing to do so, or not meeting minimum quality standards, were excluded.

\begin{table}[htbp]
\centering
\caption{Articles returned by the queried digital libraries}
\label{tab:dl-counts}
\resizebox{0.35\textwidth}{!}{%
\begin{tabular}{@{}l r@{}}
\toprule
\textbf{Source} & \textbf{Returned Articles} \\
\midrule
ACM Digital Library            & 2,157 \\
Elsevier ScienceDirect         & 1,372 \\
Scopus                         &   905 \\
Springer Link Online Library   &   775 \\
IEEE Xplore Digital Library    &   375  \\
Wiley Online Library           &   155 \\
\midrule
\textbf{Total (including duplicates)} & \textbf{5,739} \\
\bottomrule
\end{tabular}
}
\end{table}

\tabref{tab:dl-counts}~ reports the number of articles retrieved from each digital library during Phase 1, which focused exclusively on code summarization search queries. The counts represent the raw results returned by each source before any screening or duplicate removal.

\subsubsection{Study Selection}\label{sec:Study-Selection}

After running the automated search (5,739 papers) we performed \faArrowCircleRight~Publication Type Filtering, retaining only peer--reviewed journal articles and full conference papers, and excluding book chapters, early-access articles, magazines, standards, prefaces, editorials, and preprints (\eg arXiv) lacking rigorous peer review or empirical evaluation (-1608). We then applying our Python-based screening script to exclude irrelevant papers (-3946). Next, in the \faArrowCircleRight~ Venue Quality Assessment stage, we got 185 papers, and we prioritized publications from reputable, high-impact conferences and journals (as defined by \textbf{CSRankings}\footnote{\url{https://csrankings.org}} and field consensus) (-92). We then \faArrowCircleRight removed duplicates to ensure a unique set of studies (-20). After completing all screening and filtering steps, a total of 73 primary studies were finally selected for inclusion in the manual review.

\begin{table}[t]
\centering
\caption{Inclusion and exclusion criteria}
\label{tab:inclusion-exclusion}
\setlength{\tabcolsep}{6pt}         % tighter left-right padding
\renewcommand{\arraystretch}{1.05}  % tighter row height
\small                               % smaller font for compactness
\begin{tabularx}{\linewidth}{lL}
\toprule
\multicolumn{2}{l}{\textbf{Inclusion Criteria}}\\
\midrule
IC1 & The article is a peer-reviewed \emph{full} conference paper or journal article in computer science. During snowballing/manual addition, highly cited and field-relevant arXiv preprints may be included as exceptions if they satisfy IC2–IC4.\\
IC2 & The PDF of the article is available online (queried libraries and, if needed, Google).\\
IC3 & The article must apply prompt engineering techniques for code summarization. Generic code summarization papers are excluded unless the authors explicitly state that their approach is designed for prompt-based code summarization.\\
IC4 & The paper reports an empirical evaluation and provides sufficient methodological details.\\
\midrule
\multicolumn{2}{l}{\textbf{Exclusion Criteria}}\\
\midrule
EC1 & Not written in English.\\
EC2 & Not peer-reviewed (\eg posters, abstracts, editorials, magazine articles); arXiv preprints are excluded \emph{unless} explicitly added via snowballing/manual addition per IC1.\\

EC3 & The article is not a full research publication. We exclude all articles having less than six
pages with the goal of removing articles that may not have been subject to the same peer-review process typical of full research articles.\\

EC4 & Secondary studies (surveys, SLRs, mapping studies) retained only as sources for snowballing, not as primary studies.\\

EC5 & Duplicates of already included works; when both workshop/conference and journal versions exist, the most complete version is retained.\\
\bottomrule
\end{tabularx}
\end{table}

Following this, \faArrowCircleRight~ Manual screening was performed on 73 papers independently by one author against the exclusion criteria, leading to our final set of studies (\textbf{21}). Inclusion and exclusion criteria are listed in \tabref{tab:inclusion-exclusion}. \textbf{21} papers satisfied all predefined requirements and were selected for in-depth analysis during the data synthesis phase. To strengthen the reliability and validity of this stage, first author independently conducted a full-text review of each study, carefully examining the abstract, methodology, and stated contributions of each study were systematically reviewed to assess whether the work explicitly implemented advanced prompt engineering techniques for code summarization. After collecting all relevant information, a second author was involved to double-check the extracted data and ensure consistency and accuracy across all reviewed studies.  During this phase, 3 disagreements arose--two related to exclusion. These cases were resolved through open discussion, with the outcome being a consensus to retain 1 studies and discard 2 studies in the corpus.

\textbf{Snowballing and Manual addition} After identifying the primary studies from the selected venues and applying the inclusion criteria, we conducted a snowballing process on the resulting set of papers. This step ensured the inclusion of additional relevant works that were not captured through the initial search process. During snowballing, we carefully examined the reference lists of all studies that met our inclusion criteria to determine whether any cited works also qualified for inclusion.
In addition, leveraging our prior expertise in this research area, we manually incorporated studies that may have been overlooked but were recognized as relevant based on our domain knowledge.
Following Kitchenham’s \cite{kitchenham2007guidelines} recommendation to prioritize peer-reviewed evidence, only peer-reviewed studies were primarily considered. However, consistent with the extended SLR guidance by Garousi \etal \cite{garousi2019guidelines} on including grey literature when transparently justified, we incorporated three arXiv preprints, one identified during the snowballing phase and another two during manual collection. These papers were included because they (i) directly addressed our research questions, (ii) demonstrated methodological clarity and empirical grounding, and (iii) did not have corresponding peer-reviewed versions available at the time of screening. All of them are explicitly marked as non-peer-reviewed in the data extraction table. The included arXiv papers are~\cite{sun2023automatic,sun2025commenting,zhu2025reposummary}.
All queries were executed on 28 October 2025 across the six selected digital libraries. To ensure the inclusion of the most recent advancements in this rapidly evolving field, after considering all inclusion criteria, we manually added three high-quality studies to the corpus based on targeted searches conducted in October 2025. This resulted in a total of \textbf{29 primary studies} for the final analysis.

% , two additional papers were manually collected from venues not covered by CSRankings. These papers were peer-reviewed and were included because they directly address the research questions and are closely aligned with the topic of this study. Following the principles outlined by Kitchenham \etal \cite{kitchenham2007guidelines} and Garousi \etal \cite{garousi2019guidelines}, inclusion decisions were guided by relevance and contribution rather than venue prestige. Consequently, the final corpus consisted of 29 primary studies, of which 5 originated from snowballing, and 3 were collected by manual addition.

\subsection{Data Extraction and Analysis}\label{sec:Data Extraction and Analysis}
\begin{table}[ht]
\centering
\caption{Data extraction questionnaire}
\label{tab:data-extraction}
\renewcommand{\arraystretch}{1.1}
\begin{tabular}{p{0.08\linewidth} p{0.75\linewidth} p{0.1\linewidth}}
\hline
\textbf{No.} & \textbf{Question} & \textbf{Focus} \\
\hline
Q1 & What code summarization granularity level does the study address? & RQ1 \\

Q2 & Does the study involve multiple granularity levels (\eg combining function- and project-level contexts)?& RQ1 \\
\hline
Q3 & Which prompt engineering paradigm has been employed?& RQ2 \\

Q4 & If yes to Q2, (\romannumeral 1) which specific techniques are used? After getting answer for ( \romannumeral 1 ), how are they combined or compared?  & RQ2 \\
Q5 & If no to Q2, summarize the approach functioning. & RQ2 \\ \hline

Q6 & Which large language model(s) have been employed in the study (\eg GPT, CodeLlama, StarCoder, DeepSeek-Coder)? & RQ3 \\

Q7 & Are certain large language model (LLM) families consistently adopted across different code-summarization tasks or prompting paradigms, and do studies justify their model choices (\eg performance, accessibility, reproducibility) or compare multiple models? & RQ3 \\ \hline

Q8 & Which programming language has been exploited to build the technique? & RQ4 \\

Q9 & What dataset(s) or benchmark(s) have been used for evaluation? Collect information related to evaluation metrics, including overlapping-based metrics have been applied? & RQ4 \\
Q10 & Did the study incorporate any human or LLM-based evaluation? If yes, capture the evaluation protocol and whether LLMs were used as judges. & RQ4\\

Q11 & To what extent do automated metrics align with human or LLM-based judgments, and what discrepancies, if any, were reported? & RQ4\\

Q12 & What evidence is provided about human judgment of summary quality, including readability, usefulness, and practical adequacy in real development contexts? & RQ4 \\ \hline
Q13 & Is a link to a replication package available? Is the link still working? & RQ5\\
Q14 & Is the implementation of the proposed solution publicly available? & RQ5\\
Q15 & Are the datasets used for training and/or evaluating the technique publicly available? & RQ5 \\
 \\
\hline
\end{tabular}
\end{table}

The 29 primary studies have been inspected one last time with the goal of extracting the information needed to answer our RQs. All extracted information was documented in the master table provided in our replication package. The first author was responsible for extracting the required data guided by the questionnaire in \tabref{tab:data-extraction}. The questions are clustered according to the RQ they serve. Q1–Q2 collect the data needed to answer RQ1 (\ie the specific code summarization granularity addressed in the literature). Q3–Q5 classify the identified approaches according to the prompt engineering paradigms adopted (\eg zero-shot, few-shot, RAG, and chain-of-thought) and synthesize findings regarding their underlying techniques and findings under RQ2. Q6–Q7 explore model-level prominence (RQ3), identifying which LLM families have been adopted across tasks and prompting paradigms, and whether studies justify their model selection or provide comparative reasoning across different models. Q8–Q12 capture evaluation practices under RQ4, covering datasets and benchmarks applied, the use of human or LLM-based evaluation protocols, the degree of alignment between automated metrics and qualitative judgments, and evidence on human assessment of readability, usefulness, and practical adequacy. Finally, Q13–Q15 investigate reproducibility by assessing the availability of replication packages, the types of artifacts they shared, implementations, and datasets (RQ5). Whenever categorical distinctions were required—such as granularity levels in RQ1 or paradigm families in RQ2—an open-coding procedure was conducted, where both authors independently labeled key characteristics from each study and refined categories through discussion until agreement was reached.

\section{Results and Discussion}
\label{sec:results}

In this section, we report and interpret the findings of our systematic literature review by answering the research questions and synthesizing cross-study evidence.

\subsection{\RQ{1}: How do code granularity and prompting paradigm choices influence prompt-based code summarization performance?}
\label{sec:RQ1}

\begin{figure}[ht]
    \centering
   \includegraphics[width=1.0\textwidth]{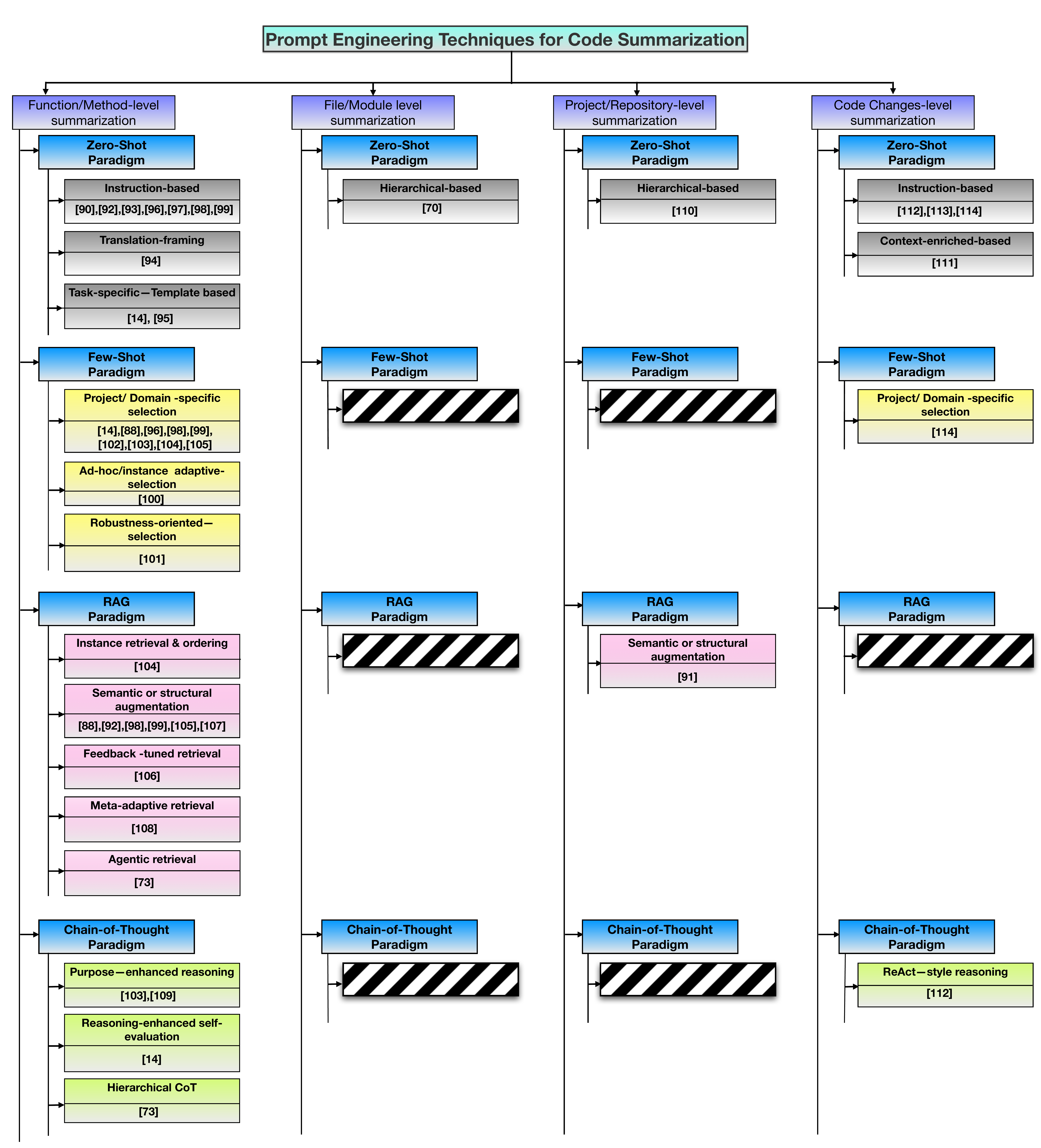}
    \caption{A Taxonomy of Prompt Engineering Techniques for Code Summarization Across Granularity Levels and Prompt Paradigms.}
    \label{fig:taxonomy}
\end{figure}

To systematically position the identified studies, we developed a taxonomy of prompt-engineering techniques for code summarization organized along two dimensions: (i) the granularity of the summarization task (function/method, file/module/class, project/repository, and code changes) and (ii) the prompting paradigms employed (\eg Zero-shot, Few-shot, Retrieval-Augmented, Chain-of-thought). The taxonomy depicted in \figref{fig:taxonomy} is organized hierarchically: prompt engineering for code summarization forms the root, from which four granularity levels branch out; these levels are subsequently divided by prompting paradigms, and the leaf nodes indicate the primary studies linked to each paradigm.

We begin by clarifying the different granularity levels in code summarization tasks, providing the foundation for understanding how prompting paradigms operate across these levels.

Based on our detailed analysis of the primary studies, we identified four (4) distinct granularity levels at which code summarization tasks have been addressed:

\begin{itemize}
    \item Function/Method--level Summarization,
    \item File/Module--level Summarization,
    \item Project/Repository--level Summarization, and
    \item Code changes--level Summarization
\end{itemize}

and at the same time, we have identified four (4) prompting paradigms employed across these granularity levels:
\begin{itemize}
\item \textit{Zero-Shot prompting},
\item \textit{Few-Shot prompting},
\item \textit{Chain-of-Thought prompting}, and
\item \textit{Retrieval-Augmented prompting}
\end{itemize}

Once a granularity level is defined, various prompting paradigms can be applied to address the summarization task at that level. For this reason, our discussion first examines each granularity level and then analyzes the prompting paradigms employed within it. This structure reflects a natural hierarchy: granularity determines the scope and complexity of the task, while prompting paradigms represent the techniques used to tackle it. Organizing the discussion by paradigm instead would be less meaningful, as each granularity level introduces distinct challenges--ranging from summarizing individual functions to capturing repository-wide architecture--that require specialized adaptations of prompting strategies. To systematically characterize the techniques reported across studies, we performed open coding and consolidated related approaches into four paradigms, each distinguished by differences in prompt input format. In this framework, any variation in the structure or composition of the input prompt is treated as a distinct paradigm.

\subsubsection{Function/Method-level Summarization.}  
At the most fine-grained level, function- or method-level code summarization focuses on producing concise, context-aware descriptions of individual functions or methods. The function-level code summarization task has been investigated in 22 primary studies featured in our SLR with the application of the full range of prompting paradigms--\textit{Zero-shot} \cite{sun2023automatic,geng2024large,wang2024natural,chai2022ernie,shin2023prompt,sun2024source,khan2022automatic,fried2022incoder,arakelyan2023exploring,makharev2025code}; \textit{Few-Shot} \cite{kruse2024can,zhang2024attacks,sun2024source,ahmed2022few,khan2022automatic,makharev2025code,zhang2025dlcog,ahmed2024automatic,gao2023makes,arakelyan2023exploring,YUN2024112149}; \textit{Retrieval-Augmented (RAG)} \cite{geng2024large,lu2024instructive,ahmed2024automatic,gao2023makes,arakelyan2023exploring,YUN2024112149,lomshakov2024proconsul,zhou2023towards,makharev2025code,yang2025docagent}; and Chain-of-Thought \cite{wang2024purpose,sun2024source,yang2025docagent,zhang2025dlcog}. 

At the function level, research identifies the individual method as the primary unit of analysis because it represents the standard logical ``snippet'' encountered during program maintenance \cite{sun2024source,sun2023automatic}. While this granularity establishes a baseline for general code understanding, it often lacks the specific ``intent'' found in larger contexts. This granularity is strategically chosen for its compatibility with model context windows; its small token footprint allows researchers to utilize the remaining prompt space for context-heavy strategies, such as the retrieval-augmented (RAG) approach used by Geng \etal \cite{geng2024large} to capture multiple intents like rationale and usage. Furthermore, the function level serves as an ideal scale for evaluating the ``raw'' multilingual and bidirectional logic of LLMs, as demonstrated by Chai \etal \cite{chai2022ernie} and Fried \etal \cite{fried2022incoder}, who highlighted that a function’s internal context is often sufficient for tasks like cross-language summarization and code infilling. However, the inherent ``contextual isolation'' of the function remains a challenge. Khan and Uddin \etal \cite{khan2022automatic} showed that at this granularity, moving from zero-shot to one-shot prompting is vital to help the model align technical code structures with proper documentation standards required by developers. Despite this, Shin \etal \cite{shin2023prompt} proved that when a function is decoupled from its project, prompting alone reaches a performance ceiling where the model cannot improve further without access to the broader logic found in the entire codebase. To resolve this, studies either optimize the unit itself through simplification to improve processing speed \cite{wang2024natural}.

While zero-shot approaches establish a baseline for general logic, research at the function level increasingly shifts toward the few-shot paradigm to overcome the limitations of ``contextual isolation''. Ahmed and Devanbu \etal \cite{ahmed2022few} demonstrated that because code is highly project-specific, providing a few function-level examples as a prefix allows models to capture local naming patterns and logic that zero-shot prompts miss. This transition to few-shot prompting is not only a technical necessity but also a practical one for users; Kruse \etal \cite{kruse2024can} found that developers often struggle to write effective ad-hoc prompts, making predefined few-shot templates essential for maintaining professional documentation standards. Furthermore, this paradigm serves as a security layer at the function level; Zhang \etal \cite{zhang2024attacks} indicated that functions are vulnerable to syntactic ``noise'' or adversarial attacks, but few-shot examples can effectively ``defend'' the model by teaching it to ignore irrelevant code structures. Collectively, these studies indicate that while the function remains the primary unit of analysis, the shift from zero-shot to few-shot prompting is critical for aligning the model with the specific project logic, developer needs, and security requirements of the codebase.

Research at the method level establishes that performance is fundamentally limited by ``contextual isolation'', requiring paradigms that bridge specific code logic with broader project intelligence. Zhang \etal \cite{zhang2025dlcog} optimized this by decomposing functions into logical segments through semantic segmentation, a strategy that captures both high-level logic and low-level details without the ``long token'' bottleneck. Building on this, research has increasingly adopted RAG to transform static examples into dynamic prompts. While basic few-shot learning aligns models with standards, Gao \etal \cite{gao2023makes} and Geng \etal \cite{geng2024large} demonstrated that instance-level retrieval allows models to shift between generating high-level rationale or low-level usage based on tailored context.

To further enhance performance, studies focus on the depth of retrieved information. Ahmed \etal \cite{ahmed2024automatic} enriched prompts with external variable and library definitions, while Yun \etal \cite{YUN2024112149} and Lomshakov \etal \cite{lomshakov2024proconsul} argued that project-specific accuracy relies on structural dependencies like call graphs. Similarly, Makharev \etal \cite{makharev2025code} found that function summaries benefit more from abstracted ``skeletons'' over raw code to reduce noise, and repository-level success depends more on few-shot example quality than retrieved data volume, highlighting the importance of structured prompts over raw context.

Advanced paradigms have moved toward model adaptation and reasoning. Lu \etal \cite{lu2024instructive} used LLM feedback to select ``instructive'' examples, while, Zhou \etal \cite{zhou2023towards} introduced a meta-learning framework (MLCS) where the summarization of a target function is treated as a unique task; the model first ``adapts'' to retrieved examples via meta-learning before generating a final summary. This approach, combined with the ability of RAG to handle distribution shifts in new projects \cite{arakelyan2023exploring}, suggested that at the function level, the combination of structural retrieval and local adaptation is the most effective way to provide the model with the ``big picture'' logic it needs without the high cost of full retraining. Systems like DocAgent \cite{yang2025docagent} exemplified this by using repository-wide topological analysis to inform component-level documentation. Collectively, these studies illustrate that at the method level, performance is dictated by a paradigm's ability to integrate high-level repository context into local function logic.
Beyond retrieving external context, research also transitioned toward enhancing internal reasoning through the Chain-of-Thought (CoT) paradigm; for instance, Wang \etal \cite{wang2024purpose} utilized a system called Chain-of-Structure that breaks down code into three levels: fine, middle, and coarse, depending on how much detail is shown. They chose the coarse-grained level because it gives a clear, high-level map of the code's design while saving space in the AI's memory.

\subsubsection{File/Module-level Summarization.}
At the file and module levels, code summarization captures high-level abstractions and complex inter-unit relationships that are often invisible at the function level. Current research in this area is predominantly grounded in the Zero-Shot prompting paradigm \cite{sun2025commenting}, as other paradigms such as Few-Shot, Retrieval-Augmented or Chain-of-Thought prompting remain largely underexplored for these specific granularities. Unlike function-level work, summarizing these higher-level units requires managing cross-unit dependencies and significant input lengths, making effective decomposition essential.

Sun \etal \cite{sun2025commenting} explored prompt-based LLM summarization for higher-level code units, specifically file- and module-level scenarios. Sun \etal \cite{sun2025commenting} investigated zero-shot, prompt-based LLM summarization for higher-level code units, distinguishing between file-level (a single source file containing one or more classes) and module-level (a collection of related files,\eg a Java package).They compared three strategies--(i) full-code, (ii) reduced-code, and (iii) hierarchical summarization--showing that full-code input yields the most accurate summaries at the file level, while hierarchical summarization, which composes file summaries into a module-level overview, performs best for larger modules due to scalability and context-window constraints. The study also demonstrates that prompt-based LLMs  can be effectively guided to generate such summaries through well-designed prompts, and that LLMs themselves can serve as scalable evaluators when human assessment is impractical. Some systems generate documentation at the module/component level while leveraging repository-wide analysis for accuracy. 

\subsubsection{Project/Repository-level Summarization.}
At the repository level, summarization generates global descriptions of entire projects, capturing intent, functionality, and architecture. This task is more complex due to heterogeneous artifacts, dependencies, and organizational structures. Research has focused on \textit{Zero-Shot} \cite{dhulshette2025hierarchical,zhu2025reposummary}, while \textit{Retrieval-Augmented}, \textit{Few-Shot} and \textit{Chain-of-Thought } reasoning remain unexplored, as shown in \figref{fig:taxonomy}.

%\textsuperscript{\ref{fn:hierarchical}}
In the repo-level, zero-shot summarization paradigm, the choice of code granularity is a primary driver of documentation quality. A central challenge highlighted by these works is that directly summarizing a large repository often leads to the omission of critical business logic or technical details due to context window constraints. To overcome this, recent research favors a fine-grained decomposition strategy. For instance, Dhulshette \etal \cite{dhulshette2025hierarchical} broke code into atomic units like functions and variables, recursively aggregating them into higher-level summaries to ensure that domain-specific details are preserved. Similarly,  Zhu \etal \cite{zhu2025reposummary} refined this by focusing on how these small units relate to one another. RepoSummary \cite{zhu2025reposummary} clustered individual methods into ``functional features'' moving away from arbitrary folder structures to align documentation with actual developer tasks. By shifting to these fine-grained, method-level analyses, these frameworks substantially improved feature coverage and traceability, proving that effective repository summarization requires deep, granular understanding rather than high-level scanning.

%In the RAG-based and few-shot sectors, Makharev and Ivanov \etal \cite{makharev2025code} extended this granularity to the class and repository levels. This study bridges function-level summarization to class and repository levels using RAG and few-shot learning. Their study reveals that while broader context improves summaries, class-level benefits from abstracted ``skeletons'' over raw code to reduce noise, and repository-level success depends more on few-shot example quality than retrieved data volume, highlighting the importance of structured prompts over raw context.

\subsubsection{Code changes-level summarization.} 
At the code change level, summarization tasks generate a short natural-language description of a software change (\textit{commit}, \textit{Pull Request}, or \textit{diff}). It takes the ``delta'' between two versions--added/removed lines and nearby context--and explains what was changed, why it was changed, and its likely impact. The output typically looks like a concise commit message or PR summary, aiding code review, release notes, and traceability. Despite preliminary investigations, research at this level remains limited, with advanced prompting paradigms like \textit{RAG} largely unexplored.

At the code change granularity level, prompt-based summarization performance is distinctly shaped by the requirement to interpret and synthesize code deltas alongside their surrounding semantic context. Studies by Imani \etal \cite{imani2024context} and Li \etal \cite{li2024only} demonstrate that integrating rich commit-level signals—such as associated issues, pull requests, and file importance—via frameworks like OMEGA or ReAct-based ``Omniscient Message Generators'' significantly improves the model's ability to capture the semantic ``why'' behind a change. While Zhang \etal \cite{zhang2024commit} confirm that these LLM-generated summaries are consistently preferred by developers and better aligned with code semantics than human-written baselines, recent empirical evidence from Wu \etal \cite{wu2025empirical} reveals that performance at this level is highly sensitive to the chosen prompting paradigm. Specifically, zero-shot prompting exhibits significant volatility, as the lack of structural examples makes the model vulnerable to minor changes in role descriptions or task constraints. In contrast, few-shot paradigms, particularly those utilizing retrieval-based selection to provide semantically similar demonstrations, markedly stabilize outputs and statistically improve summary quality \cite{wu2025empirical}. Thus, at the code change level, performance depends heavily on prompting designs that can both manage cross-file dependencies and strategically utilize in-context examples to bridge the gap between raw diffs and developer-oriented intent.

%\mbox{}\\[0.04em] %
%This task, commonly referred to as commit message generation in the literature, involves summarizing code changes into concise natural language descriptions. Recent literature \cite{li2024only,imani2024context, zhang2024commit} has marked a distinct paradigm shift in commit message generation (CMG) by advancing from template--and retrieval--based methods toward LLM--driven, context--aware approaches. Earlier generation--based and retrieval--based models, while effective in pattern repetition and surface--level summarization, consistently fell short in informativeness, contextual fidelity, and developer alignment. Prompt--based LLMs have been increasingly applied to commit message generation, a key to code changes level summarization scenario.

 %\textit{Open-source contextual summarization with diff augmentation.} 

\begin{tcolorbox}[
    colback=gray!15, % background color
    colframe=black, % border color
    arc=6pt, % corner rounding
    boxrule=0.8pt, % border thickness
    left=5pt, % left padding
    right=5pt, % right padding
    top=8pt, % top padding
    bottom=8pt, % bottom padding
    fonttitle=\bfseries,
    coltitle=black, % title font
    title=Summary of Results for \RQ{1}:,
    enhanced,
    attach boxed title to top left={yshift=-3mm, xshift=5mm},
    boxed title style={
        colback=gray!40,
        boxrule=0.7pt,
        arc=8pt,
        outer arc=8pt,
        left=5pt,
        right=5pt,
        top=0.5pt,
        bottom=0.5pt,
    }
]

Our analysis shows that prompt-based code summarization has been explored across four code granularities--function/method, file/module/class, project/repository, and code changes. Function-level summarization remains dominant, with over twenty (20) studies applying diverse prompting paradigms such as Zero-Shot, Few-Shot, RAG, and CoT. This body of work illustrates clear improvements when context-enriched or project-aware prompts are used. file/module/class-level and project/repository-level summarization have received comparatively limited attention, with most existing research relying on zero-shot or RAG approaches. From our in-depth analysis, progress at these levels is mainly achieved through hierarchical and reduced-code prompting, which help models to manage larger code contexts more efficiently. Code changes-level work, focused on commits and pull requests, shows promising results using context-rich and reasoning-based prompting, though advanced RAG applications remain unexplored. Overall, task granularity strongly shapes prompt design, and further progress on higher-level and change-centric summarization needs hierarchy-aware retrieval-sensitive prompting and standardized, human-aligned evaluation.

\end{tcolorbox}

\begin{tcolorbox}[
    colback=cyan!8, % background color
    colframe=black, % border color
    arc=6pt, % corner rounding
    boxrule=0.8pt, % border thickness
    left=5pt, % left padding
    right=5pt, % right padding
    top=8pt, % top padding
    bottom=8pt, % bottom padding
    fonttitle=\bfseries,
    coltitle=black, % title font
    title= Future Directions for \RQ{1}:,
    enhanced,
    attach boxed title to top left={yshift=-3mm, xshift=5mm},
    boxed title style={
        colback=gray!40,
        boxrule=0.7pt,
        arc=8pt,
        outer arc=8pt,
        left=5pt,
        right=5pt,
        top=0.5pt,
        bottom=0.5pt,
    }
]
Future work should expand prompt-based summarization beyond the dominant function-level focus to systematically address broader code granularities. There is a need to develop scalable prompting frameworks for file/module/class and project/repository levels that can capture cross-file dependencies and architectural intent without overwhelming model context limits. The effectiveness of hierarchical and reduced-code prompting observed in current studies suggests that combining these strategies with retrieval-augmented or memory-efficient techniques (\eg long-context transformers) can enhance coherence across large systems. There is also a need to explore more Few-Shot and RAG-based prompting at the code changes level to better capture developer intent and rationale behind commits. The emergence of long-context and adaptive transformer architectures presents new opportunities to extend summarization to complex, large-scale repositories. Finally, establishing standardized, human-aligned evaluation benchmarks across different granularities will be critical to improving comparability, reproducibility, and practical applicability in real-world software engineering environments.
\end{tcolorbox}
\subsection{\RQ{2}: How are prompt engineering paradigms instantiated in LLM-driven code summarization, and which design dimensions characterize their implementations?}
\label{sec:RQ2}

%\subsection{\textbf{\textit{Mapping design choices and implementation strategies across prompt–engineering paradigms for code summarization}}}
% \label{sec:RQ2A}

Before discussing the results related to \RQ{2}, we introduce \tabref{tab:taxonomy}, which summarizes how prior studies instantiate each prompting paradigm across different code-granularity levels. The table includes: (i) \textbf{Granularity} (function/method, file/module, repository, code changes), (ii) \textbf{Paradigm} (\eg \textit{Zero-shot}, \textit{Few-shot}, \textit{RAG}, \textit{Chain-of-Thought}), (iii) \textbf{Instantiation} (\eg prompt template, retrieval strategy, exemplar selection), (iv) \textbf{LLM}, and (v) \textbf{Reference}. This organization clarifies the concrete design and implementation strategies adopted in the reviewed studies.

% \newpage

\newcolumntype{C}[1]{>{\centering\arraybackslash}p{#1}}

{\footnotesize
\begin{longtable}{|p{2.5cm}|>{\raggedright\arraybackslash}p{2.3cm}|p{3.8cm}|C{4.8cm}|C{1.4cm}|}
% {|p{2.5cm}|p{2.3cm}|p{3.8cm}|C{4.8cm}|C{1.4cm}|}
\caption{Analysis of Prompt Engineering Paradigms and their Implementation Instantiations across Multiple Code Granularities.}
\label{tab:taxonomy}\\
\hline
\rowcolor{gray!15}
\textbf{Granularity} & \textbf{Paradigm} & \textbf{Instantiation} & \textbf{LLM} & \textbf{Reference} \\
\hline
\endfirsthead
\hline
\rowcolor{gray!15}
\textbf{Granularity} & \textbf{Paradigm} & \textbf{Instantiation} & \textbf{LLM} & \textbf{Reference} \\
\hline
\endhead
\specialrule{1pt}{0pt}{0pt}
\multicolumn{5}{r}{\footnotesize Continued on next page}\\
\endfoot
\specialrule{1pt}{0pt}{0pt}
\endlastfoot

% ===== Function-Level =====
\multirow{1}{*}{\cellcolor{FunctionLevel!14}\strut \textbf{Function-Level}}
  & \textit{Zero-shot}
  & \makecell[l]{Instruction-based\\(No in-context demonstrations)}
  & \makecell[l]{ChatGPT (GPT-3.5 series)}
  & \cite{sun2023automatic} \\ \cline{4-5}

\paintFL 
  & 
  & 
  & \makecell[l]{Codex}
  & \cite{geng2024large} \\ \cline{4-5}

\paintFL
  & 
  & 
  & \makecell[l]{ GPT-4}
  & \cite{wang2024natural} \\ \cline{4-5}

\paintFL
  &
  &
  & \makecell[l]{OpenAI Codex (GPT-3-based)}
  & \cite{khan2022automatic} \\ \cline{4-5}

\paintFL
  & 
  & 
  & \makecell[l]{InCoder-6.7B}
  & \cite{fried2022incoder} \\ \cline{4-5}

\paintFL
     & 
     & 
     & \makecell[l]{ Codex \\ ChatGPT}
     & \cite{arakelyan2023exploring} \\ \cline{4-5}

\paintFL
     & 
     & 
     & \makecell[l]{DeepSeek-Coder-1.3B \\ DeepSeek-Coder-6.7B \\ DeepSeek-Coder-33B \\ StarCoder2-15B\\Llama-3-8B}
     & \cite{makharev2025code} \\ \cline{3-5}

\paintFL 
  & 
  & Translation-framing
  & \makecell[l]{ERNIE-Code (T5 architecture)}
  & \cite{chai2022ernie} \\ \cline{3-5}

\paintFL 
   &
   & Task-specific Template-based
   & \makecell[l]{CodeLlama-Instruct-7B \\ StarChat-$\beta$ (16B) \\ GPT-3.5 (gpt-3.5-turbo) \\ GPT-4 (gpt-4-1106-preview)}
   & \cite{sun2024source} \\ \cline{4-5}

\paintFL 
  & 
  & 
  & \makecell[l]{GPT-4}
  & \cite{shin2023prompt} \\ \cline{2-5}

% % \paintFL
%   & 
%   & 
%   & \makecell[l]{INCODER-6.7B}
%   & \cite{fried2022incoder} \\ \cline{4-5}

% \paintFL
%   & 
%   & Context-enriched-based
%   & \makecell[l]{ ChatGPT-3.5(for prompt variant)}
%   & \cite{Pan2024ContextPromptTuning} \\ \cline{2-5}

% ===== Few-shot =====

\paintFL 
   & \textit{Few-shot}
   & Project and Domain-Specific Selection (incl.\ one-shot)
   & \makecell[l]{CodeLlama-Instruct-7B \\ StarChat-$\beta$ (16B) \\ GPT-3.5 (gpt-3.5-turbo) \\ GPT-4 (gpt-4-1106-preview)}
   & \cite{sun2024source} \\ \cline{4-5}

\paintFL
   &  
   & 
   & \makecell[l]{code-davinci-002 \\ text-davinci-003 \\ GPT-3.5-turbo}
   & \cite{ahmed2024automatic} \\ \cline{4-5}

\paintFL 
    &
    & 
    & \makecell[l]{OpenAI Codex (GPT-3-based)}
    & \cite{khan2022automatic} \\ \cline{4-5}

\paintFL
     & 
     & 
     & \makecell[l]{ Codex \\ ChatGPT}
     & \cite{arakelyan2023exploring} \\ \cline{4-5}

\paintFL
        & 
        & 
        & \makecell[l]{DeepSeek-Coder-1.3B \\ DeepSeek-Coder-6.7B \\ DeepSeek-Coder-33B \\ StarCoder2-15B\\Llama-3-8B}
        & \cite{makharev2025code} \\ \cline{4-5}

\paintFL
      &
      & 
      & \makecell[l]{OpenAI Codex}
      & \cite{ahmed2022few} \\ \cline{4-5}

\paintFL 
   & 
   & 
   & \makecell[l]{CodeGeeX4 \\ GPT-4 }
   & \cite{zhang2025dlcog} \\ \cline{4-5}

\paintFL
     & 
     & 
     & \makecell[l]{Codex \\ GPT-3.5}
     & \cite{gao2023makes} \\ \cline{4-5}

\paintFL     
    & 
    & 
    & \makecell[l]{gpt-3.5-turbo \\ text-davinci-003 \\ LLaMA (7B, 13B, and 33B versions)}
    & \cite{YUN2024112149} \\ \cline{3-5} 

\paintFL      
  &  
  & \parbox[t]{\linewidth}{%
      
      Ad-hoc or instance-adaptive selection;\\
       }
  & \makecell[l]{GPT-4}
  & \cite{kruse2024can} \\ \cline{3-5}

\paintFL
  &  
  & Robustness-oriented selection
  & \makecell[l]{GPT-4\\GPT-3.5\\ Claude-Instant-1 and Claude-2\\CodeLlama-7B-Instruct }
  & \cite{zhang2024attacks} \\ \cline{2-5}

% ===== RAG =====

\paintFL
     & \textit{RAG}
     & Instance retrieval and ordering
     & \makecell[l]{Codex \\ GPT-3.5}
     & \cite{gao2023makes} \\ \cline{3-5}

\paintFL
   &  
   & Semantic \& structural augmentation
   & \makecell[l]{code-davinci-002 \\ text-davinci-003 \\ GPT-3.5-turbo}
   & \cite{ahmed2024automatic} \\ \cline{4-5}

\paintFL 
  & 
  & 
  & \makecell[l]{Codex}
  & \cite{geng2024large} \\ \cline{4-5}

\paintFL
     & 
     & 
     & \makecell[l]{ Codex \\ ChatGPT}
     & \cite{arakelyan2023exploring} \\ \cline{4-5}

\paintFL
        & 
        & 
        & \makecell[l]{DeepSeek-Coder-1.3B \\ DeepSeek-Coder-6.7B \\ DeepSeek-Coder-33B \\ StarCoder2-15B \\Llama-3-8B}
        & \cite{makharev2025code} \\ \cline{4-5}

\paintFL     
    & 
    & 
    & \makecell[l]{gpt-3.5-turbo \\ text-davinci-003 \\ LLaMA (7B, 13B, and 33B versions)}
    & \cite{YUN2024112149} \\ \cline{4-5} 

\paintFL
     & 
     & 
     & \makecell[l]{CodeLlama-7B/34B-Instruct \\ GPT-4o}
     & \cite{lomshakov2024proconsul} \\ \cline{3-5}

\paintFL
   & 
   & Feedback-tuned retrieval
   & \makecell[l]{GPT-Neo-2.7B \\ Code Llama-13B}
   & \cite{lu2024instructive} \\ \cline{3-5}

\paintFL
  & 
  & Meta-adaptive retrieval
  & \makecell[l]{2-layer Seq2Seq (LSTM) \\ Meta-learner via MAML}
  & \cite{zhou2023towards} \\ \cline{3-5}

\paintFL
        & 
        & Agentic Retrieval
        & \makecell[l]{GPT-4o-mini\\CodeLlama-34B-Instruct}
        & \cite{yang2025docagent} \\ \cline{2-5}

% ===== Chain of Thought =====

\paintFL
    & \textit{Chain of Thought}
    & Purpose-enhanced reasoning
    & \makecell[l]{CodeGeeX4 \\ GPT-4}
    & \cite{zhang2025dlcog} \\ \cline{4-5}

\paintFL
    & 
    & 
      
    & \makecell[l]{ChatGPT (GPT-3.5) \\ GPT-4}
    & \cite{wang2024purpose} \\ \cline{3-5}

\paintFL 
    &
    & Reasoning-enhanced self-evaluation
    & \makecell[l]{CodeLlama-Instruct-7B \\ StarChat-$\beta$ (16B) \\ GPT-3.5 (gpt-3.5-turbo) \\ GPT-4 (gpt-4-1106-preview)}
    & \cite{sun2024source} \\ \cline{3-5}

\paintFL
        & 
        & Hierarchical CoT
        & \makecell[l]{GPT-4o-mini\\CodeLlama-34B-Instruct}
        & \cite{yang2025docagent} \\ 
\hline

% ===== Module-Level =====
\multirow{1}{*}{\cellcolor{ModuleLevel!22}\strut \textbf{Module-Level}}
        & \textit{Zero-shot}
        & Hierarchical-based
        & \makecell[l]{CodeLlama-Instruct-7B \\ CodeGemma-7B \\ GPT-4}
        & \cite{sun2025commenting} \\

\hline

% ===== Repository-Level =====
\multirow{1}{*}{\cellcolor{RepoLevel!28}\strut \textbf{Repository-Level}}
        & \textit{Zero-shot}
        & Hierarchical-based
        & \makecell[l]{Llama-3 \\ StarChat2 \\ Codestral}
        & \cite{dhulshette2025hierarchical} \\ \cline{2-5}

\paintRL
      & RAG
      &  Semantic \& structural augmentation
      & \makecell[l]{ GPT-4o-mini\\Claude 3}
      & \cite{zhu2025reposummary} \\         

\hline

% ===== Commit-Level =====
\multirow{1}{*}{\cellcolor{CommitLevel!18} \strut \textbf{Code changes-Level}}
     & \textit{Zero-shot}
     & Context-enriched-based
     & \makecell[l]{GPT-4 (proprietary baseline) \\ Llama3 70B Instruct (AWQ-quantized) \\ Llama3 8B Instruct}
     & \cite{imani2024context} \\ \cline{3-5}

\paintCL
      & 
      & Instruction-based(no in-context demontration)
      & \makecell[l]{GPT-4}
      & \cite{li2024only} \\ \cline{4-5}
\paintCL
      &
      & 
      & \makecell[l]{ChatGPT (gpt-3.5-turbo) \\ Llama 2 (7B) \\ Llama 2 (70B)}
      & \cite{zhang2024commit} \\ \cline{4-5}

\paintCL
      & 
      & 
      & \makecell[l]{GPT-3.5-Turbo\\ Claude-3-Haiku \\Qwen1.5-7B-Chat\\DeepSeek-V2-Chat\\CodeQwen1.5-7B-Chat\\DeepSeek-Coder-V2-Instruct}
      & \cite{wu2025empirical} \\ \cline{2-5}

\paintCL
      & \textit{Few-shot}
      &  Project and Domain-Specific Selection (including one-shot)
      & \makecell[l]{GPT-3.5-Turbo\\ Claude-3-Haiku \\Qwen1.5-7B-Chat\\DeepSeek-V2-Chat\\CodeQwen1.5-7B-Chat\\DeepSeek-Coder-V2-Instruct}
      & \cite{wu2025empirical} \\ \cline{2-5}

\paintCL
      & \textit{Chain of Thought}
      & ReAct-style reasoning
      & \makecell[l]{GPT-4}
      & \cite{li2024only} \\

\bottomrule

\end{longtable}
}

\subsubsection{Zero-Shot Paradigm.} Out of the total corpus of 29 studies identified in our SLR, 16 papers ($\approx$55.17\% of all included works) explicitly evaluated zero-shot prompting-related strategies for code summarization. This makes zero-shot prompting the most widely tested paradigm in the literature, reflecting both its simplicity of application and its alignment with how LLMs are often deployed in practice without task-specific fine-tuning or few-shot exemplars.

Across the zero-shot papers, we observe the following recurring technique families: (i) Instruction-based – uses direct task instructions (\eg ``Summarize this function'') without examples to guide the model's response\cite{geng2024large,sun2023automatic,zhang2024commit,wang2024natural,li2024only,fried2022incoder,khan2022automatic,wu2025empirical,arakelyan2023exploring,makharev2025code}; (ii) Translation-framing – reframes code summarization as a code-to-text ``translation'' task to leverage cross-lingual semantic alignment \cite{chai2022ernie}; (iii) Hierarchical-based  – recursively summarizes smaller units (\eg functions, classes) into higher-level entities (\eg modules, repositories) to overcome context-window limitations \cite{sun2025commenting,dhulshette2025hierarchical}; (iv) Task-specific Template-based–leverages predefined docstring or summary formats to enforce consistent structure while refining instructions to highlight specific documentation goals, ensuring that outputs align with specialized project or developer needs \cite{shin2023prompt,sun2024source};  and (v) Context-enriched-based – augments the input content by inserting intrinsic contextual signals (\eg commit metadata, caller/class details) directly into the prompt. Unlike task-specific prompting, which changes the goal of summarization through customized instructions, this technique expands the information available to the model while keeping the same instruction. Therefore it remains purely zero-shot, as no examples or retrieved data are added \cite{imani2024context}. Below we explain the principal technique families observed within zero-shot followed by brief definitions and notes on typical prompt patterns.

\textbf{\textit{Instruction-based.}} Studies such as \cite{geng2024large,sun2023automatic,khan2022automatic,zhang2024commit,li2024only,fried2022incoder,wu2025empirical,arakelyan2023exploring,makharev2025code} performed code summarization via direct, task-specific natural-language instructions, for example, ``Summarize the following function'' or ``Describe the change in this commit'' with performance hinging on the precision and clarity of the prompt. For instance, recent studies by Geng \etal \cite{geng2024large} observed that zero-shot performance remains significantly lower than supervised baselines, as models fail to capture diverse developer intents without a few-shot ``warm-up''. Similarly, Makharev \etal \cite{makharev2025code} established that zero-shot LLMs exhibit higher \textit{SIDE} scores, indicating a technical divergence from ground-truth logic due to a lack of class-level skeletons or repository-wide context. This was used to show that without examples, LLMs often produce overly explanatory or ``noisy'' summaries that don't match the ground truth.Their findings reveal that without enough information, AI models frequently hit an ``adequacy threshold'' where they begin to hallucinate or miss important technical details. To address this lack of context, Fried \etal \cite{fried2022incoder} introduced \textit{InCoder}, which moves beyond simple instructions by using a ``fill-in-the-blanks'' (causal masking) approach. They forced the model to use the bidirectional context (the code before and after the mask) to ``infill'' the summary. \textit{InCoder} looks at the code both before and after the summary location, providing the model with the exact structural context that Geng \etal and Shin \etal identified as missing. This implementation allows \textit{InCoder} to generate accurate summaries in a zero-shot setting, proving that while basic instructions may fail, providing the model with the surrounding code structure effectively bridges the context gap. Similarly, Khan \etal \cite{khan2022automatic} established that while large-scale models like GPT-3 Codex can generate meaningful documentation from simple zero-shot directives, the lack of formatting guidance often leads to sub-optimal results compared to methods that include at least one example. Furthermore, Wang \etal \cite{wang2024natural} proposed \textit{SlimCode}, which simplifies source input before prompting; then they run GPT-4 in a zero-shot setting with two instruction prompts (one for code search, one for summarization). Under zero-shot prompting, \textit{SlimCode} reduces token usage and cost while maintaining and even modestly improving effectiveness (\eg $\approx$2.35\% precision gain at 50\% reduction). Moving from function-level code to code changes, Li \etal \cite{li2024only} and Zhang \etal \cite{zhang2024commit} both apply zero-shot prompting to the task of commit message generation. Li \etal \cite{li2024only} and Zhang \etal \cite{zhang2024commit} both instantiated the zero-shot prompting paradigm by relying on instruction-only task framing, where the model receives a textual directive (\eg `` Generate a commit message for this diff'') without any exemplars. In this study \cite{li2024only}, the approach is extended through \textit{ReAct} prompting, enabling GPT-4 to reason step-by-step over commit diffs and related artifacts within the same inference sequence. In contrast, Zhang \etal \cite{zhang2024commit} employed a straightforward, single-turn instruction format for ChatGPT and Llama 2 and these models can still infer the developer's intent directly from the code diffs, though they perform best on simpler changes. Together, these papers demonstrate that zero-shot prompting for code changes can be successful either through advanced reasoning chains or direct task instructions, depending on the complexity of the change.
 
\textbf{\textit{Translation-framing.}} We observed translation framing (zero-shot) in \cite{chai2022ernie}, their strategy reframes code summarization as a code→natural-language ``translation'' task using instruction prompts without demonstrations. The translation framing leverages strong semantic alignment and can be advantageous for multilingual or heterogeneous codebases.

\textbf{\textit{Hierarchical-based.}} \label{fn:hierarchical} Sun \etal and Dhulshette \etal applied \cite{sun2025commenting,dhulshette2025hierarchical}, a hierarchical-based technique, which addresses context-window limitations by recursively synthesizing summaries for larger program entities (\eg files, modules, repositories) from their constituent components (\eg functions, classes) using instruction-only prompts. These studies report improved scalability and better documentation of multi-level systems where single-step strategies struggle due to fragmented context.

\textbf{\textit{Task-specific instruction-based.}} Beyond basic prompting, researchers have explored instructional and structural task Framing to improve zero-shot reliability. Khan \etal \cite{khan2022automatic} established that their results are often sub-optimal because the model lacks the structural guidance needed to maintain a consistent format. To resolve this formatting gap, Sun \etal \cite{sun2024source} introduced structured zero-shot techniques that utilize predefined docstring frameworks and standardized templates to guide model responses. By providing fixed fields and consistent output forms, this method improves informativeness and readability while reducing variation compared to free-form prompting where the output style can vary. It is especially useful in tasks that require standardized documentation across many functions. Furthermore, Shin \etal \cite{shin2023prompt} extended this by engineering templates to capture specialized requirements—such as design intent and rationale—ensuring that the generated summaries are not just well-formatted, but also technically relevant to developer needs. Together, these studies illustrate that effective zero-shot summarization requires moving beyond simple task descriptions toward a structured framework that defines both the form and the professional depth of the documentation.

% Masked infill technique, as implemented in \cite{fried2022incoder}, exploits the capabilities of bidirectional LLMs to infill missing documentation segments or code comments within partially annotated code files, thus offering an efficient solution for the incremental maintenance of legacy codebases.

\textbf{\textit{Context-enriched-based.}} In addition, we identify context-enriched techniques, where task-intrinsic artifacts are appended to the input without retrieval. For example, Imani \etal \cite{imani2024context} constructed a single zero-shot prompt that packages all available commit context together with the instruction, reporting substantial gains over more complex prompting schemes; this is clearly zero-shot, context-enriched rather than RAG (no demonstrations, no retrieval).

\subsubsection{Few-Shot Paradigm.}

 Guided by our taxonomy--we classified few-shot prompting as cases where the prompt supplied one or more labeled demonstrations in addition to task instructions. Under this scheme, 12 of the 29 studies ($\approx$41.38\%; see \tabref{tab:taxonomy}) evaluated few-shot prompting for code summarization or closely allied documentation tasks. A closer examination of these studies revealed a clear typology of techniques, each serving distinct methodological purposes.

Across these five studies, we identified four recurring patterns of techniques within the few-shot paradigm: (i) Project and or domain-specific selection-few-shot examples tailored to the project or domain (with one-shot as the minimal case)- fixed, carefully curated set of examples reused across inputs \cite{ahmed2022few,sun2024source,wu2025empirical,zhang2025dlcog, khan2022automatic,ahmed2024automatic,gao2023makes,arakelyan2023exploring,YUN2024112149,makharev2025code}; (ii) Ad-hoc /instance-adaptive selection-examples chosen on the fly for each input, often via similarity search and light reranking \cite{kruse2024can}; and (iii) Robustness-oriented selection- few-shot examples used as defensive cases to make the model resilient to adversarial code changes \cite{zhang2024attacks}. Below we summarize representative findings for each pattern.

 \textbf{\textit{Project and Domain-Specific Selection.}} Project- and domain-specific selection represents a shift from static, generic demonstrations toward context-aware few-shot prompting, where exemplar proximity to the target code’s technical environment is treated as a primary driver of performance. Early work by Ahmed and Devanbu~\cite{ahmed2022few} demonstrated that embedding a small set of code–comment pairs from the same project enables LLMs to adapt to local naming conventions and coding styles, allowing few-shot prompting to outperform specialized fine-tuned models. 
 Subsequent studies explored practical instantiations of this idea, showing that both the number and relevance of exemplars matter. Recent research highlights the scalability and granularity of this selection strategy. While Sun \etal \cite{sun2024source} and Khan \etal \cite{khan2022automatic} emphasized the cost-efficiency of minimal shots ($k=4$ or $k=1$) to resolve style expectations. In this study, one-shot markedly outperformed zero-shot (best overall BLEU 20.63 \textit{vs.} 12.81). Wu \etal \cite{wu2025empirical} demonstrated that larger retrieved sets ($k=16$) can capture complex commit styles without model tuning and consistently outperform zero- and one-shot settings across projects. More recent work further refines this paradigm by emphasizing instance- and project-level relevance. Gao \etal~\cite{gao2023makes} empirically validated that similarity-aware demonstration selection is critical for effective in-context learning, while Arakelyan \etal~\cite{arakelyan2023exploring} showed that adapting models with retrieved in-domain or near-domain examples mitigates distribution shifts between general pre-training data and specialized software environments. This contextual grounding is most explicitly realized in P-CodeSum~\cite{YUN2024112149}, which leveraged repository-level example pools to anchor generation in project-specific coding idioms. Moreover, Makharev \etal \cite{makharev2025code} further enriched the input context by incorporating class-level structure and RAG-retrieved repository information, showing that providing higher-level contextual cues beyond isolated functions leads to more semantically acurate and informative summaries. Zhang \etal \cite{zhang2025dlcog}, and Ahmed \etal \cite{ahmed2024automatic} enhanced selection by utilizing dual-level pairs or augmenting shots with structural semantic facts. Collectively, these studies show that carefully selected, locally relevant exemplars provide a powerful and low-cost alternative to model fine-tuning for domain-adapted code summarization.

\textbf{\textit{Ad-hoc / instance-adaptive selection.}} We next noted that the same predefined--vs--ad-hoc contrast appeared in a controlled developer study of documentation prompts by Kruse \etal \cite{kruse2024can}. Their control group executed a predefined few-shot docstring prompt, whereas the experimental group issued ad-hoc prompts in a ChatGPT-like IDE. Students rated predefined few-shot outputs higher on readability, conciseness, and helpfulness, whereas professionals valued ad-hoc flexibility evidence that predefined exemplars can raise baseline quality, particularly for less experienced users. 

\textbf{\textit{Robustness-oriented selection.}} We also observed robustness-oriented few-shot, where demonstrations were used defensively. Zhang \etal \cite{zhang2024attacks} designed few-shot defenses (FSD) by adding adversarially perturbed and clean code examples with correct labels to the prompt, and complemented this with inverse-transformation instructions and meta-prompting (LLM-generated defense prompts). In a code summarization-style labeling task (``choose one word to summarize the code''), these prompt-only defenses lowered attack success rates compared to no defense, showing that adding a few carefully chosen examples in the prompt doesn't just improve summary quality, it also helps the model resist being misled by attacks.

\subsubsection{Retrieval-Augmented Paradigm.}

While RAG originated as a retrieval-based generation architecture, in prompt-based code summarization, it is treated as a prompting paradigm because it fundamentally alters the construction of the prompt by dynamically injecting externally retrieved context (\eg examples or knowledge artifacts) into the model's context window at inference time. Thus, the generation is still guided through an augmented prompt, aligning it conceptually with other prompt-engineering strategies such as zero-shot, few-shot, and chain-of-thought. This approach addresses unique challenges in code summarization tasks, including high domain specificity and the need for rapid adaptation to low-resource or project-specific settings. Retrieval-Augmented Generation (RAG) encompassed 11 of the 29 studies ($\approx$37.9\%) in our SLR.

Specifically, RAG methods can be grouped into four main families, each serving a distinct role:
(i) Instance retrieval and ordering– retrieve example code–summary pairs that are most similar to the input query and position the closest ones near the prompt, helping the model imitate their structure and wording during generation \cite{gao2023makes};
(ii) Semantic and structural augmentation – expand the prompt with richer project and code information, such as repository metadata, role labels (\eg class, method, variable), or call graphs, allowing the model to reason about relationships beyond the local snippet \cite{ahmed2024automatic,YUN2024112149,makharev2025code,lomshakov2024proconsul,arakelyan2023exploring,geng2024large,zhu2025reposummary};
(iii) Feedback-tuned retrieval– improve the retriever itself using feedback from the language model or human preferences, so it learns which examples lead to better generations and adjusts its selection strategy accordingly \cite{lu2024instructive} ; (iv) Meta-adaptive retrieval – focus on the model rather than the retriever, enabling the LLM to adapt its internal behavior dynamically for each query. Using the retrieved examples, the model performs lightweight, on-the-fly adaptation to better handle rare, domain-specific, or unseen terminology \cite{zhou2023towards}; and (v) Agentic retrieval— involves the use of autonomous agents that iteratively reason, select, and refine the most relevant external knowledge to dynamically optimize the context provided to the model \cite{yang2025docagent}. These specific retrieval strategies and their applications across studies are analyzed in detail in the following part.

\textbf{\textit{Instance retrieval and ordering.}} In the Instance Retrieval and Ordering instantiation, Gao \etal \cite{gao2023makes} systematically explore how the selection and arrangement of demonstration examples influence model performance. The authors investigated the impact of exemplar sequencing by testing random, similarity, and reverse-similarity ordering within the prompt. Their findings reveal that these RAG-based techniques lead to substantial performance gains, specifically improving code summarization BLEU-4 scores by at least $\approx$9.90\% compared to standard construction methods. Additionally, the study established that placing the most similar retrieved samples at the end of the prompt, closest to the query, generally achieves the highest accuracy.

\textbf{\textit{ Semantic and structural augmentation.}} To bridge the ``cognitive gap'' between raw code and project intent, RAG frameworks now use semantic and structural augmentation. This consensus suggests that raw code alone is insufficient for high-quality summarization. Rather than relying on simple textual similarity.

Studies by Ahmed \etal \cite{ahmed2024automatic} and Makharev \etal \cite{makharev2025code} demonstrated that injecting structured metadata, such as identifier roles, data-flow graphs, and class-level skeletons, improves factual accuracy by helping the model look beyond the immediate function scope. To evaluate this at the repository level, \cite{makharev2025code} utilized a Naive RAG pipeline to provide models with context consisting of 12, 25, or 50 retrieved code chunks alongside few-shot examples, finding that while context chunks alone did not significantly impact performance, their inclusion alongside few-shot examples led to noticeable improvements in summary quality. Similarly, Yun \etal \cite{YUN2024112149} employed a trained neural selector to classify whether a retrieved set of examples will effectively assist the LLM in generating a high-quality summary. By utilizing this neural selection process, the framework achieved a BLEU score of 20.31, demonstrating that high-quality, project-aligned shots allow frozen LLMs to outperform traditional fine-tuned tools while maintaining significantly lower computational overhead. Moreover, Lomshakov \etal \cite{lomshakov2024proconsul} argued that integrating structural dependencies (\eg call graphs) allows for structural reasoning, which effectively reduces hallucinations by grounding the model in inter-functional relationships. Arakelyan \etal \cite{arakelyan2023exploring} and Geng \etal \cite{geng2024large} proved that  using \textit{BM25} (lexical) and \textit{UniXcoder} (semantic) matching to retrieve localized exemplars allows LLMs to adapt to project-specific styles without retraining. This evolution is epitomized by Zhu \etal \cite{zhu2025reposummary}, who implemented a feature-oriented RAG paradigm that establishes traceability links between high-level features and code elements. Their findings show that this structural alignment significantly improves documentation quality, increasing manual feature coverage from $\approx$61.2\% to $\approx$71.1\%. Collectively, these studies confirm that RAG achieves its highest precision when the prompt is enriched with semantically and structurally aligned repository information.

\textbf{\textit{Feedback-tuned retrieval.}} This approach improves the retriever by incorporating feedback from the LLM's own outputs. Instead of training only on fixed similarity scores, the retriever learns from the LLM's preferences--favoring examples that lead to higher-quality or more relevant generations. In this way, the retrieval process becomes directly aligned with the goals of the generative model \cite{lu2024instructive}.

\textbf{\textit{Meta-adaptive retrieval.}} Using this meta-adaptive strategy, the model achieved notable improvements in summary quality and contextual accuracy. Unlike traditional models that remain static after training, this approach dynamically adjusts its parameters during the inference stage using retrieved examples. By performing a rapid, gradient-based adaptation at the moment of prediction, the model produces an ``exclusive model'' for each specific code snippet. This ensures that the summaries are more precise, fluent, and sensitive to task-specific nuances. The results showed consistent gains in BLEU and ROUGE scores, confirming that on-the-fly adaptation at inference time allows the model to handle rare terms and domain-specific expressions more effectively \cite{zhou2023towards}.

\textbf{\textit{Agentic Retrieval.}} Some systems combine retrieval augmentation with explicit reasoning chains. Yang \etal \cite{yang2025docagent} integrated RAG-like context retrieval (via the Searcher agent) within a multi-step Chain-of-Thought process, where retrieval is triggered based on analysis of information needs rather than being applied uniformly. This demonstrates how RAG can be embedded within larger reasoning workflows rather than serving as a standalone paradigm.

\subsubsection{Chain-of-Thought Paradigm.}

The Chain-of-Thought (CoT) paradigm is characterized by iterative or multi-step reasoning, prompting large language models to decompose summarization into sequential logical steps (shown in \tabref{tab:taxonomy}. In our corpus, 5/29 studies ($\approx$17.2\%) fell under this paradigm. 

To synthesize the evidence, we identify four representative techniques for CoT designs, each with a distinct objective and prevalence: (i) Purpose-enhanced reasoning technique – the model first figures out why a piece of code exists (its purpose) before summarizing it. This helps the model reason in steps and produce summaries that stay correct even when the code is slightly changed, such as when variable names are altered \cite{wang2024purpose,zhang2025dlcog};
(ii) Reasoning-enhanced technique – the model is asked to think through and review its own draft, comparing it against the actual meaning of the code to fix factual or logical mistakes before finalizing the summary \cite{sun2024source};
(iii) Hierarchical CoT technique - the model reasons in layers, starting from small code parts like functions, then combining those ideas to understand larger structures like classes or modules. This makes summaries of big components more coherent \cite{yang2025docagent}; and
(iv) ReAct-style reasoning technique – the model doesn't just think step by step; it also acts by looking up related information or using tools between reasoning steps. This helps it create more accurate and grounded summaries, especially for code-change or commit-level tasks \cite{li2024only}.The individual CoT variants and their empirical findings are discussed in detail in the subsequent analysis.
eee

\textbf{\textit{Purpose-enhanced reasoning technique.}} In the seminal work by Wang \etal ~\cite{wang2024purpose}, the authors introduced Purpose-Enhanced Reasoning technique, where ChatGPT is guided through iterative reasoning prompts to uncover the latent robustness of code summaries. Their method, \textit{Perthept}, mitigates brittleness against perturbations (\eg variable renaming) by combining Chain-of-Structure and Reasoning-Enhancement prompts. Evaluation results showed, \textit{Perthept} produced more robust, higher-quality summaries than standard CoT (baseline). Similarly, \textit{DLCoG} \cite{zhang2025dlcog} employs a coordinated CoT prompt that forces the model to generate local inline comments as an intermediate reasoning step before deriving the final method-level summary to improve accuracy.

\textbf{\textit{Reasoning-enhanced technique.}} As previously outlined, this technique encourages models to reason and self-correct during inference. Building on this direction, Sun \etal ~\cite{sun2024source} applied this technique, which drives models to explicitly self-evaluate summaries against the code's semantic intent. Here, the LLM asks whether its generated explanation reflects the code's underlying purpose and, if not, continues reasoning at a higher level. This iterative refinement illustrates how CoT improves beyond shallow token-level outputs by ensuring semantic fidelity.

\textbf{\textit{ Hierarchical CoT technique.}} In \textit{DocAgent}, Yang \etal \cite{yang2025docagent} exemplified this by orchestrating specialized agents, including a Reader, Searcher, and Verifier, to perform incremental, dependency-aware documentation via topological code processing. By integrating CoT reasoning, the system guides agents to logically identify context gaps, retrieve structural metadata, and verify technical facts against a rubric. This dual-paradigm approach effectively bridges the ``cognitive gap'' of large repositories, achieving a $\approx$95.74\% ``Truthfulness'' existence ratio. Collectively, these findings confirm that the combination of autonomous retrieval and step-by-step reasoning significantly outperforms traditional, single-pass RAG methods.

\textbf{\textit{ReAct-style reasoning technique.}} At the code change level, Li \etal \cite{li2024only} apply the \textit{ReAct} prompting framework, which combines step-by-step reasoning with external context retrieval (\eg pull requests, issue reports, and code structure information) to generate commit messages. Their results show that grounding reasoning in contextual actions produces more accurate and informative summaries, outperforming prior baselines and, in some aspects, human-written messages. Overall, these findings suggest that reasoning-based prompting improves the quality and robustness of code summarization, though it introduces challenges such as higher computational cost and careful prompt design.

\begin{tcolorbox}[
    colback=gray!15, % background color
    colframe=black, % border color
    arc=6pt, % corner rounding
    boxrule=0.8pt, % border thickness
    left=5pt, % left padding
    right=5pt, % right padding
    top=8pt, % top padding
    bottom=8pt, % bottom padding
    fonttitle=\bfseries,
    coltitle=black, % title font
    title=Summary of Results for {\RQ{2}}:,
    enhanced,
    attach boxed title to top left={yshift=-3mm, xshift=5mm},
    boxed title style={
        colback=gray!40,
        boxrule=0.7pt,
        arc=8pt,
        outer arc=8pt,
        left=5pt,
        right=5pt,
        top=0.5pt,
        bottom=0.5pt,
    }
]
We classify prompting methods into four paradigms based on input format, treating any structural change as a distinct category. Zero-shot prompting is the most prevalent method, appearing in $\approx$55.17\% (16/29) of the corpus. This high frequency reflects its ease of deployment; research indicates that templating and input simplification can enhance consistency and reduce token costs without the need for demonstrations. Following the recent shift toward context-aware learning, Few-shot prompting is the second most prevalent paradigm ($\approx$41.4\%), where success is driven by example relevance rather than quantity. Evidence from the corpus indicates that performance is maximized through domain-aligned or similarity-matched selection, often using the same-project examples. Notably, even minimal interventions like one-shot prompting provide substantial gains over zero-shot by effectively establishing clear style and formatting expectations. Retrieval-Augmented Generation follows in $\approx$37.9\% (11/29) of studies, consistently improving accuracy and robustness by grounding the model on real facts. Performance in RAG is further strengthened by principled retrieval ordering and the inclusion of structural data, such as repository or graph information. Finally, Chain-of-Thought (CoT) is utilized in $\approx$17.2\% (5/29) of the literature. While CoT improves semantic alignment through step-by-step reasoning, it introduces higher computational overhead.

\end{tcolorbox}

\begin{tcolorbox}[
    colback=cyan!8, % background color
    colframe=black, % border color
    arc=6pt, % corner rounding
    boxrule=0.8pt, % border thickness
    left=5pt, % left padding
    right=5pt, % right padding
    top=8pt, % top padding
    bottom=8pt, % bottom padding
    fonttitle=\bfseries,
    coltitle=black, % title font
    title= Future Directions for {\RQ{2}}:,
    enhanced,
    attach boxed title to top left={yshift=-3mm, xshift=5mm},
    boxed title style={
        colback=gray!40,
        boxrule=0.7pt,
        arc=8pt,
        outer arc=8pt,
        left=5pt,
        right=5pt,
        top=0.5pt,
        bottom=0.5pt,
    }
]
Future research should develop more automated and context-sensitive prompting methods. A key priority is to create systems that automatically select and order demonstrations, since domain-aligned and similarity-based examples consistently achieve better results than generic ones. Researchers should also refine the balance between quality and cost by using compact prompts, structured templates, and simplified inputs to reduce token usage without losing accuracy. Another direction is to strengthen retrieval pipelines by combining instance-level retrieval with richer semantic and structural context and by applying adaptive retrievers that learn from feedback or meta-learning signals. For reasoning-based paradigms, future work should focus on controlling the number and scope of reasoning steps and on integrating CoT with feedback mechanisms to improve both efficiency and robustness. %Finally, future research must address structural and generalization constraints by developing token-efficient, project-aware prompt designs that maintain high performance across diverse programming languages and constrained resource environments.
\end{tcolorbox}

\subsection{\RQ{3}: Are certain large language models consistently adopted across different code-summarization tasks and prompting paradigms?}
\label{sec:RQ3}

\begin{figure}[th!]
  \centering
  \includegraphics[width=0.95\textwidth]{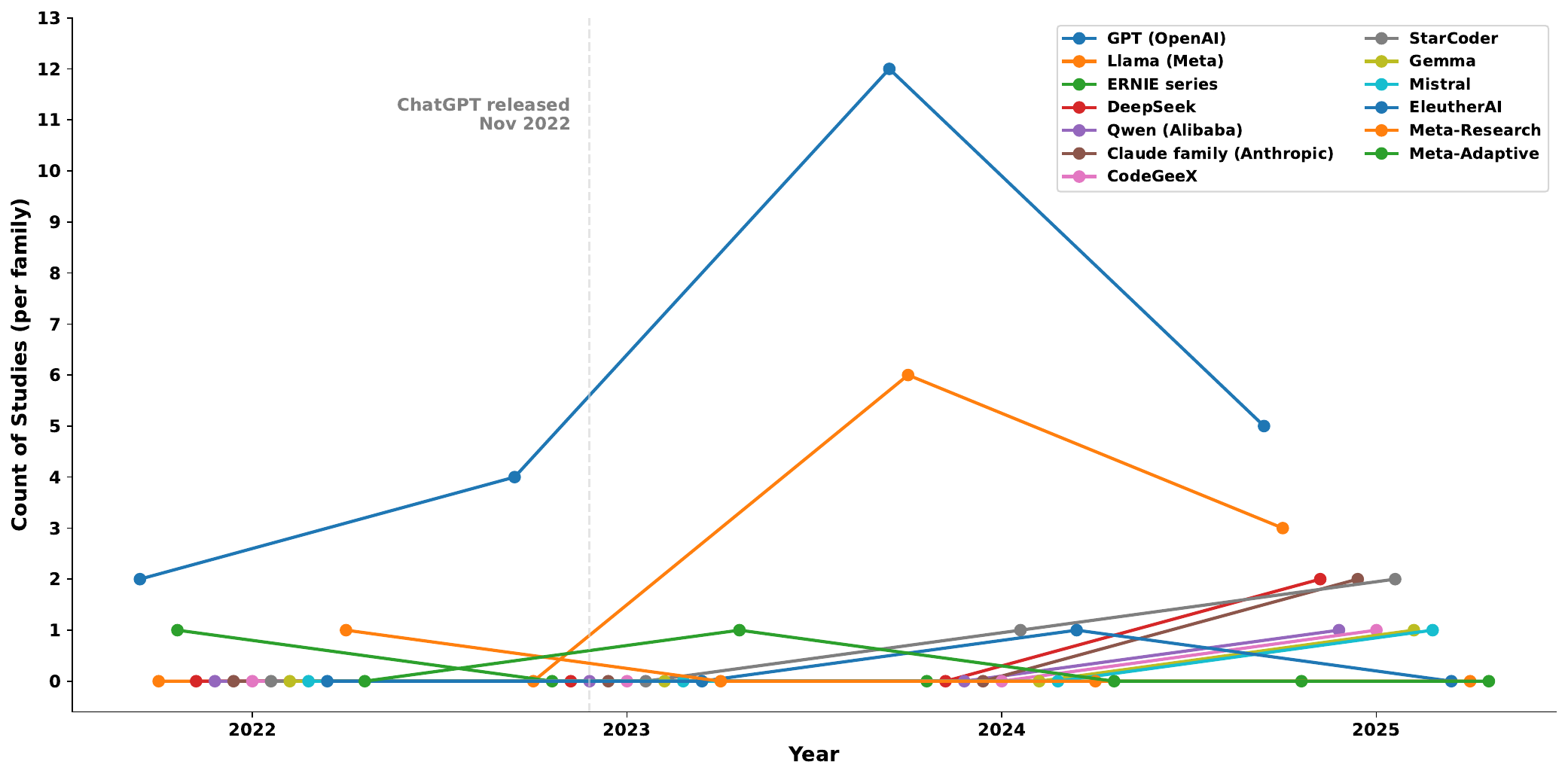}
  \caption{Trend of LLM family adoption across code-summarization studies (2020–2025). The trend highlights a gradual diversification of model adoption, reflecting the community's increasing emphasis on transparency, reproducibility, and open-source alternatives.}
  \label{fig:llmtrend}
\end{figure}

We examine whether particular model families (\eg GPT, CodeLlama, DeepSeek) consistently appear across diverse studies and methodologies, potentially suggesting that research practices are converging toward specific architectures despite the wide range of available LLMs. To address this, we analyzed all LLM occurrences in Table \ref{tab:taxonomy} and visualized their temporal adoption trend in \figref{fig:llmtrend}.

We group the models into families based on their foundational lineage (\ie their ``family tree''). Instead of looking at every individual version, we group them by the Base Model they were built from (such as GPT, Llama). This includes all different sizes and specialized versions of that specific model. For example, even though CodeLlama is a specialized version for coding, we group it under the Llama (Meta) family because Llama is its foundation. We report non-transformer models, like the Meta-Adaptive LSTM \cite{zhou2023towards}, as a separate category. Although it is a non-transformer baseline, it is included because it utilizes Model-Agnostic Meta-Learning (MAML) to adapt to retrieved code snippets at inference time.

We define the significance of column as follows from \tabref{tab:llm_classification} :

\begin{itemize}
  \item \textbf{Model Family}: Categorizes the Large Language Models into their originating organizations or series, such as the GPT series from OpenAI or the Llama family from Meta.

    \item \textbf{Key Variants:} Identifies the specific model versions and parameter scales (\eg Llama-3 70B or GPT-3.5 Turbo) used within the cited literature.

     \item \textbf{Primary Architecture:} Specifies the underlying neural architecture of the models, distinguishing between Decoder-only, Encoder-Decoder, and RNN-based structures.

     \item \textbf{Release Type:} Details the accessibility of the model, ranging from Proprietary (Closed-source) to Open-weights and Open-source. Open-weights grants access to the model's parameters, while Open-source provides the full training code and data.
\end{itemize}

\figref{fig:llmtrend} summarizes LLM usage in code summarization from 2020–2025. To capture adoption at the model-family level, we consolidate individual variants into broader families and count each family at most once per study per year, avoiding double counting. The resulting trend reveals a clear dominance of the GPT family across prompting paradigms: usage accelerates after ChatGPT's late-2022 release and peaks in 2024, coinciding with the period when GPT-3.5 and GPT-4 became the \textit{de-facto} choice for both practitioners and researchers, often favored for stable performance even without domain-specific fine-tuning. This sustained rise also highlights how proprietary, high-capacity models have anchored prompt-driven code summarization research by offering dependable baselines with minimal setup overhead.

\begin{table}[h!]
\centering
\caption{Classification of LLM Families in Code Summarization Research}
\label{tab:llm_classification}
\small % Slightly smaller font to fit more content comfortably
\renewcommand{\arraystretch}{1.4} % Adds breathing room between rows
\begin{tabularx}{\textwidth}{@{} l >{\RaggedRight\arraybackslash}X l l @{}}
\toprule
\textbf{Model Family} & \textbf{Key Variants} & \textbf{Primary Architecture} & \textbf{License Type} \\ \midrule

\textbf{GPT (OpenAI)} & GPT-4 (o/1106), GPT-3.5 (Turbo), Codex, text-davinci-003, code-davinci-002 & Decoder-only & Proprietary (Closed-source) \\

\textbf{Llama (Meta)} & Llama-2 (7B/70B), Llama-3 (8B/70B), CodeLlama (Instruct/7B/13B/34B) & Decoder-only & Open-weights \\

\textbf{ERNIE series} & ERNIE-Code & Encoder–Decoder (Seq2Seq) & Restricted / Research-only \\

\textbf{DeepSeek} & DeepSeek-Coder (1.3B/6.7B/33B) & Decoder-only (MoE*) & Open-weights \\

\textbf{Qwen (Alibaba)} & Qwen1.5-7B-Chat, CodeQwen1.5-7B-Chat & Decoder-only  & Open-weights \\

\textbf{Claude family (Anthropic)} & Claude-3-Haiku & Decoder-only  & Proprietary(Closed-source) \\

\textbf{CodeGeeX} & CodeGeeX4 & Decoder-only  & Restricted / Research-only \\

\textbf{StarCoder} & StarCoder2-15B, StarChat-$\beta$-16B, StarChat2 & Decoder-only & Restricted / Research-only \\

\textbf{Gemma} & CodeGemma-7B & Decoder-only & Open-weights \\

\textbf{Mistral} & Codestral & Decoder-only & Open-weights \\

\textbf{EleutherAI} & GPT-Neo (2.7B) & Decoder-only & Open-source \\

\textbf{Meta-Research} & InCoder (6.7B) & Decoder-only (Causal) & Open-weights \\

\textbf{Meta-Adaptive} & 2-layer Seq2Seq (LSTM) & Encoder–Decoder (Seq2Seq) & Open-source \\ 

 \bottomrule

\end{tabularx}
\end{table}
% \footnote{In this review, 'Open-weights' refers to models where parameters are publicly accessible but training data/code remain proprietary, whereas 'Open-source' denotes models with fully transparent training pipelines and datasets,while 'Restricted' refers to models with open weights available only for non-commercial research purposes.}

Before this surge, the trend graph highlights Codex (a GPT-3 variant) which played a pivotal transitional role, bridging fine-tuned encoder and encoder-decoder architectures such as CodeBERT, CodeT5 and PLBART with the newer instruction-tuned GPT series. While Codex itself is not represented as a separate curve in the graph, being subsumed under the GPT family, we can safely consider it as demarcation point where the research community shifted from domain-specific fine-tuning toward general-purpose instruction-tuned models, paving the way for models that could handle a wider variety of summarization tasks without re-training of the model's parameters. Following this transition, the \figref{fig:llmtrend} visualization shows a gradual diversification of model adoption, particularly from 2024 to 2025. While GPT usage shows a decline in 2025, families such as DeepSeek, StarCoder, and Claude maintain or increase their presence. The Llama, StarCoder, and DeepSeek-Coder families begin to show clear upward trends, reflecting a growing reliance on open-source alternatives. These models are most often employed in retrieval-augmented systems,  or context-enriched paradigms, where external information or architectural signals complement the base model's reasoning.

% Requires:
% \usepackage{booktabs,tabularx}
% (Optional, for better spacing) \usepackage{array}

\begin{table}[t]
\centering
\caption{Distribution of prompting paradigms by LLM family in code-summarization studies. Each cell reports the number of studies in which at least one model from the family was used under the corresponding prompting paradigm. A study contributes at most once per (family, paradigm) cell to avoid double counting.}
\label{tab:family_paradigm_distribution}
\small
\renewcommand{\arraystretch}{1.25}

\newcolumntype{C}{>{\centering\arraybackslash}X}
\begin{tabularx}{\textwidth}{@{} l C C C C C >{\raggedright\arraybackslash\nohyphens}p{4.0cm} @{}}
\toprule
\textbf{Model Family} & \textbf{Zero-shot} & \textbf{Few-shot} & \textbf{RAG} & \textbf{CoT} & \textbf{Total} & \textbf{Most frequent paradigm} \\
\midrule
GPT (OpenAI)        & \cellcolor{yellow!25}\textbf{11} & \cellcolor{yellow!25}\textbf{18} & \cellcolor{yellow!25}\textbf{13} & \cellcolor{yellow!25}\textbf{9} & \cellcolor{yellow!25}\textbf{51} & Few-shot  \\

LLaMA (Meta)       & 8  & 4 & 5 & 2 & 19  & Zero-shot \\

StarCoder          &  3 & 2 & 1 & 1 & 7 &  Zero-shot \&Few-shot    \\

DeepSeek           &  5 & 5 & 3 & 0 & 13 & Zero-shot \& Few-shot \\

Qwen (Alibaba)     &  2 & 2 & 0 & 0 & 4 & Zero-shot \& Few-shot \\

Claude family      &  1 & 3 & 1 & 0 & 5 &  Few-shot \\

CodeGeeX Family    &  0 & 1 & 0 & 1 &  2 & Few-shot \& CoT \\
Gemma              &  1 & 0 & 0 & 0 & 1 & Zero-shot  \\

Mistral            &  1 & 0 & 0 & 0 & 1 & Zero-shot   \\

ERNIE series       &  1 & 0 & 0 & 0 & 1 & Zero-shot \\

EleutherAI         & 0  & 0  & 1 & 0  & 1 & RAG \\
Meta-Research      &  1 & 0 & 1 & 0 & 2 &   Zero-shot \& RAG \\

Meta-Adaptive      &  0 & 0 & 1 & 0 &  1  & RAG \\
\midrule
\textbf{All families (sum)} & 34 & 35 & 26 & 13 & 108 & Few-shot \\

\bottomrule
\end{tabularx}
\end{table}

This movement toward open-source models indicates a methodological and cultural shift. While GPT-series models remain the reference standard for consistency and evaluation, the increasing use of open-source families underscores the research community's drive toward reproducibility, transparency, and adaptability. This transition began with early foundational work in 2022 \cite{khan2022automatic} and accelerated into a shift toward hybrid prompting strategies in studies published after 2023 \cite{geng2024large, li2024only, sun2024source}. These approaches integrate the model's instruction-following capabilities with external logic, such as analyzing the code's structure (structural augmentation) or using feedback loops to retrieve more accurate documentation (feedback-tuned retrieval).

The analysis summarized in \tabref{tab:family_paradigm_distribution} shows that the GPT family is the most frequently adopted solutions, appearing 51 times across all paradigms. A critical distinction emerges between study prevalence and experimental volume: while Zero-shot prompting is the most prevalent strategy in terms of unique studies (as shown in \tabref{tab:taxonomy}), Few-shot prompting represents the highest volume of experimental application with 35 instances. This indicates that although many researchers include Zero-shot as a baseline to test built-in knowledge, the majority of intensive testing and optimization, particularly for models like GPT and DeepSeek- is focused on Few-shot configurations. Conversely, Chain-of-Thought (CoT) remains the most underrepresented paradigm with only 13 instances, followed by Retrieval-Augmented Generation (RAG) with 26, suggesting that multi-step reasoning is still less common in the current code summarization literature than example-based or direct instruction approaches.

\begin{tcolorbox}[
    colback=gray!15, % background color
    colframe=black, % border color
    arc=6pt, % corner rounding
    boxrule=0.8pt, % border thickness
    left=5pt, % left padding
    right=5pt, % right padding
    top=8pt, % top padding
    bottom=8pt, % bottom padding
    fonttitle=\bfseries,
    coltitle=black, % title font
    title=Summary of Results for {\RQ{3}}:,
    enhanced,
    attach boxed title to top left={yshift=-3mm, xshift=5mm},
    boxed title style={
        colback=gray!40,
        boxrule=0.7pt,
        arc=8pt,
        outer arc=8pt,
        left=5pt,
        right=5pt,
        top=0.5pt,
        bottom=0.5pt,
    }
]
The analysis shows that the GPT family overwhelmingly dominates recent code-summarization research, establishing a grounded baseline  across prompting paradigms. The sharp post-2022 rise corresponds to ChatGPT's release, after which GPT-3.5 and GPT-4 became standard choices for zero-shot, few-shot, and reasoning-based tasks.
Meanwhile, open-source families such as Llama, StarCoder, and DeepSeek-Coder exhibit a gradual upward trend after 2023, signaling growing confidence in transparent, reproducible alternatives.

\end{tcolorbox}

\begin{tcolorbox}[
    colback=cyan!8, % background color
    colframe=black, % border color
    arc=6pt, % corner rounding
    boxrule=0.8pt, % border thickness
    left=5pt, % left padding
    right=5pt, % right padding
    top=8pt, % top padding
    bottom=8pt, % bottom padding
    fonttitle=\bfseries,
    coltitle=black, % title font
    title= Future Directions for {\RQ{3}}:,
    enhanced,
    attach boxed title to top left={yshift=-3mm, xshift=5mm},
    boxed title style={
        colback=gray!40,
        boxrule=0.7pt,
        arc=8pt,
        outer arc=8pt,
        left=5pt,
        right=5pt,
        top=0.5pt,
        bottom=0.5pt,
    }
]
Future work should (i) decouple \emph{adoption} from \emph{capability} via controlled cross-family baselines (same prompts, budgets, decoding), and (ii) quantify paradigm--family interaction effects (family $\times$ paradigm $\times$ context budget).  
To reduce the bias triggered by ``common adoption'', studies should report reproducibility metadata (prompt templates, retrieval corpora, inference settings) and cost/performance trade-offs.  
\end{tcolorbox}
\subsection{\RQ{4}:Which benchmarks and metrics have been used to assess the quality of prompt-based code summarizers, and how do these measures align with human and LLM evaluations?}
\label{sec:RQ4}

\tabref{tab:benchmarks-metrics} provides a structured overview of how different studies have evaluated prompt-based code summarization. 
Each column in the table serves a specific purpose:
\begin{itemize}
  \item \textbf{Paradigm}: describes the prompting method or paradigm applied in each study (\eg zero-shot learning).

    \item \textbf{Benchmarks}: lists the datasets or corpora used to test the summarization approach.
    
    \item \textbf{NL Metrics}: identifies which automatic evaluation scores were applied (\eg BLEU, ROUGE, METEOR, embedding-based metrics).
    
    \item \textbf{Human Evaluation}: indicates whether a study included human judgments of summary quality (\ding{51}for yes, \ding{55} for no).

     \item \textbf{LLM Evaluation}: indicates whether a study included LLM-based assessments of summary quality (\ding{51}for yes, \ding{55} for no).
    
    \item \textbf{Ref}: provides the citation for the primary study being reported.
\end{itemize}

To assess the performance of prompt-based code summarization methods, studies employ a wide variety of benchmark datasets and evaluation metrics. This section summarizes both aspects--first outlining the commonly used benchmarks, followed by an analysis of the evaluation metrics and practices. 

\textit{Benchmark Selection and Coverage.}
In code summarization, benchmarks serve as standardized datasets that enable consistent evaluation and cross-comparison of code summarization models. By fixing the task formulation, language coverage, and the granularity of the code units, benchmarks strongly influence how well models generalize across programming languages and project domains. Our analysis indicates that evaluation in prompt-based code summarization research is still heavily concentrated at the function/method (sub-routine) level: we found 21/29 papers (72.41\%) evaluating at this granularity, while only a smaller subset of 8 studies(27.59\%) extended evaluation to larger artifacts such as files, classes, or full repositories.

With respect to the overall pool of studies included in our SLR, $\sim$55\% relied on large-scale, multi-language benchmarks such as CodeSearchNet (CSN), CodeXGLUE, TL-CodeSum, and the recently integrated CommitBench. These datasets support systematic comparisons and cross-language generalization. About 24\% of studies employed specialized or domain-oriented resources, including FunCom, TLC, OMG, and TypeWriter. Here, ``domain-oriented'' refers to datasets curated around narrower application contexts or data-generation assumptions—for example, restricting the corpus to a specific programming language, API ecosystem, documentation style, or project type—thereby enabling targeted stress tests that broad multi-language benchmarks may not capture. For instance, FunCom and TLC are function-level code comment benchmarks used in a single study in our corpus \cite{geng2024large}. While the datasets differ in corpus scale and annotation focus, with FunCom emphasizing broad project coverage and TLC emphasizing curated high quality comments, the summarization target remains individual methods, consistent with our function level definition.

% \ANTONIO{Please adjust the FunCom/TLC characterization to match your exact definitions and citations.}

Finally, roughly 14\% of studies constructed custom benchmarks at varying levels of code granularity, typically to capture artifacts or tasks not covered by public datasets. This is particularly common for class- or file-level summarization, where authors assemble corpora by mining repositories, applying structural filters (\eg minimum size, complexity, or dependency constraints), and pairing code artifacts with documentation signals including, inline comments, class headers, README sections, or commit messages).  Sun \etal \cite{sun2025commenting} address class-level code summarization by building a custom benchmark of 10 hand-selected Java projects, where class artifacts are paired with documentation signals and evaluated using expert review and LLM-based judgment.
At the repository level, Zhu \etal \cite{zhu2025reposummary} construct a custom multi-repository dataset from Dronology, eTour, and iTrust, augmented with 26 curated commits, to evaluate summaries that capture cross-file and system-level intent.
Finally, at the commit level, Zhang \etal \cite{zhang2024commit} curate a dataset of 7,661 Java commit diffs mined from top GitHub repositories and evaluate zero-shot LLM-based commit message generation using BLEU and ROUGE-L, complemented by controlled human evaluation.

% A representative example for class-level code summarization is \textcolor{red}{[CITE]}, where the authors built a benchmark by \textcolor{red}{[DESCRIBE SOURCE + FILTERING + PAIRING STRATEGY]} to evaluate \textcolor{red}{[MODEL/PARADIGM]} under \textcolor{red}{[EVALUATION SETUP]}. With a different granularity, it is also worth mentioning that \textcolor{red}{[CITE]} evaluates summarization over \textcolor{red}{[FILES/REPOSITORIES/COMMITS]} by constructing \textcolor{red}{[DATASET NAME]} from \textcolor{red}{[SOURCE]}, highlighting an emerging shift toward higher-level summaries that better reflect realistic comprehension and maintenance workflows.

Ultimately, only a small fraction of studies ($\approx$7\%) employ controlled or adversarial evaluation settings. A notable exception is Zhang \etal \cite{zhang2024attacks}, who systematically assess LLM robustness in code summarization by applying semantics-preserving adversarial perturbations to a controlled subset of 1,000 Python programs from the Srikant \etal \cite{srikant2021generating} dataset. Using white-box attack methods, they measure robustness through attack success rate (ASR) and accuracy across multiple LLMs, revealing vulnerabilities that standard benchmarking overlooks -- namely, that models can maintain high overlap-based scores on perturbed code while producing semantically incorrect summaries. 

\begin{figure}[ht]
    \centering
    \includegraphics[width=0.65\textwidth]{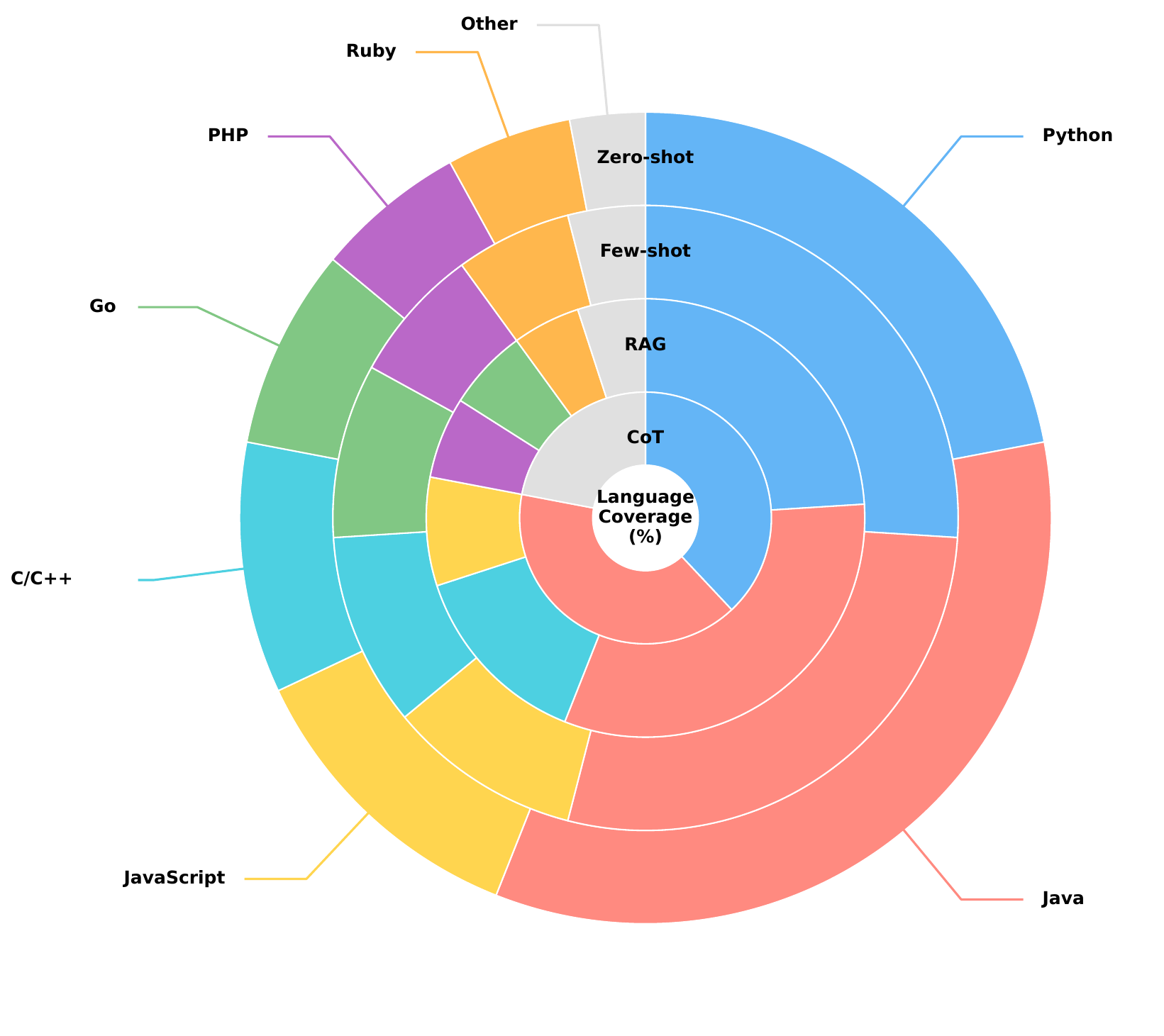}
    \caption{Distribution of programming languages across 29 prompt-based code summarization studies. Each ring represents a prompting paradigm (Zero-shot, Few-shot, RAG, Chain-of-Thought), and each colored segment corresponds to a programming language, illustrating the dominance of Python and Java benchmarks across paradigms.}
    \label{fig:benchmark}
\end{figure}

The limited adoption of adversarial evaluation is compounded by another form of narrowness: the choice of programming languages, which concentrates heavily on just two languages across all prompting paradigms. \figref{fig:benchmark} illustrates this phenomenon through a concentric breakdown of \emph{language coverage} across the four prompting paradigms in this SLR (Zero-shot, Few-shot, RAG, and CoT). Each ring corresponds to a paradigm, and each colored slice represents the proportion of studies evaluating that paradigm on a given programming language.
The figure reveals a dominant focus on \texttt{Java} and \texttt{Python} across all paradigms, while \texttt{JavaScript}, \texttt{C/C++}, and \texttt{Go} appear substantially less often and are typically evaluated only in few-shot or retrieval-enhanced settings. The long tail of languages (\eg, \texttt{PHP}, \texttt{Ruby}, and ``Other'') underscores that cross-language generalization remains largely unexplored: there is limited evidence that prompting strategies perform consistently beyond the Java/Python benchmarks that dominate current practice.

%\ANTONIO{Which is the trend??}

% Regarding the use of benchmarks and how they related to programming languages and prompt paradigms, \figref{fig:benchmark} presents a  of programming languages 
% This trend in benchmark selection is reflected in language coverage patterns shown in . Most studies rely on datasets centered around six dominant programming languages--Python, Java, JavaScript, Go, PHP, and Ruby--while only a few extend to languages such as C/C++ or domain-specific corpora. The distribution also varies by paradigm: Zero-shot and RAG studies exhibit the broadest cross-language coverage, whereas Few-shot and Chain-of-Thought experiments remain concentrated in Python and Java—accounting for approximately 70\% of evaluations in these paradigms, due to robust model support and superior dataset accessibility. Overall, these trends indicate a steady shift from narrow, function-level evaluations toward more realistic, multi-language, and context-rich benchmarks that better reflect the complexities of real software development environments.

% Regarding the distribution of programming languages, Java receives most of the attention (\figref{fig:}) -- with more than \textcolor{red}{XX\%} of studies (YY/29) evaluating prompt-based approaches implementing the whole spectrum of paradigms (\ie Zero-shot, Few-shot, RAG and CoT).  

% Similarly to Java, Python receives wide coverage both in terms of paradigms ....

% \ANTONIO{Afia, here we need more work. Provide numbers etc...}

\textit{NL Metrics Trends and Shortcomings.} Among the 29 studies in \tabref{tab:benchmarks-metrics}, evaluation remains dominated by \emph{overlap-based} metrics--scores that measure how much a generated summary \emph{lexically matches} a reference summary (\ie ground truth) through shared $n$-grams or shared subsequences. Common examples include BLEU (precision of overlapping $n$-grams), ROUGE-L (longest common subsequence), and METEOR (word-level matches with stemming and synonyms). While these metrics are simple and reproducible, they systematically favor lexical copying and can fail to credit semantically equivalent paraphrases expressed with different wording. Despite these known limitations, overlap-based scores frequently serve as the primary quantitative evidence when papers claim improvements in prompt-based code summarization.

This trend is reflected in the distribution of metrics in \tabref{tab:benchmarks-metrics}. Specifically, 22 of 29 studies ($\approx 76\%$) report at least one overlap metric, with BLEU being the most prevalent, often accompanied by ROUGE-L and METEOR. Their popularity stems from ease of computation and their long-standing role as the de facto benchmarking standard. In contrast, only 3 of 29 studies ($\approx 10\%$) incorporate embedding-based or learned semantic metrics (\eg BERTScore, BLEURT, or other semantic-similarity measures). Notably, none of these studies relies on semantic metrics alone; instead, they report them \emph{alongside} overlap-based scores, suggesting that overlap metrics remain the default reporting choice, while semantic metrics are used (when present) as complementary evidence rather than as a replacement.

While this imbalance partly reflects that embedding-based evaluation for code summaries is still not uniformly established, \tabref{tab:benchmarks-metrics} also reveals an inconsistent adoption of newer semantic metrics specifically designed to address the weaknesses of lexical overlap. For instance, SIDE, introduced by Mastropaolo \etal \cite{mastropaolo2024evaluating} to capture code--summary semantic alignment, appears in only 2 of 29 studies ($\approx 7\%$). Even among the most recent 2025 studies from, we find 4 ($\approx 57\%$) still rely primarily on BLEU/ROUGE/METEOR variants, while only 2 ($\approx 29\%$) include any embedding-based metric and just 1 ($\approx 14\%$) reports SIDE.

% Overall, \tabref{tab:benchmarks-metrics} highlights a persistent gap between recognized limitations of lexical overlap and current evaluation practice. Overlap-based metrics remain privileged for comparability and convenience, whereas semantic metrics--including SIDE--are adopted sporadically rather than systematically. This suggests that, for prompt-based code summarization, stronger evidence will likely require more consistent multi-dimensional evaluation pipelines that combine overlap metrics (for continuity with prior work) with semantic and task-aware measures that better reflect meaning preservation and code faithfulness.

A smaller but conceptually important subset of studies broadens evaluation to capture other dimensions of summary quality, including readability, informativeness, and robustness. For example, some works report readability indicators (\eg Flesch--Kincaid Grade~\cite{kincaid1975derivation}) and informativeness proxies (\eg TF--IDF weighting or summary length), while robustness-oriented evaluations explicitly test how fragile summarizers are under adversarial conditions. In particular, Zhang \etal \cite{zhang2024attacks} reports both \emph{clean} accuracy and \emph{Attack Success Rate (ASR)}, where ASR measures the fraction of instances for which adversarial perturbations successfully steer the model toward incorrect or misleading summaries (\eg changing the intended behavior, omitting key functionality, or following malicious instructions embedded in the prompt/context). Finally, several studies complement automatic metrics with human assessments (\eg informativeness, fluency, and factual/functional accuracy) or with LLM-based evaluations--often referred to as \emph{LLMs-as-Judge}~\cite{dhulshette2025hierarchical,sun2024source,wu2025empirical,li2024only,zhou2023towards,wang2024purpose,zhang2024commit}, where an LLM (\eg ChatGPT/GPT-4) is prompted to rate or compare summaries (\eg for semantic accuracy and faithfulness) yielding human-level performances.

%This trend reflects a gradual shift away from purely surface-level lexical overlap toward evaluation protocols that more directly target meaning preservation, alignment with code behavior, and human-perceived quality.

{\footnotesize
\rowcolors{2}{gray!15}{white}
\begin{longtable}{p{2.5cm} p{4cm} p{4cm} p{1cm} p{1cm} c}
\caption{Benchmarks and Metrics Used to Assess Prompt-based Code Summarization. 
\cmark\,indicates studies with human or LLM-based evaluation; 
\xmark\,indicates no such evaluation.}
\label{tab:benchmarks-metrics}
\\
\toprule
\textbf{Paradigm} & \textbf{Benchmarks / Datasets} & \textbf{NL Metrics} & \textbf{Human Eval} & \textbf{LLM Eval} & \textbf{Ref} \\
\midrule
\endfirsthead

\toprule
\textbf{Paradigm} & \textbf{Benchmarks / Datasets} & \textbf{NL Metrics} & \textbf{Human Eval} & \textbf{LLM Eval} & \textbf{Ref} \\
\midrule
\endhead

Zero-shot & 10 Java projects (hand-selected) & No automatic metrics;Study relied on human(5 experts) and LLM as judge (GPT-4) & \cmark  & \cmark  & \cite{sun2025commenting} \\

Zero-shot & CSN-Python (CodeSearchNet Python functions) & BLEU-4, METEOR, ROUGE-L  & \xmark & \xmark & \cite{sun2023automatic} \\

Zero-shot & Code search and code summarization tasks (CodeBERT, CodeT5 on CodeSearchNet) and GPT-4 evaluation on same tasks &  BLEU-4 (code summary),  & \xmark & \xmark &  \cite{wang2024natural} \\

Zero-shot & mCoNaLa (CoNaLa dataset translated into Spanish, Japanese, Russian for code→text); plus a Chinese code summary dataset (crowd-sourced) & BLEU-4, ROUGE-L, chrF (character n-gram F-score) & \xmark & \xmark & \cite{chai2022ernie} \\

Zero-shot & CodeXGLUE (CSN-Python-14,918 functions, CT), HumanEval, MBPP & BLEU & \cmark & \xmark & \cite{shin2023prompt} \\

Zero-shot   & CodeXGLUE, HumanEval, TypeWriter & BLEU-4 & \xmark & \xmark &  \cite{fried2022incoder} \\

Zero-shot & Java codebase in the telecom domain (a BSS) & ROUGE-L, BLEU, BERTScore, Semantic Similarity and LLM as judge (GPT-4) & \cmark & \cmark & \cite{dhulshette2025hierarchical} \\

Zero-shot   & OMG benchmark (381 commits from 32 Apache Java projects) & BLEU, METEOR, ROUGE-L & \cmark & \xmark & \cite{imani2024context} \\

Zero-shot & 7,661 commit diffs from top 1,000 GitHub Java projects & BLEU-Moses, BLEU-Norm, BLEU-NLTK, ROUGE-L & \cmark & \xmark & \cite{zhang2024commit} \\
\hline

Few-shot & 2 Python functions (controlled dev study) & No NL metrics  & \cmark  & \xmark & \cite{kruse2024can} \\

Few-shot  & 1,000 samples randomly selected from the Python dataset of Srikant \etal for the code summarization task & Accuracy (clean), Attack Success Rate (ASR) & \xmark & \xmark & \cite{zhang2024attacks} \\

Few-shot & CodeXGLUE (Java, Python, Ruby, JS, Go, PHP; project \& same-project split) & BLEU-4 (smoothed)	& \xmark & \xmark & \cite{ahmed2022few} \\

\hline

RAG & Repository Dataset (Dronology, eTour, iTrust)Java language and Commit Dataset(26 real-world code commits curated from three GitHub projects) & No automated metric & \cmark & \cmark & \cite{zhu2025reposummary} \\

RAG  & CodeSearchNet (CSN-Java, CSN-Python, CSN-JavaScript, CSN-PHP, CSN-Ruby) & BLEU-4, ROUGE-L, METEOR, chrF & \xmark &\xmark & \cite{lu2024instructive} \\

RAG  &  Custom C/C++ benchmark (functions from open-source C/C++ repos: ffmpeg, openssl, wrk, llvm, libuv) &  No automated metric & \cmark  & \cmark  & \cite{lomshakov2024proconsul} \\

RAG & Java (JCSD, $\sim$87k method–summary pairs) and Python (PCSD, $\sim$108k function–docstring pairs) datasets from GitHub. & BLEU-4, ROUGE-L, METEOR & \cmark & \xmark & \cite{zhou2023towards} \\

\hline

CoT &  TL-CodeSum (Java), CodeSearchNet (Python/Java), Wan \etal Python, and a reasoning-required filtered subset & Precision, F1, ChatGPT-based semantic accuracy & \xmark & \cmark & \cite{wang2024purpose}\\
\hline

Zero-shot + Few-shot+ CoT & CodeSearchNet corpus (Java, Python, C) + new datasets for Erlang, Haskell, Prolog & BLEU, ROUGE-L, METEOR, BERTScore, SIDE & \cmark & \cmark & \cite{sun2024source} \\

Zero-shot(baseline) +  RAG  & FunCom and TLC Java code-comment datasets & BLEU, ROUGE-L, METEOR & \xmark & \xmark & \cite{geng2024large}  \\

Zero-shot + Few-shot & CodeSearchNet (6 languages: Python, Java, PHP, Go, JavaScript, Ruby) & BLEU-4 (smoothed), Flesch-Kincaid Grade (readability), TF-IDF (informativeness), Doc Length & \xmark & \xmark & \cite{khan2022automatic} \\ 

Zero-shot + Few-shot+ RAG  & CodeSearchNet-JS (org/project/module split), The Vault (cross-dataset gen.) & BLEU-4 (summ.) & \xmark & \xmark & \cite{arakelyan2023exploring} \\

Zero-shot + Few-shot + RAG & Modified ClassEval, Modified CodeSearchNet (Python, class- and repo-level, see Table I/II) & BLEU-4, ROUGE-L, METEOR, BERTScoreF1, BLEURT, SIDE (contrastive) & \xmark & \xmark & \cite{makharev2025code} \\

Zero-shot + CoT & 32 Apache project commit data (curated by Mannan \etal) & BLEU, ROUGE-L, and METEOR & \cmark & \xmark & \cite{li2024only} \\

Zero-shot + Few-shot & CommitBench(Java,Python,JavaScript,C/C++,Go,Others) & BLEU, METEOR, ROUGE-L, and Cider & \cmark & \cmark & \cite{wu2025empirical} \\

Few-shot+ RAG & CodeSearchNet (Java, Python, Ruby, JavaScript, Go, PHP) (cross-project, same-project, all de-duplicated) & BLEU-4 (smoothed BLEU-CN and BLEU-DC), ROUGE-L, METEOR & \xmark & \xmark & \cite{ahmed2024automatic} \\

Few-shot+ RAG  & CodeSearchNet (Java), TLCodeSum (Java) & BLEU-4, ROUGE-L, METEOR & \xmark & \xmark & \cite{gao2023makes} \\

Few-shot+ RAG  & Six project-specific code summarization datasets (Python, Java, Go, JS, PHP, Ruby) from open-source repos, sampled from CodeXGLUE & BLEU-4, ROUGE-L & \xmark & \xmark & \cite{YUN2024112149} \\

Few-shot + CoT & Author-constructed Java datasets & BLEU, METEOR, ROUGE-L & \xmark & \xmark & \cite{zhang2025dlcog} \\

RAG  + CoT & Custom dataset(164 top-starred Python repositories) & No automated metric & \xmark & \cmark & \cite{yang2025docagent} \\

\bottomrule
\end{longtable}
}

To reduce the time and effort required to assess qualitative properties of generated summaries -- such as \textit{fluency}, \textit{conciseness}, and \textit{factual correctness or adequacy} -- several recent studies explicitly compare human-provided ratings with LLM-based judgments, examining how closely LLMs can approximate human assessments of summary quality \cite{li2024only,dhulshette2025hierarchical,sun2025commenting,crupi2025effectiveness}. 

In this regard, \tabref{tab:human-llm} offers a comprehensive view of human- and LLM-based evaluation in code summarization. The table consolidates evidence from recent studies and organizes it along two dimensions: (i) \textbf{Human/LLM Assessment Design}, which captures how evaluators—human experts, developers, students, or LLM judges—are recruited and instructed (\eg Likert-scale criteria such as readability, accuracy, informativeness); and (ii) \textbf{Level of Agreement}, which reports how closely LLM judgments and/or automatic metrics track human ratings. 

Concretely, multiple studies report \emph{strong human–LLM alignment}: for instance, Sun \etal \cite{sun2025commenting} show that GPT-4 scores mirror software engineers' ratings of class-level summaries (readability, accuracy, informativeness, and conciseness), while Lomshakov \etal \cite{lomshakov2024proconsul} find high agreement between expert annotations and GPT-4o proxy scoring for dimensions such as sufficiency and triviality. Similarly, Zhu \etal \cite{zhu2025reposummary} adopt a multi-LLM judging setup (two primary LLM judges plus a third tie-breaker) and report moderate-to-strong correlations between LLM-judge scores and human evaluation, supporting the feasibility of scalable evaluation at the repository level.

% \begin{table}[htbp!]
% \footnotesize
% \caption{\footnotesize Human/LLM Judgment Alignment with Prompt-based Summaries.}
% \label{tab:human-llm}
% \rowcolors{2}{gray!15}{white}
% \centering
% \begin{tabular}{p{3.3cm} p{3.2cm} p{2.8cm} p{5.0cm}}
% \toprule
% \textbf{Study (Citation)} & \textbf{Human/LLM Assessment Design} & \textbf{Level of Agreement} & \textbf{Key Findings} \\
% \midrule

% \cite{geng2024large} & 400 comments, rated on 3 criteria (5-point scale) & Strong alignment Metric and human scores  & LLM prompts rated more fluent; 10-shot LLMs preferred over baseline.\\
% \hline

\begingroup % Keeps the \footnotesize local to this table
\footnotesize
\rowcolors{2}{gray!15}{white}
\begin{longtable}{p{3.3cm} p{3.2cm} p{2.8cm} p{5.0cm}}
    \caption{Human/LLM Judgment Alignment with Prompt-based Summaries.} \label{tab:human-llm} \\
    
    % --- FIRST PAGE HEADER ---
    \rowcolor{gray!15} % Gray background only for the header row
    \toprule
    \textbf{Study (Citation)} & \textbf{Human/LLM Assessment Design} & \textbf{Level of Agreement} & \textbf{Key Findings} \\ 
    \midrule
    \endfirsthead

    % --- SUBSEQUENT PAGES HEADER ---
    \hiderowcolors % Ensure the "continued" line stays white
    \multicolumn{4}{c}{{\bfseries \tablename\ \thetable{} -- continued from previous page}} \\ \noalign{\smallskip}
    
    \rowcolor{gray!15} % Explicitly make the header row gray
    \toprule % Single rule to prevent double-line effect
    \textbf{Study (Citation)} & \textbf{Human/LLM Assessment Design} & \textbf{Level of Agreement} & \textbf{Key Findings} \\ 
    %\midrule
    \endhead
    \showrowcolors % Restart zebra-striping for the actual data rows

\cite{sun2024source} & 15 human evaluators (PhD + MS + UG developers) rating summaries on a 1–5 scale and GPT-4-based evaluation as an LLM-as-evaluator approach.& Strong alignment (GPT-4 eval best matches human scores); weak alignment(n-gram metrics).& GPT-4-based evaluation showed the strongest correlation with human judgments, validating it as a reliable proxy, while traditional text or semantic metrics proved inadequate and often mis-ranked models.\\ %LLM and human strong; human and NL weak ,HM->LMM: (1s), HM->NL: (1w)

\hline

\cite{sun2025commenting} & 5 software engineers(1–5 scale on readability, accuracy, informativeness, conciseness); GPT-4 as judge.& Strong correlation reported between GPT-4 and human scores.& GPT-4 evaluations reliably mirrored human judgments, showing LLMs can substitute human raters for large-scale summary assessment. \\ %LLM and human strong , HM->LLM:1(s)
\hline

\cite{yang2025docagent}& LLM-as-a-judge (with CoT) to assess qualities like helpfulness and completeness. & No human–LLM alignment analysis conducted.& It notes the significant improvement in performance over traditional baselines (FIM and Chat-based LLMs) across three specific dimensions: Completeness, Helpfulness, and Truthfulness.\\ % No check 
\hline

\cite{zhu2025reposummary} & Used GPT-5 and Claude 4 as primary judges(Relevance, Coverage) and if the two primary judges disagreed, DeepSeek V3.1 was introduced as a third ``tie-breaking '' judge and first and second authors (both with 5+ years of Java experience) manually evaluated.& Strong correlation (0.590 to 0.839) between the human scores and the LLM-judge scores.& Validated an LLM-as-a-judge framework that matches human accuracy at a fraction of the cost/time.\\ % %HM->LMM: (1s)

\hline

\cite{shin2023prompt} & 27 graduate students + 10 industry practitioners evaluated GPT-4 outputs through conversational prompts.&  Human ratings showed a consistent positive alignment with BLEU (+8.3 BLEU).& GPT-4 with task-specific prompting outperformed fine-tuned models. Human feedback prompting improved results, demonstrating that human-in-the-loop prompting enhances LLM performance. \\ %strong allignment HM->NL: 1(s)
\hline

\cite{kruse2024can} & 50 participants--20 professionals, 30 students rated summaries on correctness, readability, missing/unnecessary information, usefulness, helpfulness.& No quantitative alignment between automatic and human metrics was computed.& Ad-hoc prompting preferred by developers without prompting experience; predefined prompts yield more readable, concise, and consistent documentation preferred by students. \\ %NO
\hline

\cite{lomshakov2024proconsul} & Two expert raters annotated sufficiency, factuality, and hallucinations for 25 C/C++ functions. GPT-4o proxy metrics used for fast scoring; results cross-validated.& strong alignment( GPT-4o highly agree with humans: 94.6\% (triviality), 72\% (sufficiency)).& PROCONSUL summaries rated most accurate and complete, with lowest hallucination/factual error rate; expert acceptance high. \\ %LLM and human strong, HM->LLM:1(s)
\hline

\cite{zhou2023towards} & 12 experts rated summaries for informativeness and naturalness; compared meta-learning + retrieval (MLCS).& Strong alignment in positive correlation and consistency between human and automatic evaluations (stating that BLEU/ROUGE/METEOR trends align with human judgments).& Automatic metrics (BLEU-4~$\uparrow$112.7\%, ROUGE-L~$\uparrow$23.2\%, METEOR~$\uparrow$31.5\%) and human evaluation both confirmed that MLCS produced more contextually relevant and higher-quality summaries than baselines. \\ %strong, HM->NL:(1s) 
\hline

\cite{wang2024purpose} & ChatGPT used as an automatic evaluator for semantic accuracy.& ChatGPT-based evaluation correlated closely with metric outcomes, confirming consistent robustness improvements across datasets.& LLM-based scoring aligned with precision/F1 trends, confirming that iterative reasoning enhances summary correctness and resilience against code perturbations.\\ % strong between NL and LLM LLM->NL:(1s)
\hline 

\cite{dhulshette2025hierarchical} & LLM-as-a-judge (G-Eval), expert validation.&  Strong alignment with BERTScore and semantic similarity ($\approx$0.70–0.85), whereas their correlation with BLEU and ROUGE is notably weaker($\approx$0.25 – 0.45 (low agreement).& LLM-based (G-Eval) scores aligned strongly with semantic metrics but weakly with lexical ones, showing traditional BLEU/ROUGE fail to reflect real quality.  \\ %strong allignment ith llm and semantic and weak between LLM and NL.LLM->NL:1(s), LLM->NL:1(w)
\hline

\cite{imani2024context} & three practitioner surveys; > 30 participants with $\geq$ 2 years Java experience (Rationality, Comprehensiveness, Conciseness, Expressiveness).& Strong qualitative alignment between human preference rankings and automatic metrics.& Higher BLEU and METEOR scores aligned with higher human ratings, showing that semantic metrics reflected human preferences better than ROUGE-L.  \\ %strong, HM-> NL:1(s),
\hline

\cite{li2024only} & 4 experienced researchers rated 381 samples on four criteria (Rationality, Comprehensiveness, Conciseness, and Expressiveness) using a 5-point Likert scale and 21 Apache developers also evaluated and chose their preferred messages.& Strong alignment was observed between human ratings and OMG (a GPT-4 + ReAct prompting system), whereas alignment between human evaluations and automatic metrics (BLEU, ROUGE-L, and METEOR) was weak.& Prompt-based LLM summaries better match human quality standards and expectations, with rationality, completeness, and expressiveness judged superior.\\ %LLM and human strong; human and NL weak, HM->LMM: (1s), HM->NL: (1w)

\hline

\cite{zhang2024commit} & 2-phase: (1) Automated metrics on all; (2) Human eval by two authors, 366 samples.& Weak alignment between BLEU/ROUGE scores and human preferences.&  LLMs win overwhelmingly over human eval;	LLM-generated messages judged “best” in 78\% of cases, even preferred over human-written messages.  \\ %weak HM-> NL:1(w),HM->LLM:1(s)
\hline

\cite{wu2025empirical} & Three human evaluators(postgraduate students + industry professionals) rated on informativeness, conciseness, and expressiveness using a Likert scale, with GPT-4 used as an LLM-based evaluator.& Strong human–LLM (GPT-4) agreement observed, with weak alignment between human judgments and LLM with traditional NL metrics.& LLM-based evaluation provides a more reliable proxy for human judgment in commit message generation. \\ %HM->LMM: (1s), HM->NL: (1w)

\bottomrule

\end{longtable}
\endgroup
% \end{tabular}
% \end{table}
\bigskip
\noindent

At the same time, \tabref{tab:human-llm} highlights an important recurring limitation: \emph{traditional lexical metrics} (\eg BLEU/ROUGE) often correlate weakly with human preferences, even when LLM-as-a-judge performs well. For example, Sun \etal \cite{sun2024source} report that GPT-4-based evaluation aligns best with human scores, whereas n-gram metrics exhibit weak agreement and can mis-rank systems; similarly, Li \etal \cite{li2024only} observe strong alignment between human ratings and a GPT-4+ReAct judging setup, but only weak alignment with BLEU/ROUGE/METEOR. Dhulshette \etal \cite{dhulshette2025hierarchical} further reinforce this trend, showing that LLM-judge scores correlate strongly with semantic metrics such as BERTScore and embedding similarity, while agreement with BLEU/ROUGE is notably lower. Finally, the table also includes studies that \emph{do not} quantify human–LLM agreement \cite{yang2025docagent,kruse2024can}, underscoring that alignment analyses are still inconsistently reported despite their growing importance for interpreting evaluation results. 

This discrepancy motivates a two-step synthesis of evaluation practice. If lexical overlap can diverge from perceived quality, then it becomes essential (i) to map \emph{which} evaluation instruments the literature relies on (manual protocols, LLM-based judging, and automatic metrics), and (ii) to isolate the subset of studies that explicitly test whether automated scores are defensible proxies for human judgment.

The evaluation of code summarization is characterized by a diverse landscape of metrics and methodologies. \tabref{tab:benchmarks-metrics} provides an overall view of this landscape, summarizing the specific metrics utilized and indicating whether assessments were conducted manually, via LLMs, or through traditional automatic metrics. Building on this, \tabref{tab:human-llm} focuses specifically on correlation-oriented evaluation, examining the explicit relationship between human-provided ratings and automated judgments, including both LLM-based scoring and traditional NL metrics (\eg BLEU, ROUGE). By comparing these automated scores against human qualitative assessments of properties like readability and correctness, this table highlights the degree to which different automated methods truly align with human expertise.

Viewed through this lens, prompt design emerges as a central confounding factor. Prompts can alter summaries in ways that humans and LLM judges perceive as improvements—even when traditional metrics change only marginally—thereby shaping conclusions about ``quality'' depending on the chosen evaluation instrument.

From this characterization, studies consistently show that prompt strategies substantially influence both human and LLM-based evaluations of code summaries. Zhang \etal\cite{zhang2024commit} report that 78\% of evaluators preferred LLM-generated commit messages over human-written references despite only modest BLEU/ROUGE improvements, illustrating that surface-level overlap metrics fail to capture perceived quality. Similarly, Imani \etal\cite{imani2024context} showed that adding more contextual information to code diffs led humans to prefer summaries generated by open-source LLaMA models, indicating that relevant context can matter more than model size. Li \etal's OMG framework\cite{li2024only}, which combines ReAct prompting with GPT-4, achieved higher ratings for rationality, comprehensiveness, and expressiveness. Studies exploring structured prompt designs, such as Shin \etal\cite{shin2023prompt} and Sun \etal \cite{sun2025commenting}, report improvements in accuracy, readability, and overall human preference, while Li \etal \cite{li2024only} further demonstrates that explicit reasoning and feedback-driven prompting can enhance rationality, comprehensiveness, and expressiveness in code summaries. Across evaluation setups—from small expert panels to larger developer surveys, evaluators consistently prioritize relevance/usefulness, accuracy, and completeness, followed by fluency and informativeness. Less frequently assessed but still important criteria (\eg rationale/explanation, hallucination reduction, and style conformity) are closely tied to trustworthy summaries, reinforcing the need for evaluation frameworks that sits at the intersection of semantic fidelity and surface/token-based overlap.

\begin{tcolorbox}[
    colback=gray!15,
    colframe=black,
    arc=6pt,
    boxrule=0.8pt,
    left=5pt,
    right=5pt,
    top=8pt,
    bottom=8pt,
    fonttitle=\bfseries,
    coltitle=black,
    title=Summary of Results for \RQ{4}:,
    enhanced,
    attach boxed title to top left={yshift=-3mm, xshift=5mm},
    boxed title style={
        colback=gray!40,
        boxrule=0.7pt,
        arc=8pt,
        outer arc=8pt,
        left=5pt,
        right=5pt,
        top=0.5pt,
        bottom=0.5pt,
    }
]
Overall, prompt-based code summarization is evaluated mostly on function-level benchmarks and a narrow set of languages (primarily Python/Java): \textit{CodeSearchNet} and \textit{CodeXGLUE} appear in over half of the studies, while only a few datasets capture richer commit-, class-, or repository-level context. Automated metrics dominate, especially BLEU-4 ($\approx 80\%$), followed by ROUGE-L ($\approx 55\%$) and METEOR ($\approx 40\%$). Human or LLM-as-judge evaluation is used in 14/29 studies ($\approx 48\%$), but it often diverges from overlap scores (strong alignment in $\approx 43\%$ vs.\ weak/inconsistent alignment in $\approx 57\%$), reflecting that humans prioritize relevance, accuracy, and completeness, whereas overlap metrics largely capture lexidcal similarity. As a result, reported gains can be sensitive to benchmark and metric choice, and may not reliably translate to trustworthy summaries in realistic development settings.
\end{tcolorbox}

\begin{tcolorbox}[
    colback=cyan!8,
    colframe=black,
    arc=6pt,
    boxrule=0.8pt,
    left=5pt,
    right=5pt,
    top=8pt,
    bottom=8pt,
    fonttitle=\bfseries,
    coltitle=black,
    title=Future Directions for \RQ{4}:,
    enhanced,
    attach boxed title to top left={yshift=-3mm, xshift=5mm},
    boxed title style={
        colback=gray!40,
        boxrule=0.7pt,
        arc=8pt,
        outer arc=8pt,
        left=5pt,
        right=5pt,
        top=0.5pt,
        bottom=0.5pt,
    }
]
Future evaluations should broaden \emph{both} coverage and measurement. On the coverage side, benchmarks should extend beyond Python/Java and beyond function-level inputs by curating larger datasets that capture commit-, class-, and repository-level context, enabling more realistic assessment of prompts that exploit project history, structural dependencies, and retrieved documentation. On the measurement side, studies should reduce over-reliance on lexical overlap by adopting more semantically grounded and task-aware metrics (\eg embedding- or contrastive-based measures such as SIDE) and by standardizing human and LLM-as-judge protocols with explicit rubrics for relevance, factuality/faithfulness to code behavior, and completeness. Bringing these dimensions together--multilingual, multi-granularity benchmarks paired with consistent, multi-dimensional evaluation pipelines--would make comparisons across prompt-based code summarizers more reliable and would better predict real developer utility in practice.
\end{tcolorbox}
\subsection{\RQ{5}:What is the state of reproducibility and resource sharing in prompt-based code summarization research? }
\label{sec:RQ5}

\tabref{tab:resource-sharing} summarizes our reproducibility and resource-sharing analysis. Prompt-driven code summarization pipelines are inherently multi-stage (data curation and preprocessing, prompt construction, model configuration, and evaluation), and empirical outcomes can be highly sensitive to seemingly minor implementation choices. Small inconsistencies in dataset filtering, split construction, decoding settings, prompt templates, or metric scripts can translate into non-negligible variance in reported results. This sensitivity makes methodological transparency--and the public release of executable artifacts--a key item for credible comparison and cumulative progress in the field.

Across the 29 primary studies in \tabref{tab:resource-sharing}, 20 studies ($\approx 69\%$) provide at least one publicly available replication resource (\eg a GitHub repository or a Zenodo package), while 9 studies ($\approx 31\%$) provide no usable artifacts. Within this latter group, 4 studies share no link at all, and 5 provide links that are inaccessible, incomplete, or not reliably usable at the time of our inspection.\footnote{The systematic inspection of artifact availability and link integrity was performed in late October 2025}.

Even among the 20 studies with accessible resources, \tabref{tab:resource-sharing} suggests that ``artifact availability'' does not necessarily imply end-to-end reproducibility. Full replication is often constrained by missing or underspecified details, including: (i) learning procedures and optimization settings (loss functions, schedulers), (ii) hyperparameters (architectural choices, batch size, number of epochs, initialization, early stopping), (iii) data preprocessing and filtering rules (deduplication, tokenization, leakage controls), (iv) environment configuration (OS/driver/toolchain, pinned library versions, random seeds), and (v) evaluation commands and implementations (exact metric variants, prompts used for judging, and run scripts). Moreover, studies relying on commercial LLM APIs face additional reproducibility threats--non-determinism, model drift, version changes, and rate limits--which can prevent faithful replication even when code is shared. Overall, the results in \tabref{tab:resource-sharing} indicate that while most studies now provide \emph{some} resources, the field would benefit from stronger artifact standards (\eg containerized environments, pinned dependencies and fixed seeds) to ensure that prompt-based improvements are verifiable and comparable.

% \begin{table*}[ht]
% \footnotesize
% \caption{\footnotesize Resource Sharing and Accessibility of Prompt-based Code Summarization Studies.}
% \label{tab:resource-sharing}
% \rowcolors{2}{gray!15}{white}
% \centering
% \begin{tabular}{|p{3.8cm}|p{7cm}|p{2.0cm}|}
\begin{table*}[ht]
\small
\caption{\small Resource Sharing and Accessibility of Prompt-based Code Summarization Studies.}
\label{tab:resource-sharing}
\rowcolors{2}{gray!15}{white}
\centering
\setlength{\tabcolsep}{8pt}      % default ~6pt; increase if you like
\renewcommand{\arraystretch}{1.15}% row height
\begin{tabular}{|p{4.2cm}|p{9.0cm}|p{1.6cm}|}
\hline
 \textbf{Citation} & \textbf{Link} & \textbf{Accessible} \\
\hline

    Sun \etal\cite{sun2024source} & \url{https://github.com/wssun/LLM4CodeSummarization} & \cmark \\
\hline

    Sun \etal~\cite{sun2025commenting} & \url{https://github.com/wssun/LLM4ModuleSum} & \cmark \\
\hline

     Yang  \etal \cite{yang2025docagent} & \url{https://github.com/facebookresearch/DocAgent} & \cmark \\
\hline

    Ahmed \etal\cite{ahmed2024automatic} & \url{https://doi.org/10.5281/zenodo.7779196} & \cmark \\
\hline

    Sun \etal\cite{sun2023automatic} & \textcolor{red}{No public repo or dataset shared} & \xmark \\
\hline

    Zhu \etal \cite{zhu2025reposummary} & \textcolor{red}{No public repo or dataset shared } & \xmark \\

\hline

    Geng \etal\cite{geng2024large} & \url{https://github.com/gmy2013/LLM_Comment_Generation} & \cmark\\
\hline

    Wang \etal\cite{wang2024natural} & \url{https://github.com/gksajy/slimcode} & \cmark \\

\hline
    Chai \etal\cite{chai2022ernie}  & \url{https://github.com/PaddlePaddle/PaddleNLP/tree/develop/model_zoo/ernie-code} & \xmark \\
\hline 

    Shin \etal\cite{shin2023prompt} & \url{https://github.com/shinjh0849/gpt4-ase-tasks} & \xmark \\
\hline

    Khan \etal~\cite{khan2022automatic} & \url{https://github.com/disa-lab/CodeDoc_GPT3_ASE22}  & \xmark \\
\hline

    Fried \etal~\cite{fried2022incoder} & \url{https://sites.google.com/view/incoder-code-models/} & \cmark \\
\hline

    Arakelyan \etal\cite{arakelyan2023exploring} & \url{https://github.com/ShushanArakelyan/code_shift} & \xmark\\
\hline

    Makharev \& Ivanov \etal\cite{makharev2025code} & \url{https://github.com/kilimanj4r0/code-summarization-beyond-function-level} & \cmark \\
\hline

    Kruse \etal\cite{kruse2024can} & \url{https://zenodo.org/doi/10.5281/zenodo.13127237} & \cmark \\
\hline

    Zhang \etal\cite{zhang2024attacks} & \textcolor{red}{No public repo or dataset shared } & \xmark\\ 
\hline

    Ahmed \etal~\cite{ahmed2022few} &  \url{https://doi.org/10.5281/zenodo.6592064} & \cmark \\
\hline

    Zhang \etal \cite{zhang2025dlcog} & \url{https://github.com/Zhang20000807/DLCoG} & \cmark\\
\hline

     Gao \etal \cite{gao2023makes} & \url{https://github.com/shuzhenggao/ICL4code} & \cmark \\
\hline

     Yun \etal\cite{YUN2024112149} & \url{https://github.com/Linshuhuai/P-CodeSum} & \cmark \\
\hline

    Lu \etal\cite{lu2024instructive} & \url{https://github.com/kingofheven/ICR} & \cmark \\
\hline

     Lomshakov \etal\cite{lomshakov2024proconsul} & \url{https://github.com/trinity4ai/ProConSuL} & \cmark \\
\hline

    Zhou \etal\cite{zhou2023towards} & \url{https://github.com/zy-zhou/MLCS} & \cmark \\

\hline

    Wang \etal \cite{wang2024purpose} & \textcolor{red}{No public repo or dataset shared } & \xmark \\
\hline

    Dhulshette \etal~\cite{dhulshette2025hierarchical} & \url{https://github.com/microsoft/near-duplicate-code-detector} & \cmark \\
\hline

     Imani \etal~\cite{imani2024context} & \url{https://github.com/aaron-imani/omega} & \cmark \\
\hline

    Li \etal\cite{li2024only} & \url{https://github.com/ucirvine-se/OMG} (code, dataset, context, eval scripts) & \xmark\\
\hline

    Zhang \etal\cite{zhang2024commit} & \url{http://doi.org/10.5281/zenodo.10164356} & \cmark\\
\hline

    Wu \etal \cite{wu2025empirical} & \url{https://github.com/wuyifan18/LLM4CMG} & \cmark \\
\hline

\end{tabular}
\end{table*}

\begin{figure}[htbp]
    \centering
    \includegraphics[width=1.0\textwidth]{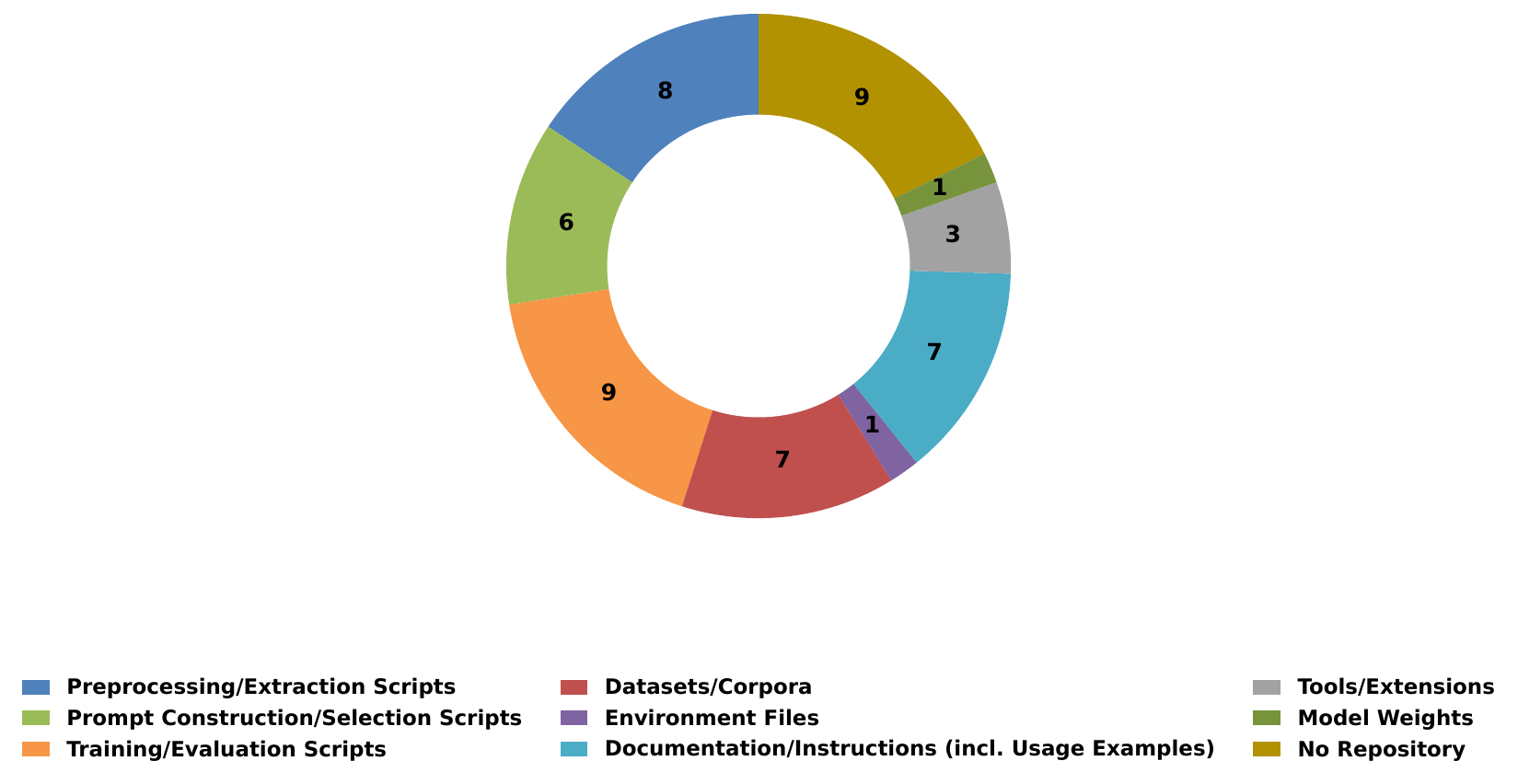}
    \caption{Artifact-sharing landscape across 29 studies.
Preprocessing and training scripts are the most frequently shared artifacts, followed by datasets and documentation resources.}
    \label{fig:replication}
\end{figure}

To assess the reproducibility landscape, we analyze the artifacts shared across the 29 primary studies. \figref{fig:replication} summarizes the availability of datasets, scripts, and related resources. In addition, \tabref{tab:resource-sharing} reports—on a per-study basis—the publicly accessible links we were able to retrieve from each paper included in our SLR.

The resulting picture is mixed: complete replication packages remain the exception, but several recent papers have begun to release end-to-end implementations and evaluation assets, indicating a clear shift toward stronger reproducibility practices.

For instance, Yang \etal \cite{yang2025docagent}, Zhang \etal \cite{zhang2025dlcog}, and Wu \etal \cite{wu2025empirical} the authors make an effort to release publicly accessible implementations, including source code, runnable experiment scripts, and evaluation components that support end-to-end execution of their methods. In addition, Zhang \etal \cite{zhang2025dlcog} provide the complete annotated dataset and experimental results via an external Google Drive link due to size constraints, while Wu \etal \cite{wu2025empirical} share result figures and dataset links hosted on Zenodo and external platforms.
Geng \etal \cite{geng2024large} release pre-processing scripts and several prompting strategies for comment generation.
Sun \etal \cite{sun2023automatic} provide a source-code summarization package that includes datasets, detailed steps for building Erlang, Haskell, and Prolog corpora, and evaluation scripts.
The ICL4Code project \cite{gao2023makes} offers preprocessing routines, prompt-construction scripts, and evaluation programs for Codex, GPT-3.5, and ChatGPT.
Similarly, Yun \etal (P-CodeSum) \cite{YUN2024112149} share data-generation scripts, a project-specific prompt selector, and run scripts for training and evaluation.
Ahmed \etal \cite{ahmed2022few} upload a Zenodo archive containing language-specific datasets and result folders, along with scripts for extracting data-flow graphs and identifiers.
Makharev \etal \cite{makharev2025code} release preprocessed datasets, notebooks, and scripts for class- and repository-level summarization, together with a conda environment file.
Lomshakov \etal (ProConSuL) \cite{lomshakov2024proconsul} provide installation instructions and scripts for two-phase training, prediction, and evaluation.
Kruse \etal \cite{kruse2024can} share data from their controlled experiment, a Visual Studio Code extension, and presentation materials.
Finally, Zhou \etal (MLCS) \cite{zhou2023towards} publish scripts for preprocessing data, training a bi-LSTM summarizer, performing retrieval, and running the meta-learning model.

As anticipated, despite recent progress, a non-trivial portion of studies still lack a public or accessible repository, highlighting persistent challenges for reproducibility and transparency in prompt-based code summarization. Reported issues range from releasing only partial artifacts—such as a test script to reproduce the reported results without the corresponding training code \cite{lu2024instructive}, to sharing tooling components (\eg a tool and language-specific tokenizers) while omitting the dataset \cite{dhulshette2025hierarchical}.

On the other extreme, reproducibility is largely absent: approximately 33\% (9/29) of the studies provide no usable replication package. In some cases, the provided links resolve to missing or private repositories (\eg \cite{shin2023prompt,li2024only}), while others explicitly state that neither code nor data are shared \cite{sun2023automatic,wang2024purpose,zhang2024attacks,zhu2025reposummary}.

We also observed link rot: for instance, Chai \etal~\cite{chai2022ernie} reference ERNIE-Code in the PaddleNLP model zoo, but the specific directory cited is no longer accessible, limiting reuse. Finally, even when repositories are available, key components are sometimes missing; \eg Ahmed \etal~\cite{ahmed2022few} release scripts and datasets but omit the preprocessing pipeline and baseline configurations/outputs (\eg CodeBERT/CodeT5), which hinders end-to-end replication. Overall, these gaps--often exacerbated by reliance on proprietary APIs--continue to constrain reproducibility in the field.

\begin{figure}[htbp]
    \centering
    \includegraphics[width=0.60\textwidth]{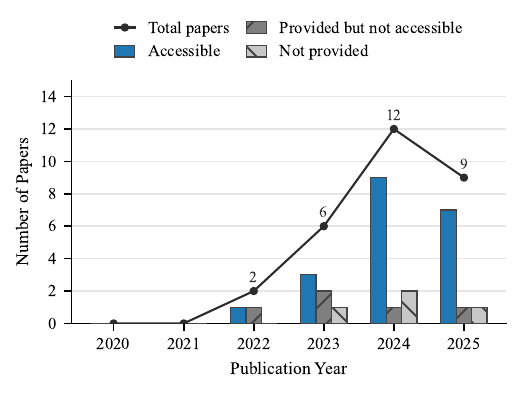}
    \caption{Year-wise availability of replication packages in primary studies (2020--2025). Bars show \textit{Accessible}, \textit{Provided but not accessible}, and \textit{Not provided} packages; the line indicates the total number of papers.}
    \label{fig:replication-availability}
\end{figure}

\figref{fig:replication-availability} further illustrates year-wise trends in replication artifacts availability.
In 2022, two studies were published, of which one ($50\%$) released accessible replication packages, while the remaining one ($50\%$) provided repository links that were no longer accessible.
In 2023, six studies were identified: three ($50\%$) shared accessible artifacts, two ($33\%$) included inaccessible links, and one ($17\%$) did not provide any replication package.
A notable improvement is observed in 2024, where nine of twelve studies ($75\%$)  made their code and associated resources publicly accessible, one ($\approx8\%$) provided an unreachable link, and two ($\approx17\%$) shared no artifacts.
In 2025, seven of nine studies ($\approx78\%$) released accessible replication packages, while one study ($\approx11\%$) provided an unreachable link and one ($\approx11\%$) did not provide a public implementation.

The increasing trend in artifact availability since 2024 reflects a growing commitment to reproducibility in LLM-based research. However, the fact that $31\%$ of surveyed studies (nine in total) still lack accessible resources indicates that open sharing practices are not yet universal. To bridge this gap, future work must prioritize the use of stable repositories (\eg GitHub, Zenodo) and provide sufficient documentation to support independent reproduction of results.

\begin{tcolorbox}[
    colback=gray!15, % background color
    colframe=black, % border color
    arc=6pt, % corner rounding
    boxrule=0.8pt, % border thickness
    left=5pt, % left padding
    right=5pt, % right padding
    top=8pt, % top padding
    bottom=8pt, % bottom padding
    fonttitle=\bfseries,
    coltitle=black, % title font
    title=Summary of Results for \RQ{5}:,
    enhanced,
    attach boxed title to top left={yshift=-3mm, xshift=5mm},
    boxed title style={
        colback=gray!40,
        boxrule=0.7pt,
        arc=8pt,
        outer arc=8pt,
        left=5pt,
        right=5pt,
        top=0.5pt,
        bottom=0.5pt,
    }
]

Our investigation highlights a positive shift toward open science in the code summarization community. While approximately 69\% of all surveyed studies provide accessible artifacts, the most significant finding is the recent momentum: accessibility rates have surged from 50\% in 2022 to over 85\% in 2025. Despite this progress, reproducibility remains fragile due to ``link rot'' (defunct URLs) and omission of essential experimental components, such as raw data preprocessing pipelines and specific prompt templates. We conclude that while artifact sharing is becoming a standard expectation, the long-term stability and completeness of these repositories are crucial for ensuring independent verifiability.
\end{tcolorbox}

\begin{tcolorbox}[
    colback=cyan!8, % background color
    colframe=black, % border color
    arc=6pt, % corner rounding
    boxrule=0.8pt, % border thickness
    left=5pt, % left padding
    right=5pt, % right padding
    top=8pt, % top padding
    bottom=8pt, % bottom padding
    fonttitle=\bfseries,
    coltitle=black, % title font
    title= Future Directions for \RQ{5}:,
    enhanced,
    attach boxed title to top left={yshift=-3mm, xshift=5mm},
    boxed title style={
        colback=gray!40,
        boxrule=0.7pt,
        arc=8pt,
        outer arc=8pt,
        left=5pt,
        right=5pt,
        top=0.5pt,
        bottom=0.5pt,
    }
]
% Looking forward, we identify two primary directions to strengthen reproducibility. First, the community needs clear, standardised guidelines defining the essential components of a replication package (including source code, datasets, prompts and preprocessing scripts) and specifying how these artifacts should be documented and archived; existing models such as structured ``ID‑Cards'' from other subfields could be adapted for prompt engineering. Second, long‑term accessibility must be prioritised: researchers should deposit artifacts in stable, DOI‑backed repositories (\eg Zenodo), ensure that private links are avoided, and provide enough detail to reconstruct experiments even when proprietary APIs are involved. 

To strengthen reproducibility, the community should first adopt standardized reporting that treats prompt templates as essential artifacts; while model configurations are common, specific prompt phrasing is often missing. Second, long-term accessibility must be prioritized by moving from unstable links to DOI-backed repositories like Zenodo. For studies using proprietary APIs, providing ``prompt-response logs'' is vital to counter model non-determinism. By following such practices, future studies can make prompt‑engineering research more transparent, which will accelerate progress in automated code summarization.
\end{tcolorbox}
\section{Threats to Validity}

Although our systematic literature review adheres to known and well-established guidelines by Kitchenham \etal \cite{kitchenham2007guidelines}, we have to acknowledge the presence of potential threats that could hinder the results.

\textbf{Construct Validity.} This threat relates to the accuracy of our identification and categorization of prompt engineering strategies. While we applied open coding and conducted multiple rounds of verification, alternative categorizations remain possible. Moreover, some inconsistencies may persist due to incomplete methodological descriptions in certain primary studies, particularly concerning how prompting techniques were reported.

\textbf{Internal Validity.} This threat concerns whether our findings accurately capture underlying trends without being distorted by our study selection or data extraction procedures.  To this extent, we queried six major digital libraries and employed snowballing to capture additional studies--yet--it is possible that relevant work was missed, particularly papers published in venues outside software engineering or using alternative terminology.  We partially mitigated this risk by designing inclusive queries and validating them through a trial-and-error process, but some omissions may remain.  

A different threat can arise from the manual coding we performed for the 29 included studies--specifically, the strategy taxonomy, granularity labels, evaluation metrics, and artifact availability. In this case, we might have introduced subjectivity. To reduce bias, both authors independently reviewed studies and reached consensus through discussion, following established SLR guidelines \cite{kitchenham2007guidelines}.

\textbf{External Validity.} This threat relates the generalizability of our findings. In this regard, we must underline that our synthesis reflects the state of research between 2020 and 2025 (October).
Given the rapid pace of advancements in LLMs and prompt engineering, future studies may quickly outdate our findings.  Nevertheless, by consolidating evidence from this formative period, we provide a baseline reference that can be extended and updated in subsequent reviews.  

We also recognize that our venue selection is not flawless, as certain studies may have been missed due to indexing issues or limitations in the search queries we designed to retrieve relevant research. To address this, we supplemented our search with snowballing and manual inclusion of influential studies which minimizes the risk.
\section{Conclusion}

In this paper, we presented the first systematic literature review of prompt engineering for code summarization. By analyzing 29 studies published between 2020 and 2025 in top-tier venues in software engineering and machine learning, we developed a taxonomy of prompt engineering strategies, examined their effectiveness across models and code granularities, and assessed reproducibility and artifact sharing practices.   Our findings indicate that while prompt engineering can significantly enhance the quality of LLM-based code summarization, the field remains fragmented, with limited standardization in evaluation, reporting, and resource sharing.  
Addressing these gaps will require community-wide efforts toward reproducible experimentation and benchmark development.
Looking forward, we encourage researchers to explore advanced prompting strategies (\eg self-consistency, retrieval augmentation, and instruction tuning) while rigorously evaluating them against both human judgments and practical development needs.   For practitioners, our synthesis underscores the value of prompt engineering as a lightweight yet powerful paradigm to distill information from LLMs for documentation and maintenance tasks. Ultimately, by consolidating the fragmented body of research in prompt-based code summarization, we build for the community a common ground on which further research can be conduced.

\bibliography{main}

\end{document}